\documentclass[%
aps,pre,%
 amsmath,amssymb,
reprint,%
superscriptaddress,showpacs]{revtex4-1}

\usepackage{color, graphicx}
\usepackage{dcolumn}
\usepackage{bm}
\usepackage{amssymb}
\usepackage{latexsym}
\usepackage{amsfonts}
\usepackage{amsmath}
\usepackage{multirow}
\usepackage{ifthen}
\usepackage{float}
\begin{document}

\title{Chaos and Anderson Localisation in Disordered Classical Chains:\\ Hertzian vs FPUT models}

\author{A.~Ngapasare}
\author{G.~Theocharis}
\author{O.~Richoux}
\affiliation{ Laboratoire d'Acoustique de l'Universit\'e du Maine, UMR CNRS 6613 Av. O. Messiaen, F-72085 LE MANS Cedex 9, France}
\author{Ch.~Skokos}
\affiliation{Department of Mathematics and Applied Mathematics, University of Cape Town, Rondebosch 7701, South Africa
}
\affiliation{Max Planck Institute for the Physics of Complex Systems, N\"othnitzer Str.~38, D-01187 Dresden, Germany}
\author{V.~Achilleos}
\affiliation{ Laboratoire d'Acoustique de l'Universit\'e du Maine, UMR CNRS 6613 Av. O. Messiaen, F-72085 LE MANS Cedex 9, France}

\begin{abstract}
We numerically investigate the dynamics of an one-dimensional disordered lattice using the Hertzian model, describing a granular chain, 
and the $\alpha+\beta $ Fermi-Pasta-Ulam-Tsingou model (FPUT). The most profound difference between the two systems is the discontinuous nonlinearity of the granular chain appearing whenever neighboring particles
are detached. We therefore sought to unravel the role of these discontinuities in the destruction of Anderson localization 
and their influence on the system's chaotic dynamics.
Our results show that both models exhibit an energy range where localization coexists with chaos. 
However, the discontinuous nonlinearity is found to be capable of triggering energy spreading of initially localized modes, at lower energies than the FPUT model. A transition from Anderson localization to chaotic dynamics and energy equipartition
is found for the granular chain and is associated with the``propagation" of the discontinuous nonlinearity in the chain.
On the contrary, the FPUT chain exhibits an alternate behavior between localized and delocalized  chaotic behavior
which is strongly dependent on the initial energy of excitation.
\end{abstract}

\maketitle

\section{Introduction}
The role of nonlinearity in disordered systems which exhibit Anderson localization~\cite{pwand,anderson2}
 is a topic that triggered a vast amount of theoretical, numerical~\cite{disordernon1,disordernon2,disordernon3,disordernon4,flachbook} and experimental 
 studies~\cite{ christoexp,Page, Page2, gran}. The two principal questions under consideration are (a) does the energy carried by 
localized wave-packets eventually spread or not and (b) what is the route to equipartition?

Among different nonlinear models, large theoretical work and progress has been made especially for the Klein-Gordon (KG) system and the Discrete Nonlinear Schr$\ddot{\text{o}}$dinger (DNLS) equation with disorder. 
For these systems, it has been found that the combined influence of disorder and nonlinearity 
leads to sub-diffusive energy transport~\cite{disordernon3}. It is also now understood that whether nonlinear Anderson localization
persists or is destroyed has probabilistic features and is directly associated with chaos~\cite{FlachProb}. Additionally a variety 
of different physical settings have been exploited in order to study this interplay between nonlinearity and disorder,
especially in optical and atomic systems~\cite{christoexp,skipetrov18}. 

Recently a classical lattice i.e.~the granular chain described by the Hertzian contact force~\cite{hertzbook}, has 
also attracted much attention in the same context~\cite{pikovskygran,vassospre2016,kev2016,Sen2017,kev2018,vassospre2018}. The considerable interest in the Hertzian chain can be attributed to the strong nonlinearity of the system which is however easily tuned (usually by the pre-compression of the chain). The Hertzian contact forces also  allow access to wave propagation in an almost linear system up to the case of a lattice where only nonlinear waves propagate (``sonic vacuum")~\cite{chapter, chiaropanos}. 
An additional interesting dynamical feature of the granular chain is that the power law nonlinearity, due to the Hertzian force, coexists with a non-smooth nonlinearity describing detached particles~\cite{granular, poly,bookgran, col}.
Recent works on both uncorrelated and correlated disorder granular chains showed that the system traverses 
from a sub-diffusive regime for sufficiently weak nonlinearities to a super-diffusive regime 
for increasing nonlinearity~\cite{kev2018}.
In strongly disordered granular chains it was found that localization coexists with chaos and 
equipartition is reached for finite times~\cite{vassospre2018}.

Furthermore, the granular chain in the weakly nonlinear regime provides an experimental setting to study
the Fermi-Pasta-Ulam-Tsingou (FPUT) model with both $\alpha$ and $\beta$ type terms~\cite{los,fpu2}. In contrast to the DNLS and KG models, which have been studied for disordered systems, 
the FPUT system has been mostly studied in the homogeneous case, although some studies regarding disorder also
exist e.g. in Refs.~\cite{fpudis1,fpudis2,fpudis3}.
%

In fact, the phenomenon of equipartition for the homogeneous case is a long standing problem, which originates from the pioneer work of Fermi, Pasta, Ulam and Tsingou~\cite{los,fpu2},
although substantial progress has been made on the subject~\cite{FPUEqui, Mula}. 
Very recent studies both in $\alpha$-FPUT (but also in the KG model) periodic lattices showed that the
thermalization is reached through high order resonant interactions leading to large timescales for equipartition~\cite{onorato1, 
onorato2}. It was also found that the fluctuations of the entropy after the system reaches equipartition are characterized by sticky 
dynamics close to $q$-breathers for the FPUT model and discrete breathers for the KG model~\cite{danieli}.

In this work we aim to expose the role of different nonlinearities in
the destruction of Anderson localization, the chaoticity of the system but also the timescales to reach equipartition.
To do so we perform a detailed comparison
between the granular chain model and the FPUT system. 
We numerically study a strongly disordered configuration with the same linear limit for both models. Our goal is to identify the mechanisms that lead to energy spreading of
an initially excited localized mode. We use chaos indicators~\cite{BGGS, S10} to quantify the total systems' chaotic behaviour.
Additionally we provide information about chaos propagation in the lattice enabling us to 
differentiate localized and extended chaos.
By tracking the mode distribution during the dynamics' evolution we provide insights regarding equipartition. 

The paper is organized as follows. In Section II we introduce the two models and establish the disorder strength for the lattice which ensures strongly localized modes in the linear limit.  Selecting a single configuration
and focusing on a highly localized mode near the center of the lattice, in Section III we study the mode's evolution in both models for 
increasing excitation energy. A thorough analysis  of the lattice dynamics is performed focusing on the spreading of the initially localized mode, on the chaoticity of the system and on monitoring the appearance of particle detachments. Finally we show results illustrating how and for which energy the two systems reach energy equipartition. In section IV we summarize our findings and discuss their significance.

\section{Hertzian and FPUT models with disorder}
Both models studied here, namely the granular chain with Hertzian interactions and the FPUT system,
are considered to be energy preserving (i.e. without losses). Their total energy for a chain with $N$ spherical homogeneous beads of radius $R_n$ and 
mass $m_n$ ($n=1,2,3,\ldots, N$) is given by the following Hamiltonian
\begin{eqnarray}
H=\sum_{n=1}^{N} H_{_{n}} = \sum_{n=1}^{N}\frac{p_{_{n}}^2}{2m_{_{n}}} +V_{n}^{(Hz,F)}.
\label{Hamlinear}
\end{eqnarray} 
Here, $p_n=m_n\dot{u}_n$ and $u_n$ 
denote respectively the momentum and displacement from equilibrium for each particle, ($\dot{ }$)  denotes the first order time derivative, while
the random radii $R_n$ are uniformly chosen in the interval $ [ \min(R_n), \max(R_n) ] $. 

The Hertzian potential $V_n^{Hz}$ for each bead due to the nearest neighbor coupling is defined as $V_{_{n}}^{Hz}=[V^{Hz}(u_{_{n}})+V^{Hz}(u_{_{n+1}})]/2$ where
\begin{equation}
\begin{aligned}
&V^{Hz}(u_{_{n}})=\frac{2}{5}A_{_{n}}[\delta_{_{n}}+u_{_{n-1}}-u_{_{n}}]_+^{5/2}-\frac{2}{5}A_{_{n}}\delta_{_{n}}^{5/2} \\
&-A_{_{n}}\delta_{_{n}}^{3/2}(u_{_{n-1}}-u_{_{n}}).
\end{aligned}
\label{Vhz}
\end{equation}
The static overlap $ \delta_n$  between two neighboring beads $ n-1$ and $ n$ is given by $ \delta_n = (F_0/ A_n)^{2/3} $ 
where $F_0$ is the pre-compression force. The coefficient  $ A_n $ for spherical beads is given as $ A_n  = (2/3) \varepsilon \sqrt{ R_{n-1} R_n /( R_{n-1}  + R_n  )  } /(1- \nu^2 )  $  where $ \varepsilon$ and $\nu $ are the elastic modulus and the Poisson ratio respectively~\cite{hertzbook}.
The plus sign in $ [ \cdot ]_{+}$ describes the fact that this term is present as long as
$\delta_n+u_{n-1}-u_n>0$ and is absent otherwise, since then the particles are no longer in contact. This is the non-smooth nonlinearity which substantially
differentiates the two models.

The FPUT model we study is described by  Eq.~(\ref{Hamlinear}) with a potential 
\begin{equation}
V^{F}(u_{_{n}}) =  \sum_{k=2}^{4}  K^{(k)}_{_{n}} (u_{_{n}} - u_{_{n-1}})^{k} .
\label{fputmodel}
\end{equation}
%
Accordingly, the  potential of the $n^{th}$ particle is written as $V_{n}^F=[V^{F}(u_{_{n}})+V^{F}(u_{_{n+1}})]/2$.

For the rest of this work we consider a chain of $ N = 40 $ particles. In our simulations we choose units corresponding to a mean radius of $\bar{R}=0.01$m, and a static pre-compression force $F =1$N. 
The mean radius is used as a reference to the uniform system with particles of radius $ R = ( \alpha + 1)\bar{R}/2 $. The disorder strength, is quantified by the parameter 
$ \alpha =  \max(R_n) / \min(R_n) $.
This choice of disorder naturally leads to a random distribution of both the masses and stiffness coefficients~\cite{vassospre2016}. 
In all calculations we use fixed boundary 
conditions with dummy beads on both ends such that $u_{_{0}}=u_{_{N+1}}=0$ and $p_{_{0}}=p_{_{N+1}}=0$. 
The corresponding equations of motion for the  Hertzian model ~(\ref{Vhz}) are
\begin{equation} \label{eq2}
\begin{split}
m_{_{n}} \ddot{ u}_{_{n}}  & =  A_{_{n}} [ \delta _{_{n}} +  u_{_{n-1}}  - u_{_{n}} ] _{+}^ {\frac{3}{2}} - A_{_{n+1}} [ \delta _{_{n+1}} +   u_{_{n}}  - u_{_{n+1}}  ] _{+}^ {\frac{3}{2}}
\end{split}
\end{equation}
%
while for the FPUT model we obtain
\begin{equation} \label{fput}
\begin{split}
m_{_{n}} \ddot{ u}_{_{n}}  & = \sum_{k=2}^{4} \big[ K^{(k)}_{_{n+1}}  (u_{_{n+1}} - u_{_{n}})^{k-1} - K^{(k)}_{_{n}} (u_{_{n}} - u_{_{n-1}})^{k-1} \big]. \\
\end{split}
\end{equation}
In order to establish a connection between the two models we note that
for sufficiently small displacements i.e. $u_n/\delta_{n,n+1}\ll 1$, the Taylor series expansion of Eq.~(\ref{eq2}) up to fourth order terms leads to Eq.~(\ref{fput}) with coefficients $K^{(2)}_{_{n}} = (3/2) A_{_{n}} \delta _{_{n}} ^{1/2} $,  $K^{(3)}_{_{n}} = -({3}/{8}) A_{_{n}} \delta _{_{n}} ^{-1/2}$ and  $K^{(4)}_{_{n}} = (3/48) A_{_{n}} \delta _{_{n}} ^{-3/2}$~\cite{chapter}.
We normalize our units such that for the linear homogeneous chain with $\alpha=1$ the frequency cut-off is $\omega_{\rm max}=\sqrt{\frac{4 K}{m}} = 1$ with $ K = K^{(2)}_{_{n}} $ and $m = m_{_{n}} = 1$.

\subsection{Linear mode analysis of the disorder chain}

In order to highlight the similarities and differences between the granular and the FPUT disordered models, we choose
to focus on strongly localized modes. This allows us to monitor 
less degrees of freedom, at least for the initial evolution of the dynamics. To identify such localized modes,  
we first perform an analysis of the linearized equation of motion
\begin{eqnarray}
m_{_{n}} \ddot{ u}_{_{n}}  &= K^{(2)}_{_{n+1}} (u_{_{n+1}} - u_{_{n}}) -  K^{(2)}_{_{n}} (u_{_{n}} - u_{_{n-1}})
\end{eqnarray}
which is common for both models. Assuming harmonic solutions of the form
$\mathbf{U} (t) = \mathbf{U_{0}} e^{i\omega t}$, where $\mathbf{U} _0$ is a column matrix with elements 
${U}_n$, $n=1,2,3,\ldots, N$. We then solve the corresponding eigenvalue problem
%
\begin{equation} \label{eigen}
- \omega^2 \mathbf{M U_0 }  = \mathbf{K} \mathbf{U_0}.
\end{equation}
The matrix $ \mathbf{M }$  is a diagonal matrix with elements $m_n$ and $ \mathbf{K }$ is a sparse diagonal matrix 
containing the stiffness coefficients $K^{(2)}_n$.
To quantify the localization properties of the disorder system, we calculate the participation number~\cite{flachbook} of the wave-packet $ P = 1/ \sum h_n^2 $ where $h_n = H_n / H $.
This quantity is defined in a way so that its maximum value equals the total number
of particles (extended mode) and its minimum value equals to 1 when only one particle is participating in a mode (strongly
localized mode).
\begin{figure} 
\includegraphics[width=8cm]{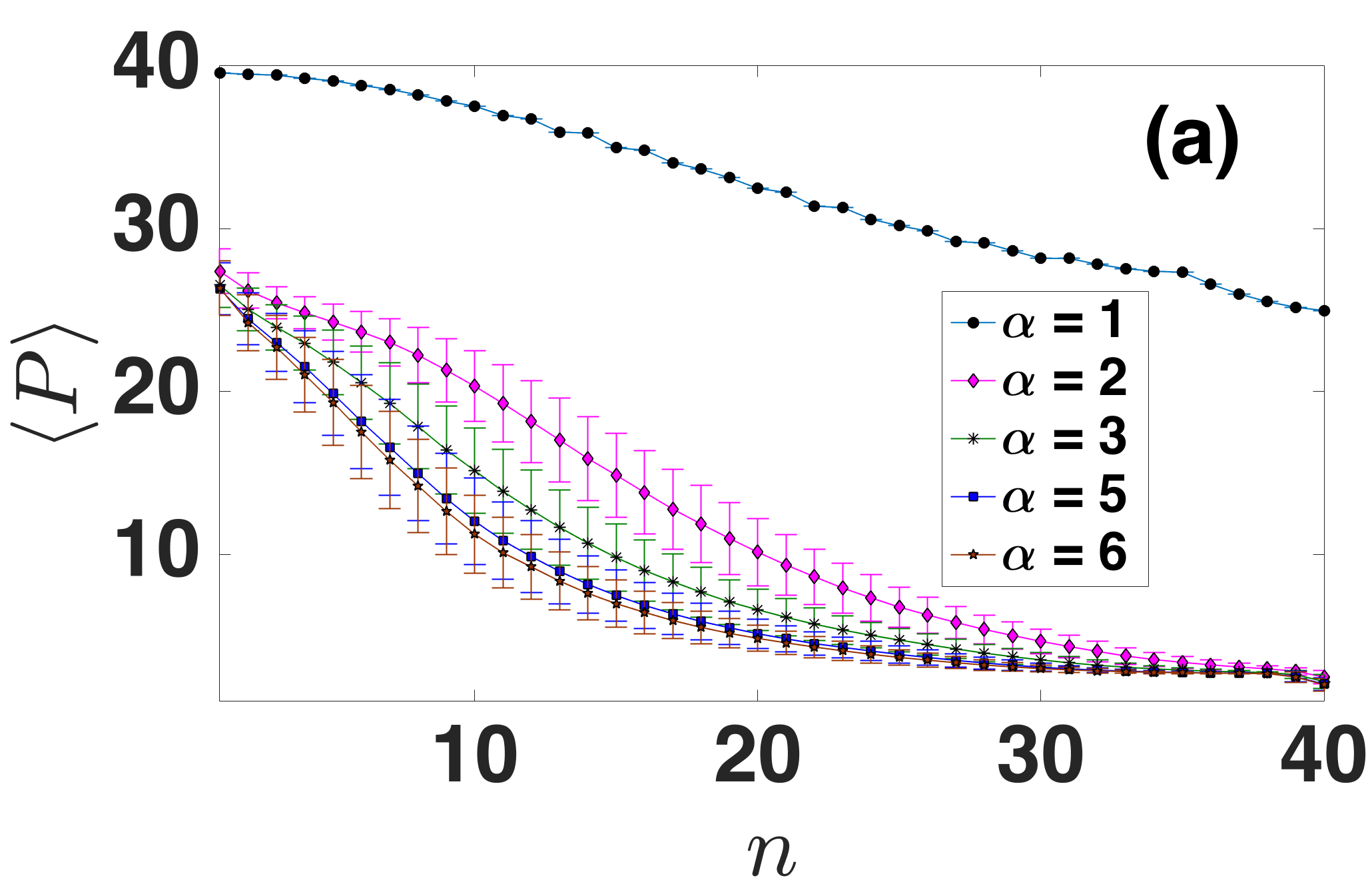}\\
\includegraphics[width=8cm]{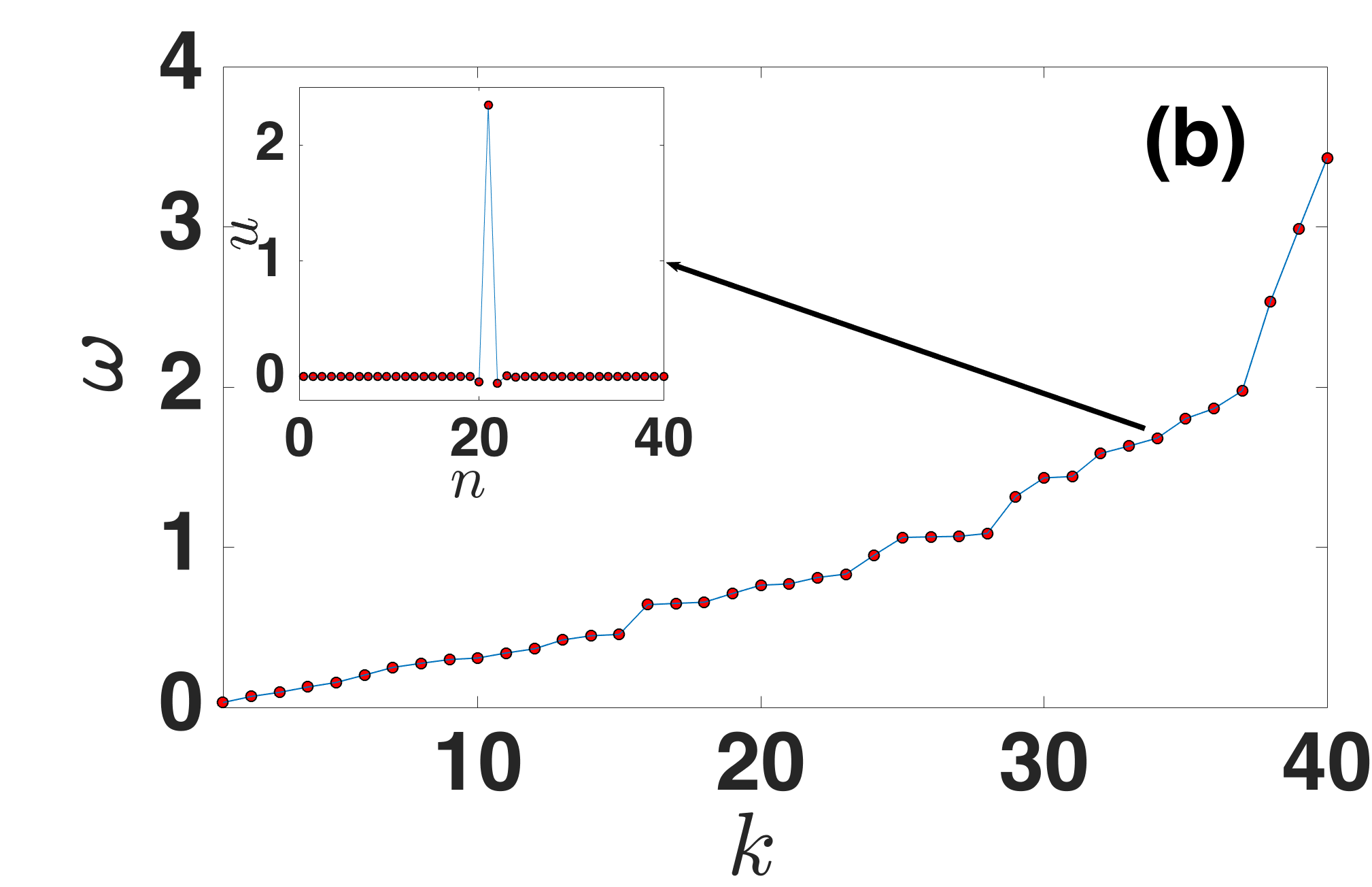}
\caption{ (a) Mean (over 1000 disorder realizations) participation number $\langle P \rangle$ of the eigenmodes 
for varying disorder strengths $ \alpha$, sorted in descending order $ k $ for each realization. 
The standard deviation at each point is shown by the error bars. 
(b) The eigenfrequencies of a particular disordered chain of $ 40 $ sites for $ \alpha = 5$ sorted by increasing frequency. The insert shows the profile of the $ 34^{th}$ mode.}
\label{stats1}
\end{figure}

In Fig.~\ref{stats1}(a), we show the mean value $\langle P \rangle$ of the participation number of the eigenmodes for different disorder strengths,
using an ensemble of $1000$ disorder realizations. The modes are sorted with descending values of
$ P $  for each realization. We first note that even for relatively weak disorder (e.g. for $\alpha=2$) $\langle P \rangle$ largely
deviates for the homogeneous case ($\alpha = 1$) and localized modes appear in the system.
On the other hand for values of $\alpha \geq 4$ the averaged participation number reaches a limiting curve
with about $10$ strongly localized modes with  $\langle P \rangle \approx 2$. 
According to the above analysis, a single disorder realization with 
$\alpha=5$ is sufficient for the chain to possess several strongly localized modes.

The eigenfrequencies corresponding to one particular disorder realization for $ \alpha = 5$, is 
shown in Fig.~\ref{stats1}(b). We remind the reader that generally low frequency modes extend over many particles,
whilst the high frequency modes are localized. The numerical simulations in the following
section will be performed for this particular disorder realization. More specifically, we identify the $ 34^{th} $ mode of this
chain as a strongly localized  mode ($P \approx 2.5$) around the middle of the chain and in particular at site $n=21$.
 The profile of this  mode is shown in the inset of Fig.~\ref{stats1}(b).

\begin{figure}
\includegraphics[width=4.2cm]{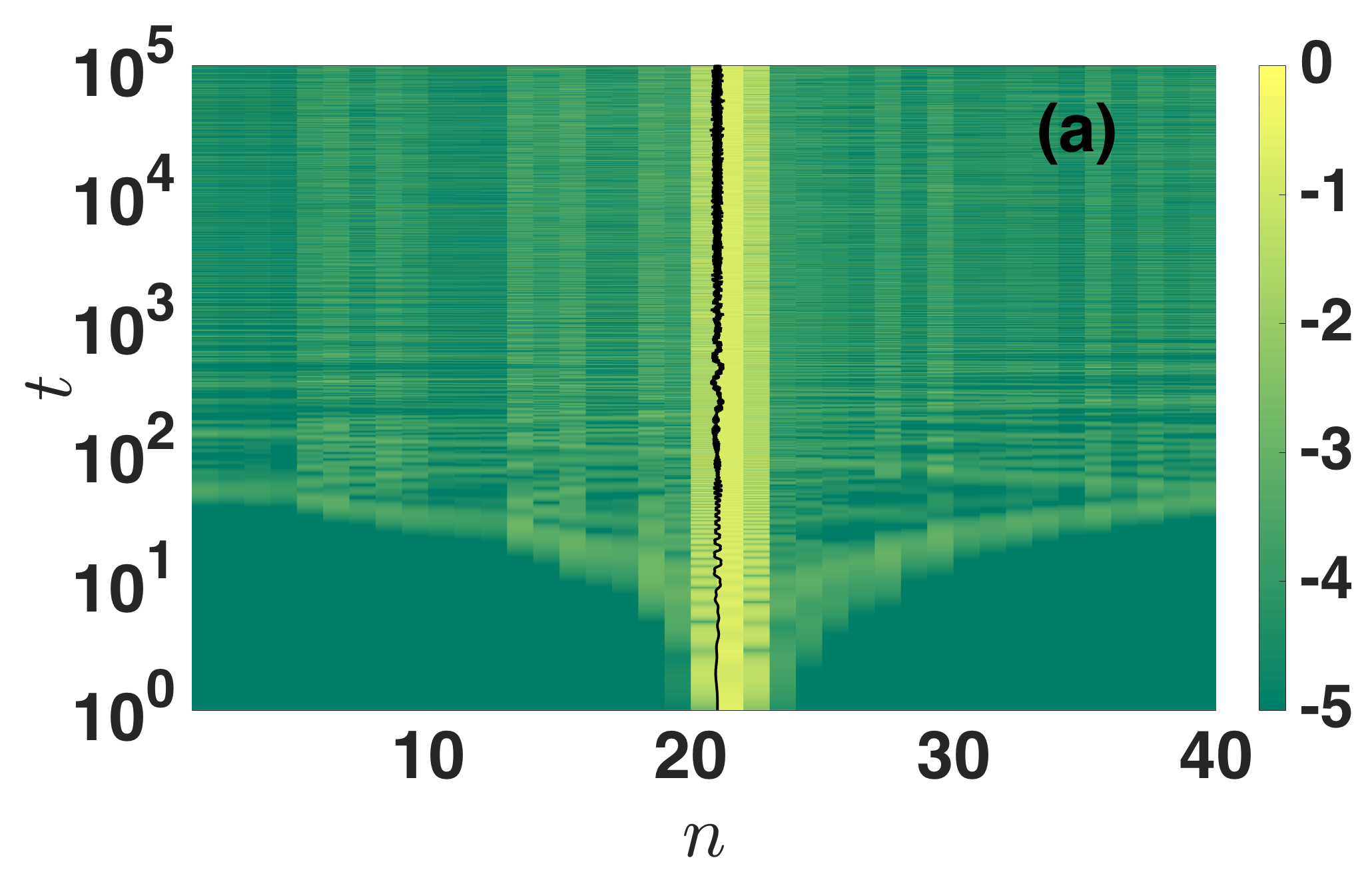}
\includegraphics[width=4.2cm]{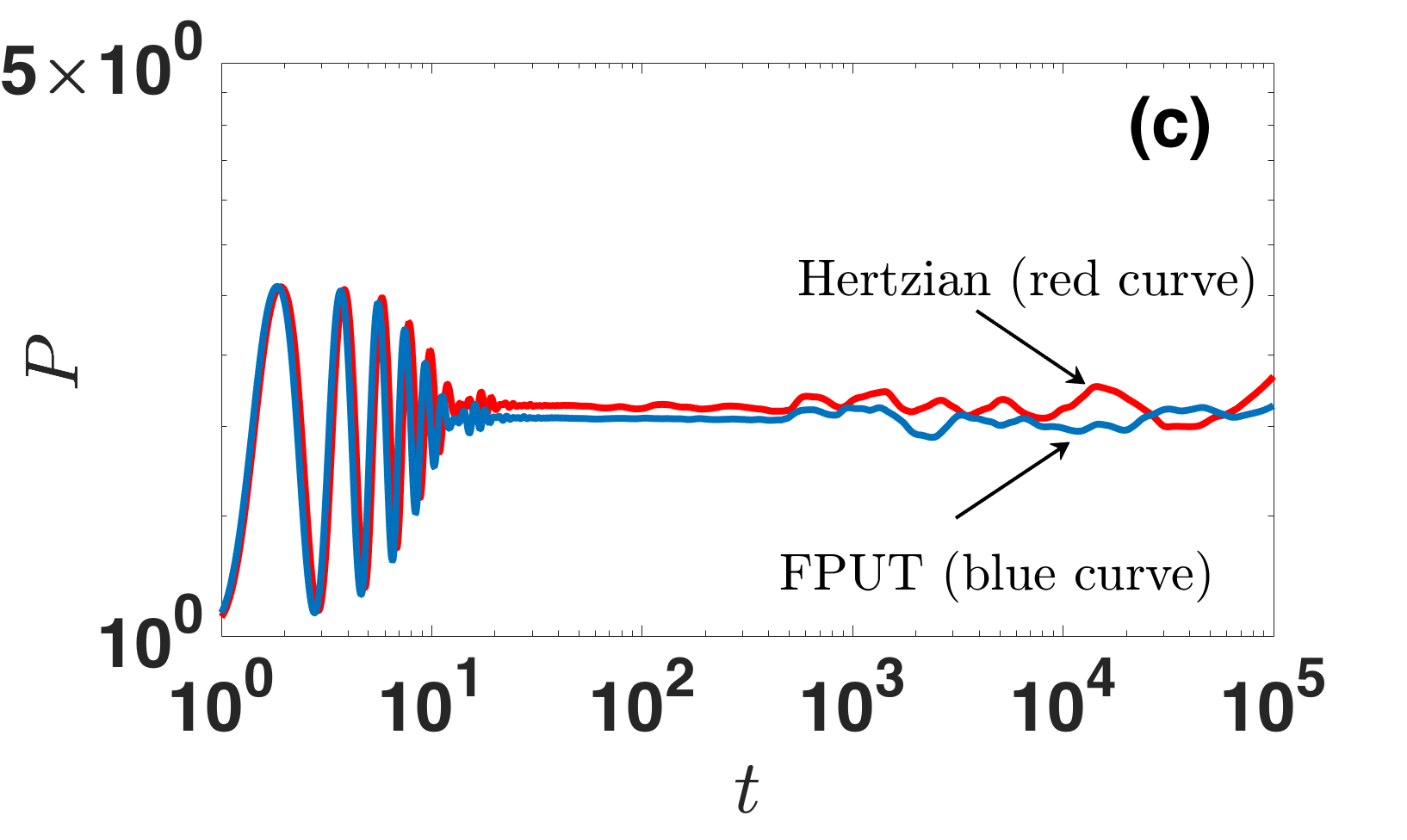}
 \includegraphics[width=4.2cm]{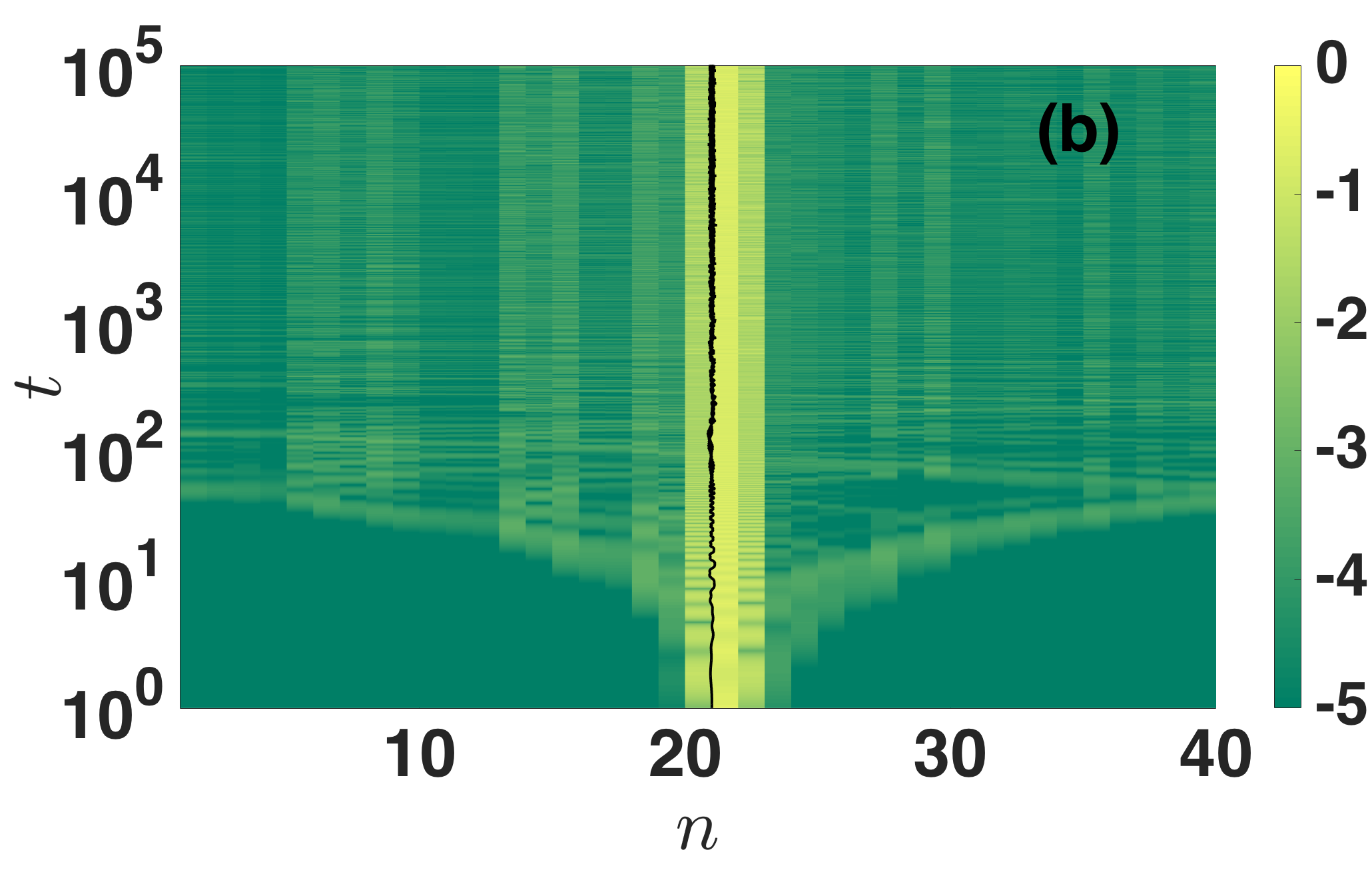} 
 \includegraphics[width=4.2cm]{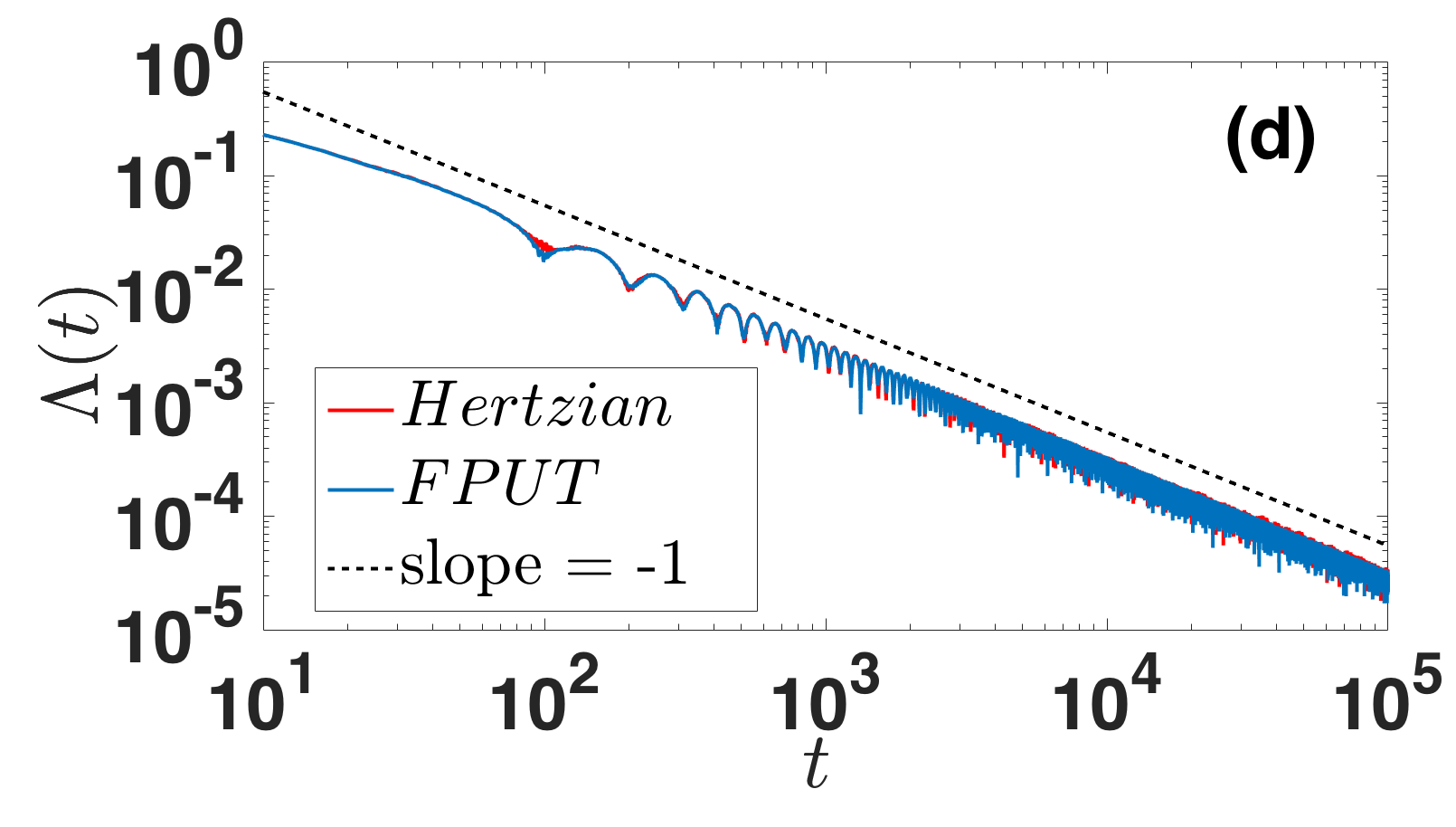} 
\caption{(a) and (b) The spatiotemporal evolution of the energy distribution for the Hertzian and FPUT chains respectively for $H=0.25$. The black curves indicate the running  mean position of the energy distributions.
The color bars on the right sides of (a) and (b) are in logarithmic scale.
(c) The locally weighted smoothed values of $P$ as a function of time for the Hertzian chain (red curve)  and the FPUT chain (blue curve). (d) The time evolution of $ \Lambda (t) $ for the Hertzian chain (red curve) and the FPUT chain (blue curve). Both lines practically overlap and the dashed line indicates the law $\Lambda(t) \propto t^{-1}$.
}
\label{dynms025a}
\end{figure}

In our numerical simulations below we use a single site initial excitation of the $21^{st}$ site, which results in the excitation of almost only the $34^{th}$ mode for sufficiently small energy $H$. 
By increasing the initial displacement we study the energy spreading due to nonlinearity as described by 
the time evolution of the energy density $ h_{n} $, and the participation number $ P $.
At the same time we identify and quantify chaos in the system using the maximum Lyapunov characteristic exponent (mLCE)~\cite{BGGS,S10},
which is obtained by numerically integrating the corresponding variational equations~\cite{tmap}.
The two sets of equations where integrated using the so called ``Tangent Map'' method with a fourth order optimal integration scheme with a marching step of $5\times10^{-4}$ in all our simulations \cite{tmap,Int1}.
The variational equations govern, at  first order of approximation, the time evolution of a deviation
vector $\vec{v}(t)=[\delta u_1, \delta u_2, \ldots, \delta u_N, \delta p_1, \delta p_2, \ldots, \delta p_N]$
where  $\delta u_n$, $\delta p_n$, $n=1,2,\ldots, N$ are respectively small perturbations in positions and momenta (see e.g.~\cite{S10}).

%

The mLCE  is given by  $\lambda=\underset{t\rightarrow \infty}{{\rm lim}}\Lambda(t)$, where
\begin{equation} \label{eqmlce}
\Lambda(t) = \frac{1}{t}  \ln \frac{|| v(t) ||}{|| v(0) ||},
\end{equation}
 is the so-called finite time mLCE~\cite{S10}.
Note that in Eq.~(\ref{eqmlce}), $|| \cdot ||$ denotes the usual Euclidean vector norm. For chaotic orbits, $\Lambda(t)$ eventually converges to a positive value, while for regular orbits it tends to zero following the power law $\Lambda(t)\propto t^{-1}$~\cite{S10}.

In order to gain more insight about the spatial properties of chaos in our system, we also 
calculate the deviation vector density (DVD) which is calculated as
\begin{equation} \label{eqdvd}
w_n = \frac{\delta u_n^2 + \delta p_n^2}{\sum_n (\delta u_n^2 + \delta p_n^2)}.
\end{equation}
The deviation vectors are known to align with the most unstable region in phase space.
They have been employed in disorder nonlinear lattices in order to visualize the spatial 
evolution of the most chaotic regions~\cite{SGF13, Bob2018, malcolmDNA}. 
We will employ the use of the DVDs in order to characterize the chaoticity of our system either as
localized or extended chaos. The initial condition used for the deviation vectors $\vec{v}(0)$ is a random uniform distribution of momentum pertubations $\delta p_i$
as for this choice, the time evolution of the finite time mLCE was found to converge faster to the  $\Lambda(t)\propto t^{-1}$ law for regular orbits.

\section{Dynamical evolution of an initially localized mode }

\subsection{Near linear limit}
For sufficiently small energies $H$, we have numerically confirmed that the two models 
behave both qualitatively and quantitatively the same. An example is given in Fig.~\ref{dynms025a}
which corresponds to $H = 0.25$. In panels (a) and (b) we notice that the energy 
density for both models is completely localized around the initially 
excited site $ n =21 $ as shown by the black solid line which indicates the mean position of the energy distribution.
This observation is complemented  by the time evolution of the participation number which gives a constant value of $ P \approx 1.8 $ for both models as 
shown in Fig.~\ref{dynms025a}(c). The curves of the Hertzian (red curve) and the FPUT (blue curve) models almost overlap.
The time evolution of $ \Lambda (t) $ is depicted in Fig.~\ref{dynms025a}(d) and confirms that the dynamics is regular as $ \Lambda (t) $ follows
 the power law $\Lambda(t)\propto t^{-1}$. 
 
\begin{figure}
 \includegraphics[width=4.2cm]{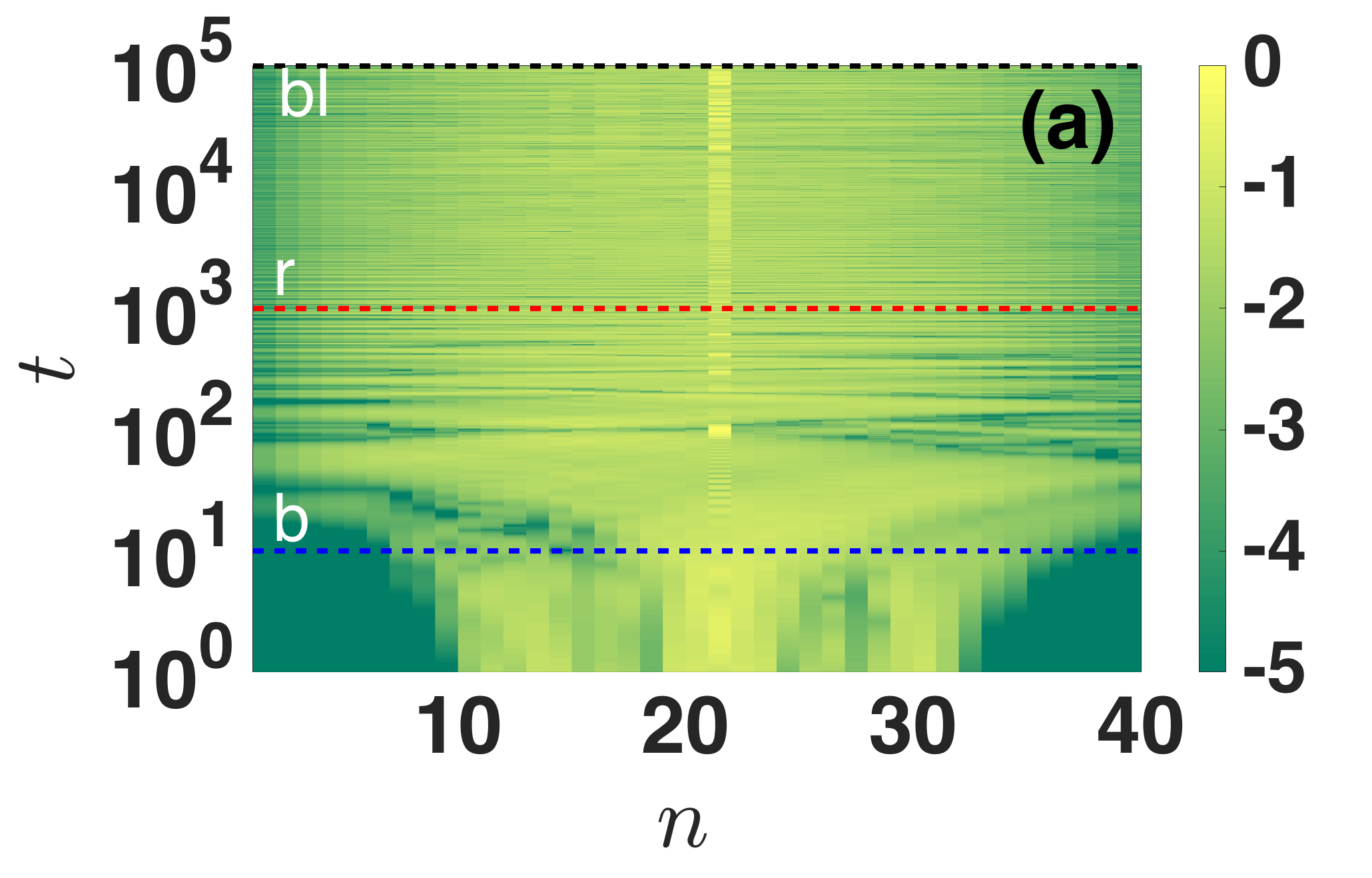} 
\includegraphics[width=4.2cm]{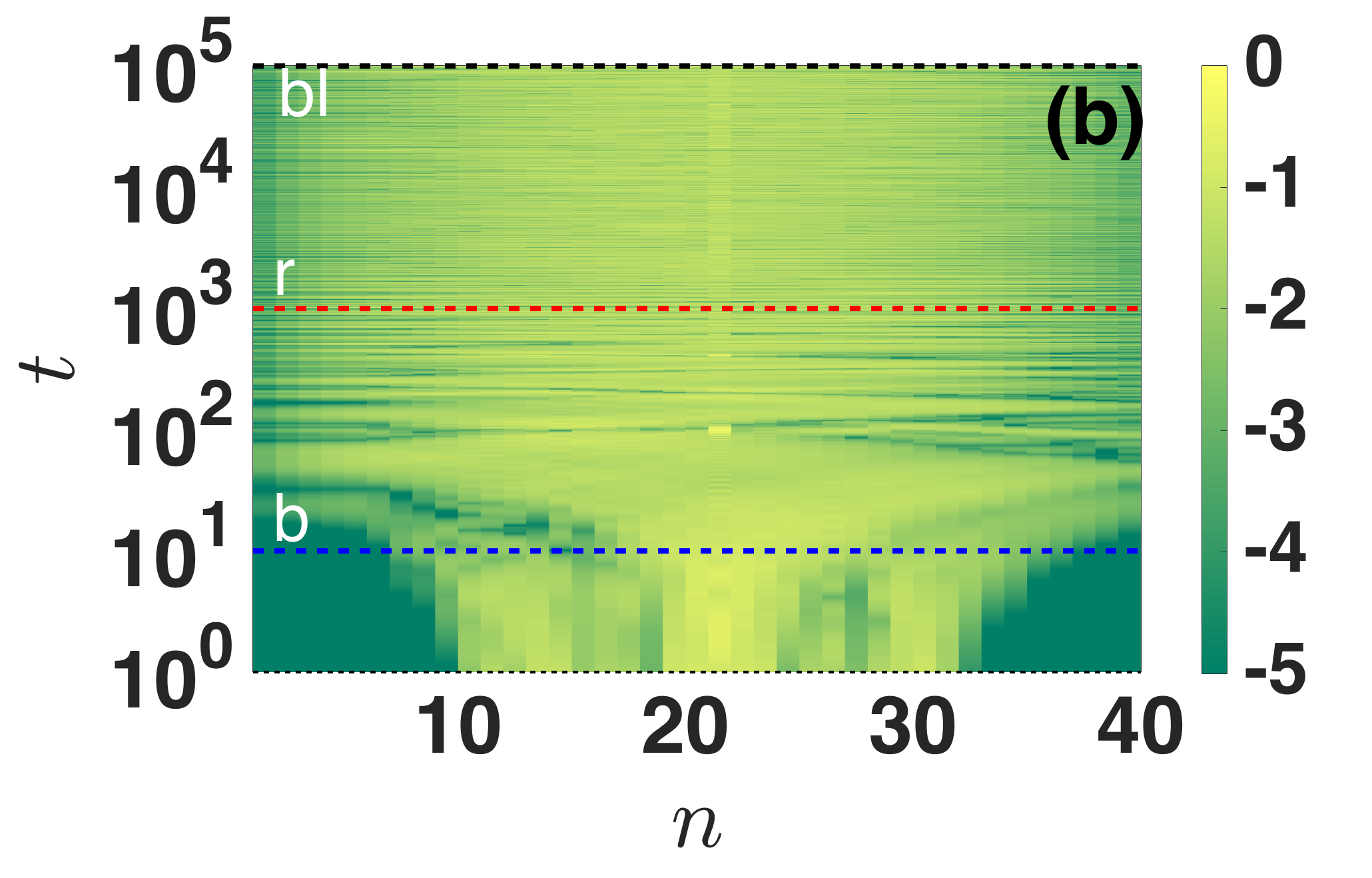} \\
 \includegraphics[width=4.2cm]{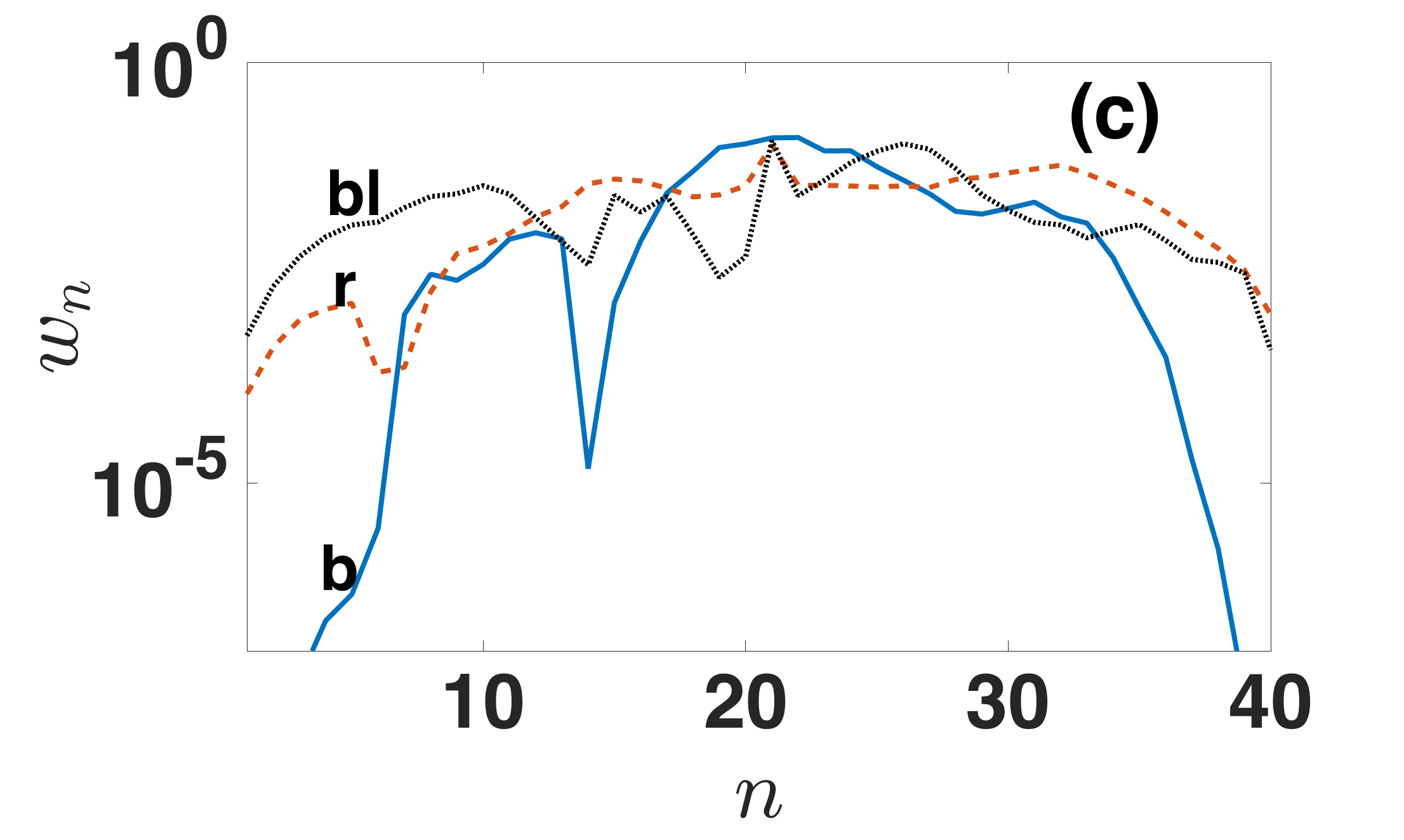}
 \includegraphics[width=4.2cm]{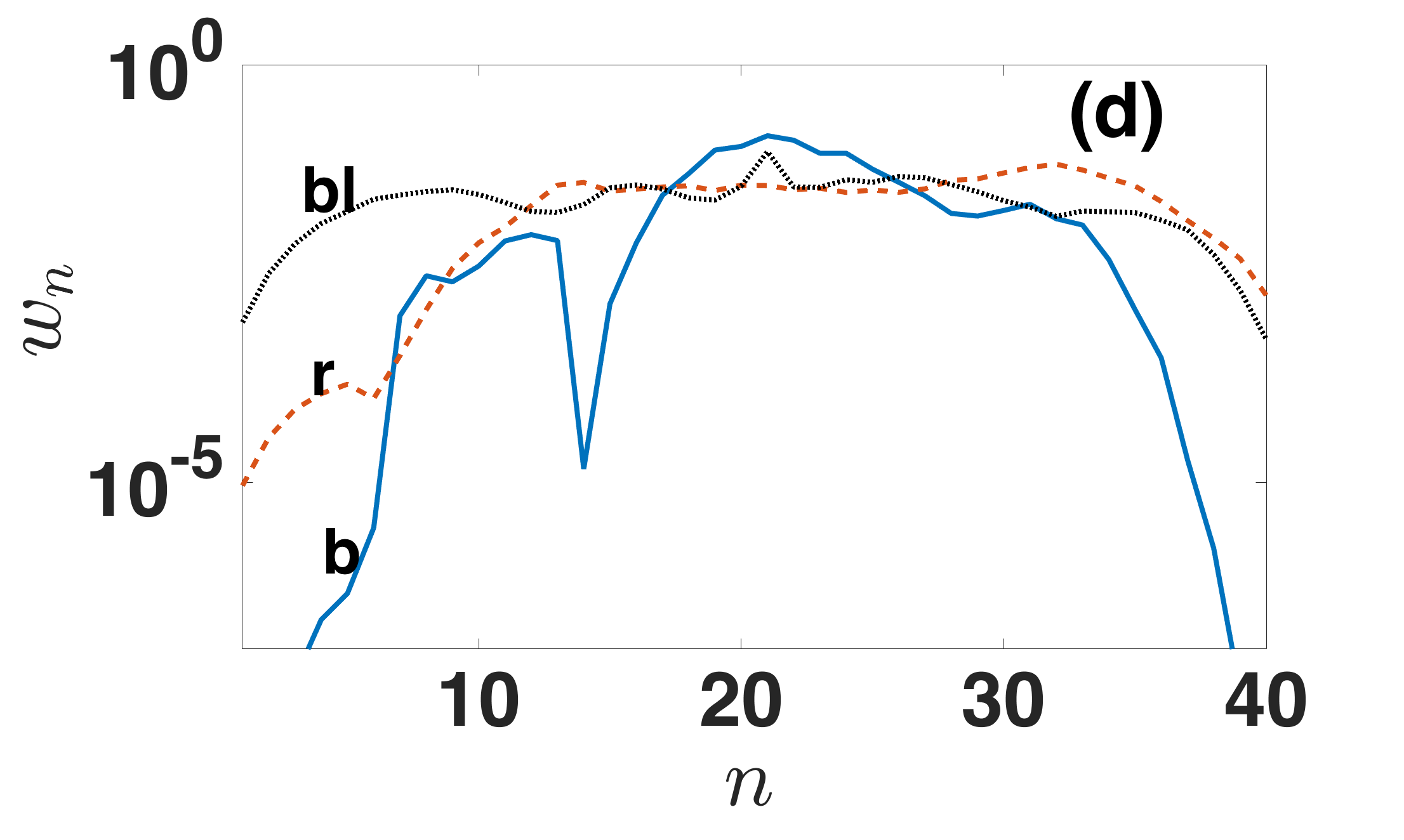}\\
\includegraphics[width=4.2cm]{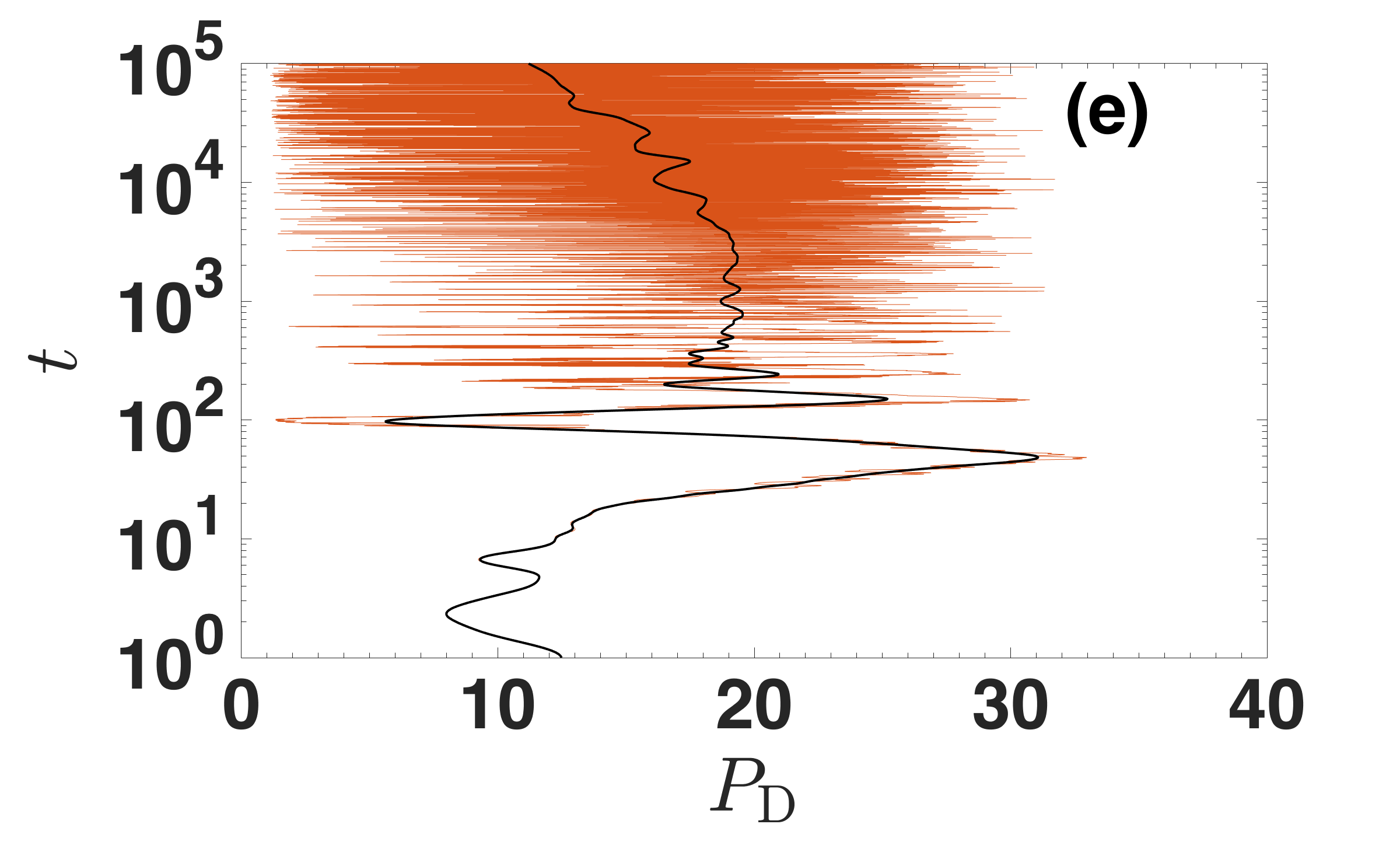}
 \includegraphics[width=4.2cm]{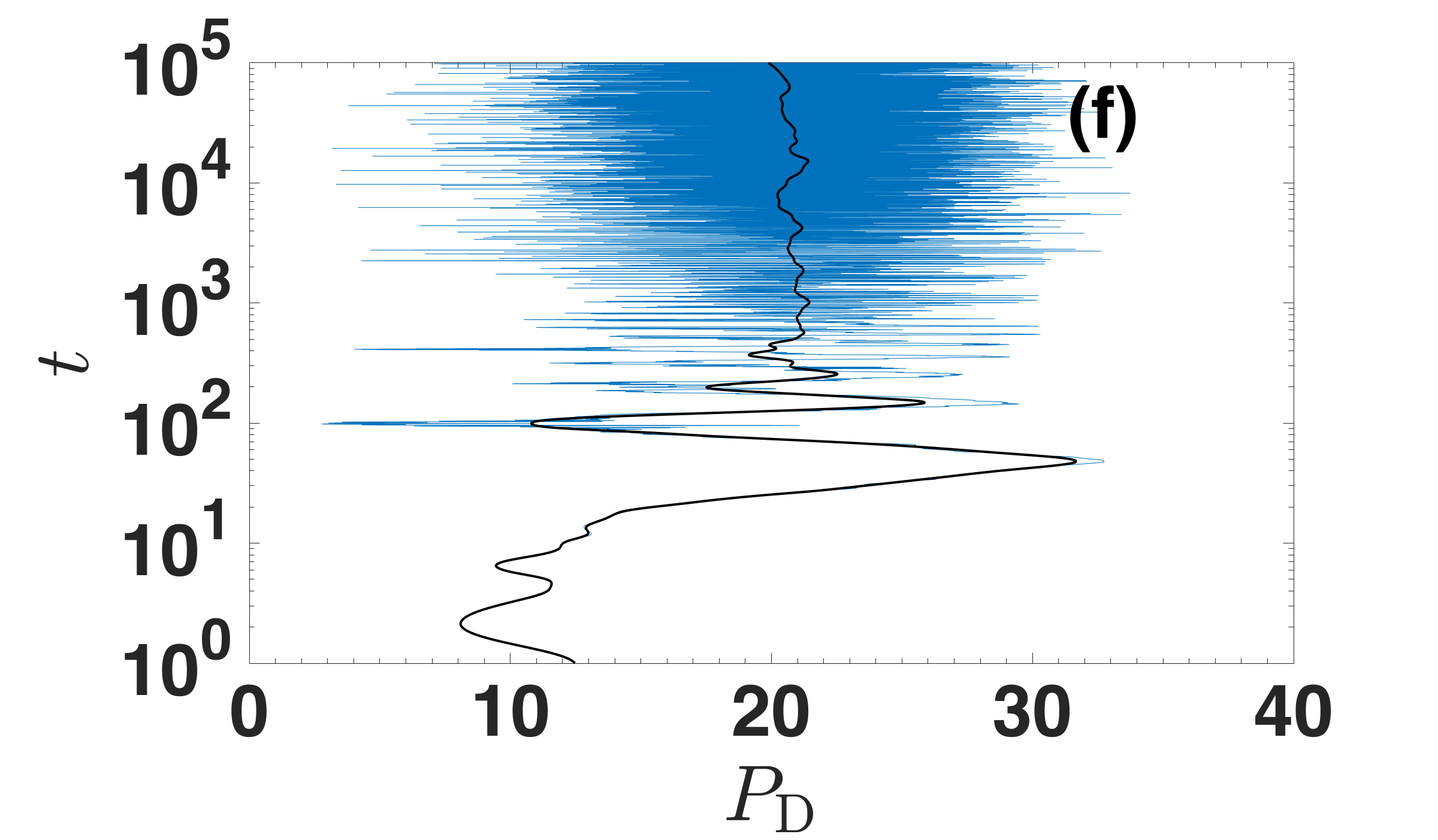}
\caption{ (a) [(b)]: The spatiotemporal evolution of the deviation vector density (DVD) for the Hertzian [FPUT] disordered chain. The 
color bars on the right sides of (a) and (b) are in logarithmic scale.
(c) [(d)]: Deviation vector profiles for three time instances of $t  \approx 10^1 $ indicated by the blue (b) curve,  $t  \approx 10^3 $ indicated by the red (r) curve and 
 $t  \approx 10^5  $ indicated by the black (bl) curve. These times correspond respectively to the blue, red and black horizontal lines in panel (a) [b].
(e) [(f)]: The time evolution of the participation number $ P_D $ of the DVD for the Hertzian [ FPUT] model.
All results are obtained for $ H = 0.25$. }
\label{dynms025b}
\end{figure}

 The spatiotemporal evolution of the corresponding DVD, plotted in Figs.~\ref{dynms025b}(a) and (b), indicates 
 that the dynamics is characterized by extended deviation vector distribution in contrast to the localized, pointy shape that DVDs exhibit for chaotic orbits~\cite{SGF13, Bob2018, malcolmDNA}.
Accordingly particular profiles of the DVDs taken at different times shown in Figs.~\ref{dynms025b}(c) and (d) are found to be extended covering the whole excited part of the lattice in a relatively smooth way.

However, a difference between the two models is found by closely inspecting the corresponding 
participation number $P_{\rm D}$ of the DVDs shown in Figs.~\ref{dynms025b}(e) and (f).
This quantity is calculated in a similar way as  for the energy density and it gives the number of 
sites that are  significantly participating in the dynamics of the DVD. In Figs.~\ref{dynms025b}(e) and (f) we observe that although
up to $t\approx 10^3$ both DVDs exhibit approximately  $P_{\rm D}\approx 20$, for the case of the Hertzian chain [panel  (e)]
it starts to drop to a smaller value. As discussed earlier, the tendency of the DVD to start to localize is a precursor of a chaotic spot that may appear 
in the dynamics over a longer timescale. 

It is interesting to note that, although the two models behave almost identically for $H=0.25$, this energy
corresponds for the Hertzian model to a initial displacement of $u_{21}(0)=1.01$ with the neighboring static overlaps being $\delta_{21,22}\approx 1.06$. These 
values are far from the small amplitude approximation ($u_n/\delta_{n,n+1}\ll 1$). The two models however show no differences
(at least for the studied time scales), mainly due to the fact that practically only a single mode is participating in the dynamics. 
 
%

\subsection{Chaos and destruction of localization}
 \begin{figure}
 \includegraphics[width=4.0cm]{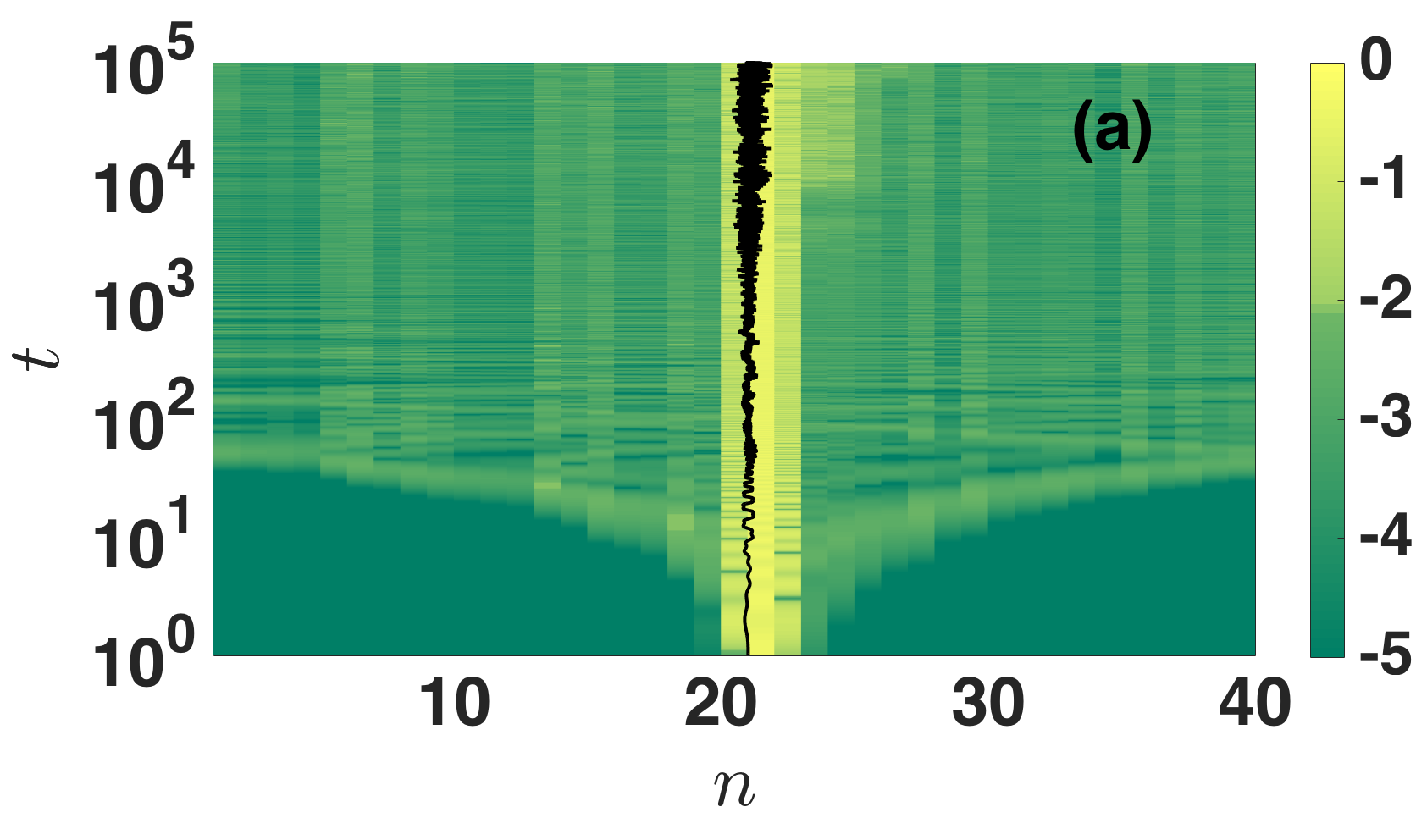}
 \includegraphics[width=4.0cm]{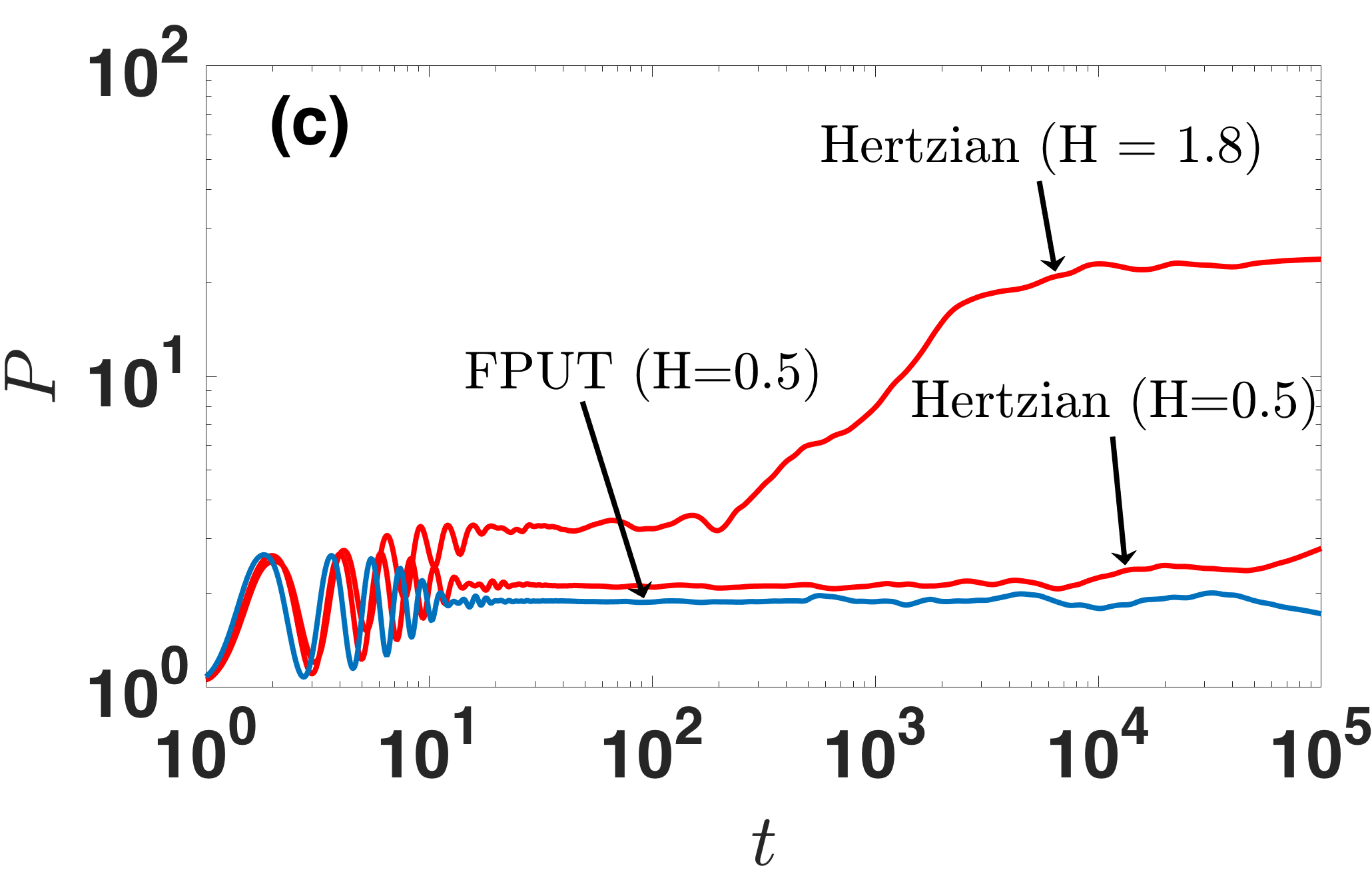} \\
 \includegraphics[width=4.0cm]{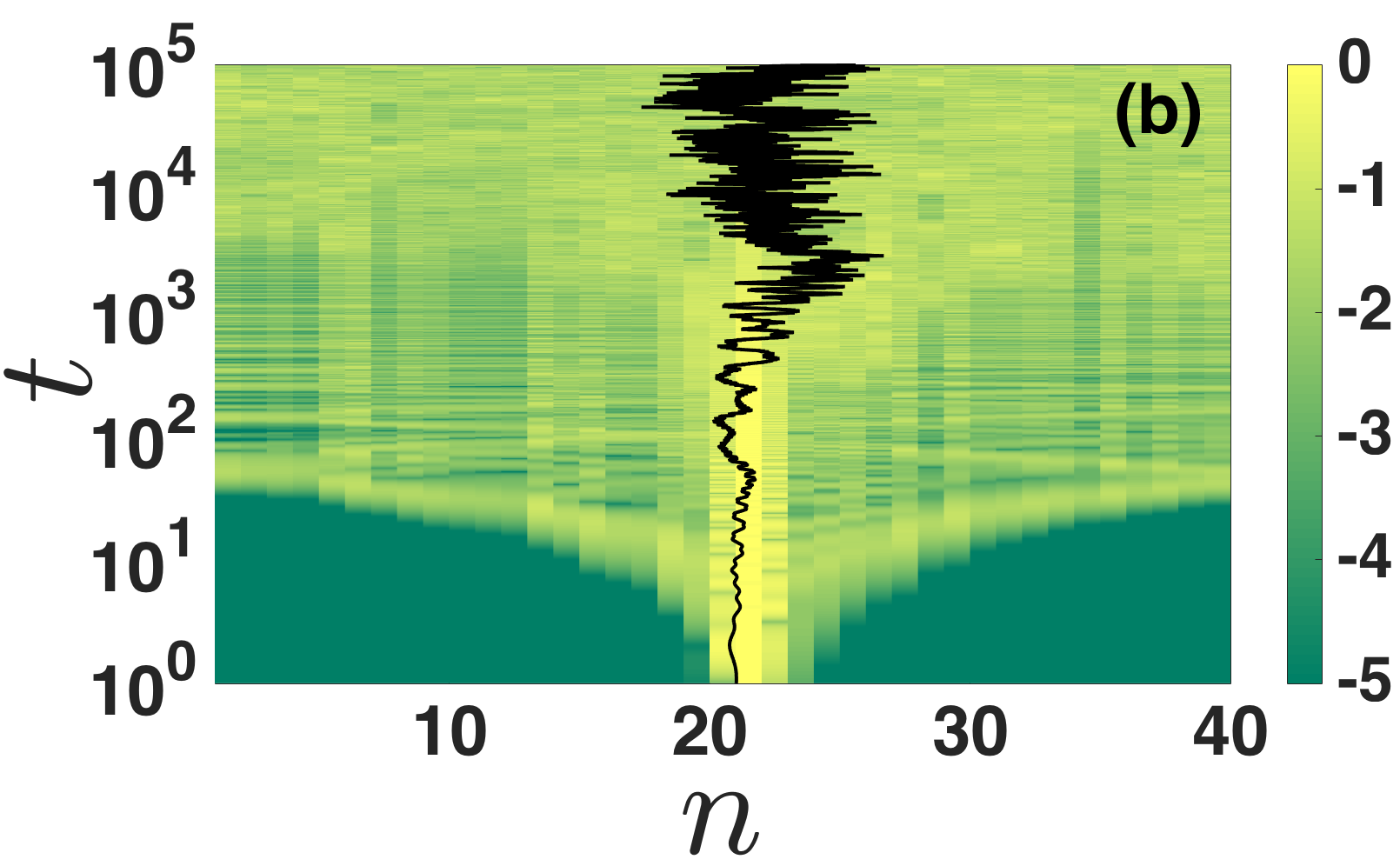} 
 \includegraphics[width=4.0cm]{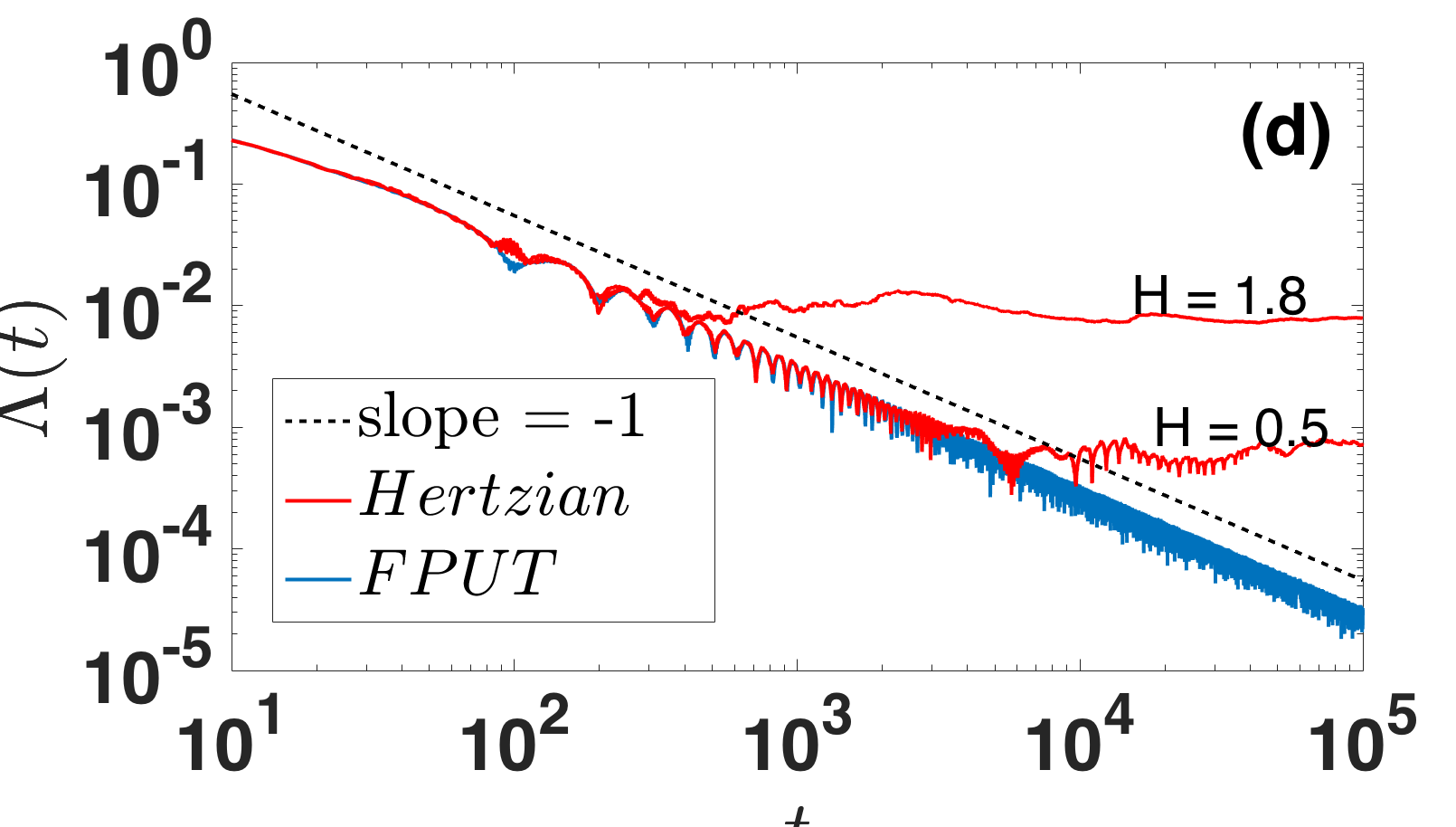} 
\caption{Panels (a) and (b) show the spatiotemporal evolution of the energy distribution for the Hertzian model with $H=0.5$ 
and $H=1.8$ respectively. Black curves indicate the running  mean position of the energy distributions. The color ~\cite{SGF13, Bob2018, malcolmDNA}bars on the right sides of (a), (b) are in logarithmic scale.
Panels (c) and (d) are the same as Figs.~\ref{dynms025a}(c) and (d); for the Hertzian model $H=0.5$ 
and $H=1.8$  and for the FPUT model with $H=1.8$. 
}
\label{dynms05a}
\end{figure}

\subsubsection{Energy density evolution and chaos}
In Fig.~\ref{dynms05a}(a) we show the energy density evolution for the Hertzian model with energy $H = 0.5$. 
Note that energy density of the FPUT model is not shown here since it is similar to Fig.~\ref{dynms025a}(b). 
The energy distribution for both models is still found to be localized for $H = 0.5$. However, there is a difference during the last decade, better 
captured by the evolution of  $ P$ as illustrated in Fig.~\ref{dynms05a}(c), since the Hertzian chain exhibits a tendency to increase the number of highly excited particles.

The most intriguing feature for this particular case is found in the system's chaoticity as quantified 
by the time evolution of $ \Lambda (t) $ shown in Fig.~\ref{dynms05a}(d). The red solid line, which corresponds to the Hertzian chain with $ H = 0.5 $,
deviates from the $ \Lambda (t) \propto t^{-1}$ curve, at the last decade, and attains an almost constant value. This signals that the system is chaotic.
In contrast, for the same energy the FPUT model's orbit remains regular and the corresponding $ \Lambda (t) \propto t^{-1}$. 
The Hertzian model therefore exhibits localized chaos whilst the FPUT model is localized and regular.
 \begin{figure}[ht!]
 \includegraphics[width=4.0cm]{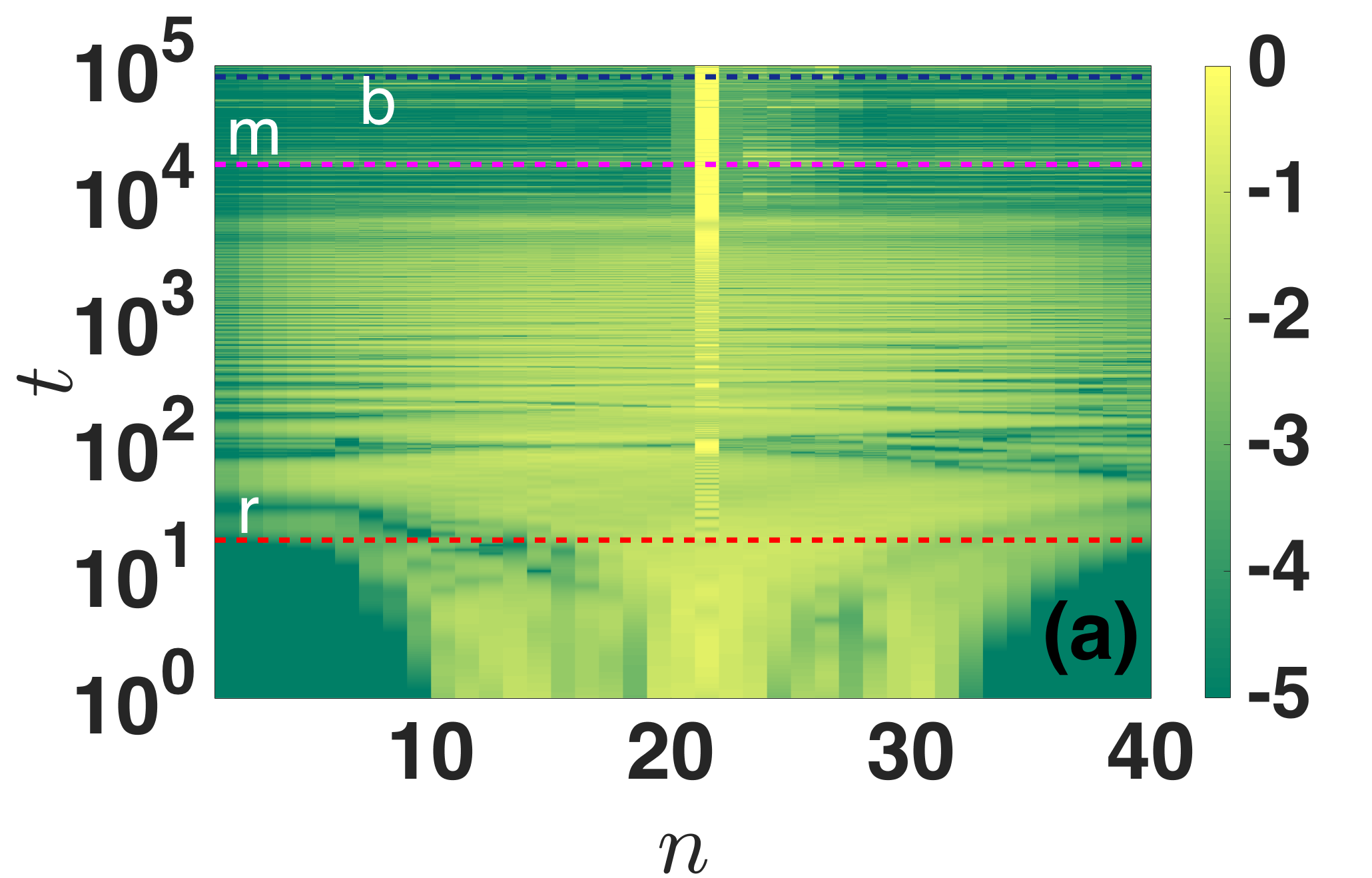} 
 \includegraphics[width=4.0cm]{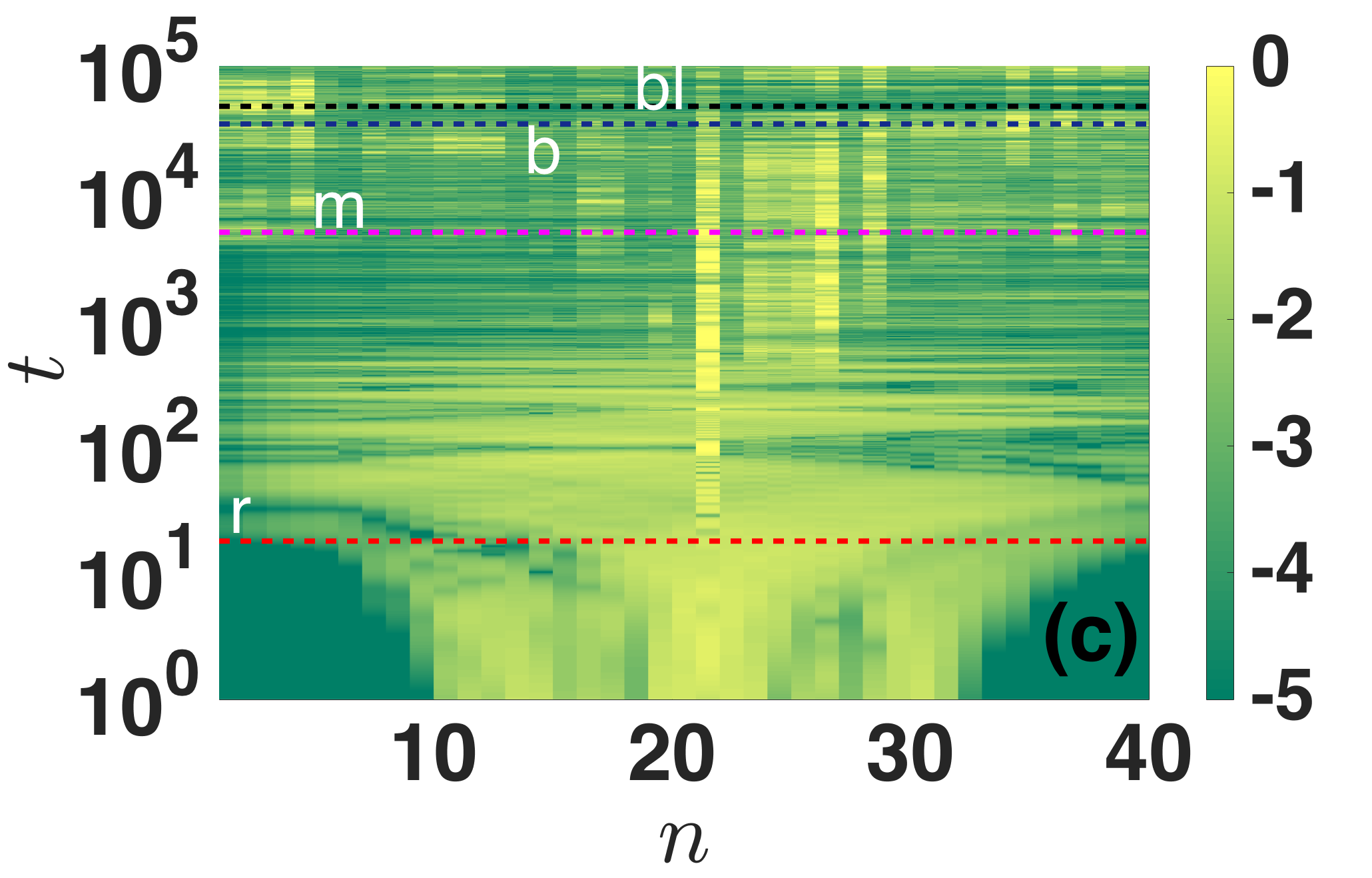}\\
  \includegraphics[width=4.0cm]{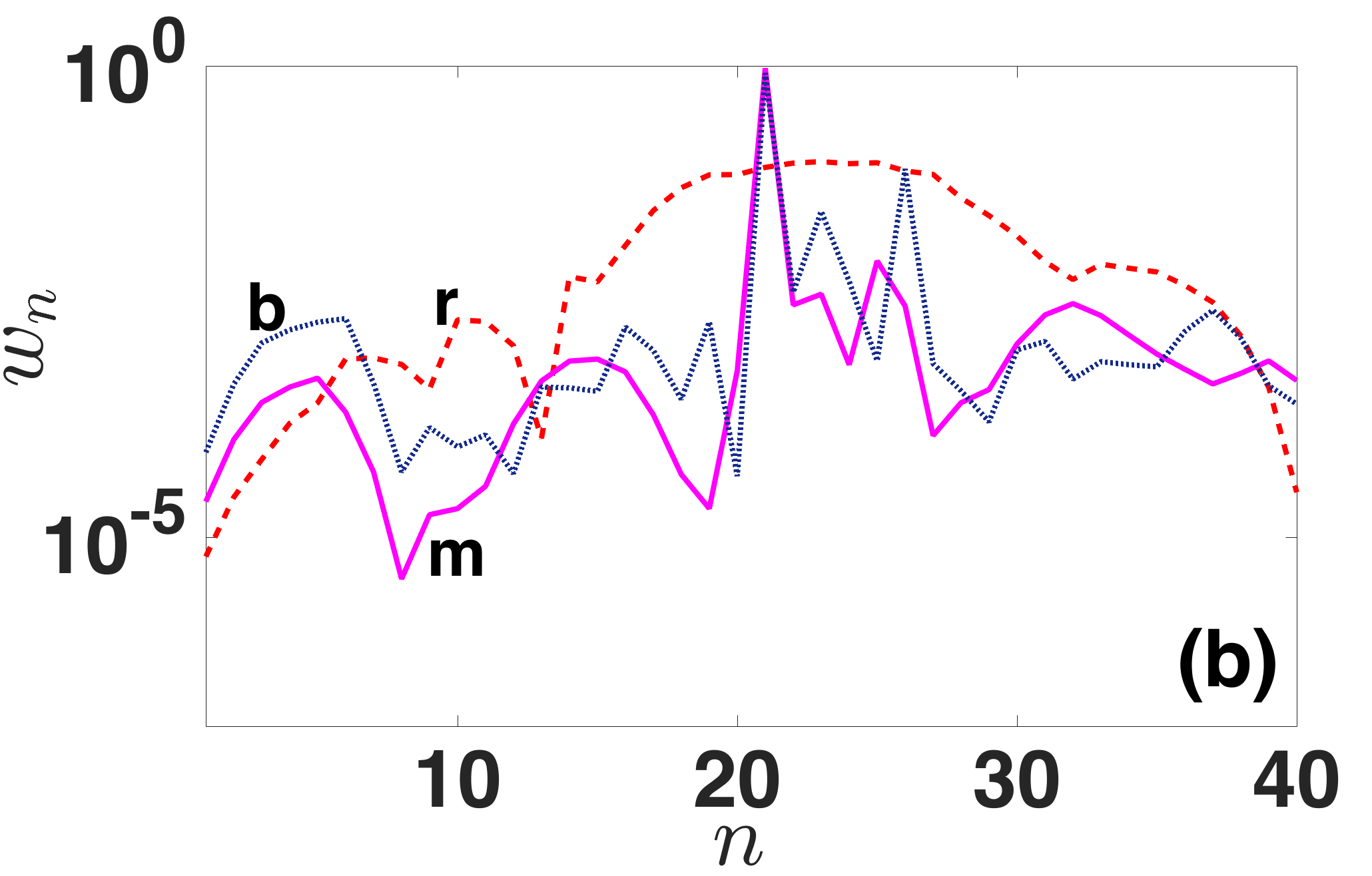} 
 \includegraphics[width=4.0cm]{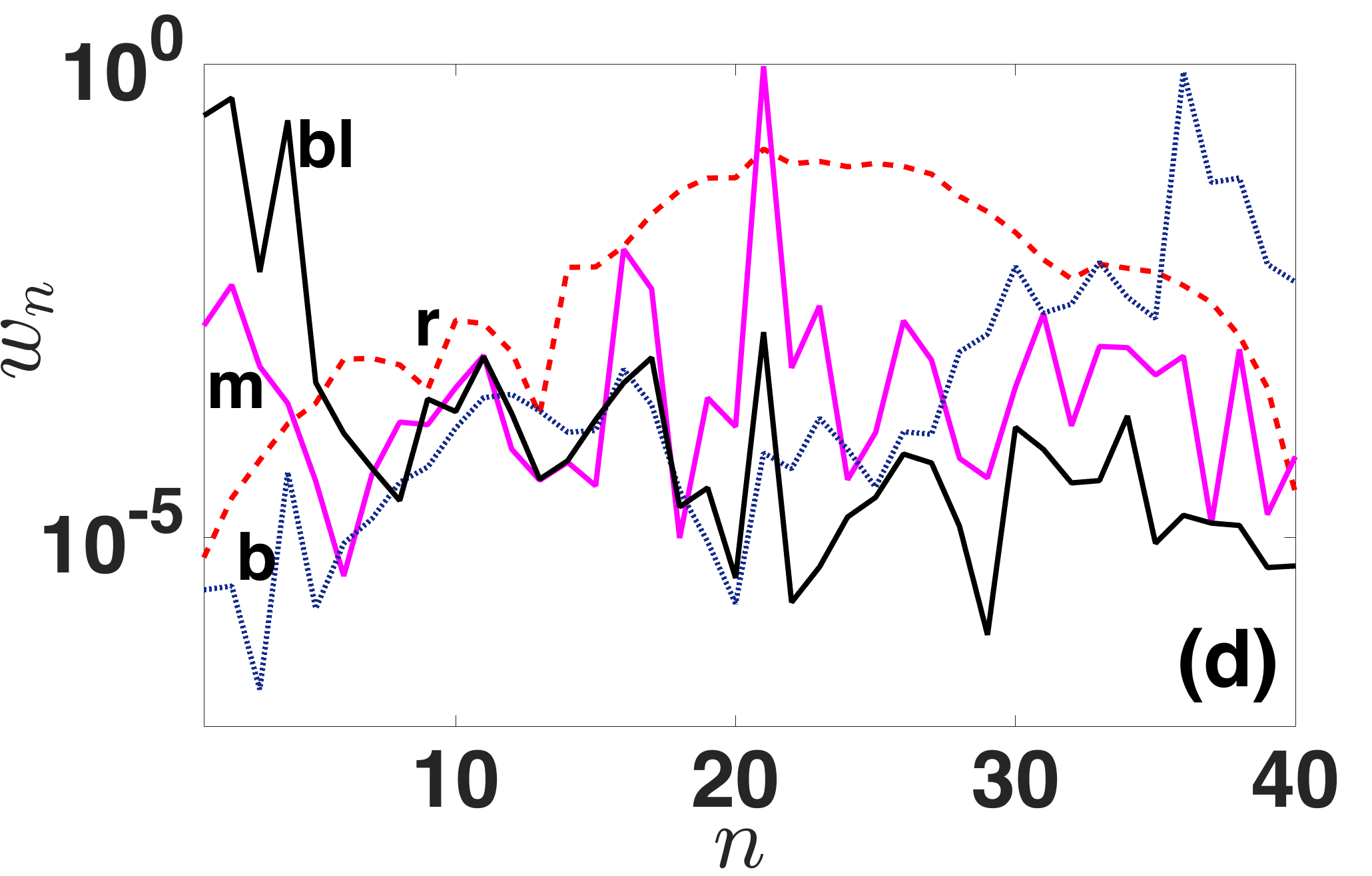} 
\caption{
Panel (a) shows the spatiotemporal evolution of the DVD for the Hertzian model at $ H = 0.5 $ whilst panel (b) shows the profiles of the DVDs at 
$t  \approx 1.7 \times 10^1 $ red (r) curve, $t  \approx 1.7 \times 10^4 $ magenta (m) curve and $t  \approx 8.2 \times 10^4  $ blue (b) curve.
(c) Same as (a) but for $ H = 1.8 $. (d) Same as (b) but for $ H = 1.8$ at $t  \approx 1.7 \times 10^1 $ red (r) curve, $t  \approx 4.9 \times 10^3$ 
magenta (m) curve, $t  \approx 3.5 \times 10^4  $ blue (b) curve and $t  \approx 4.8 \times 10^4  $ black (bl) curve. The color bars in (a) and (c) are in logarithmic scale.
.}
\label{dynms05b}
\end{figure}
%

%
%
%

For the Hertzian model, using a stronger excitation with $ H = 1.8 $, the initially localized wave-packet gradually spreads 
throughout the lattice.  In particular, although up to $t\approx 2\times 10^2$ the wave-packet is localized [see Figs.~\ref{dynms05a}
(b) and (c)]  with a participation  number $P<3$, it then rapidly spreads until $t\approx 4\times 10^{3}$, when 
eventually the participation number saturates to a value  
$ P \approx 26 $. We found that this is the maximum value of $P$ that can be obtained and it also corresponds to the participation 
number of the most extended mode of a disordered lattice [see Fig.~\ref{stats1}(a)].
Furthermore, according to the corresponding $ \Lambda (t) $ shown in Fig.~\ref{dynms05a}(d) for $H=1.8$, the system becomes  chaotic as early as $t\approx 2\times 10^2$ acquiring an almost constant positive value of $ \Lambda (t) \approx 10^{-3} $. Results for the FPUT are not shown for this energy since excitations were still found to be localized and regular.

 \begin{figure*}
 \includegraphics[width=4.2cm]{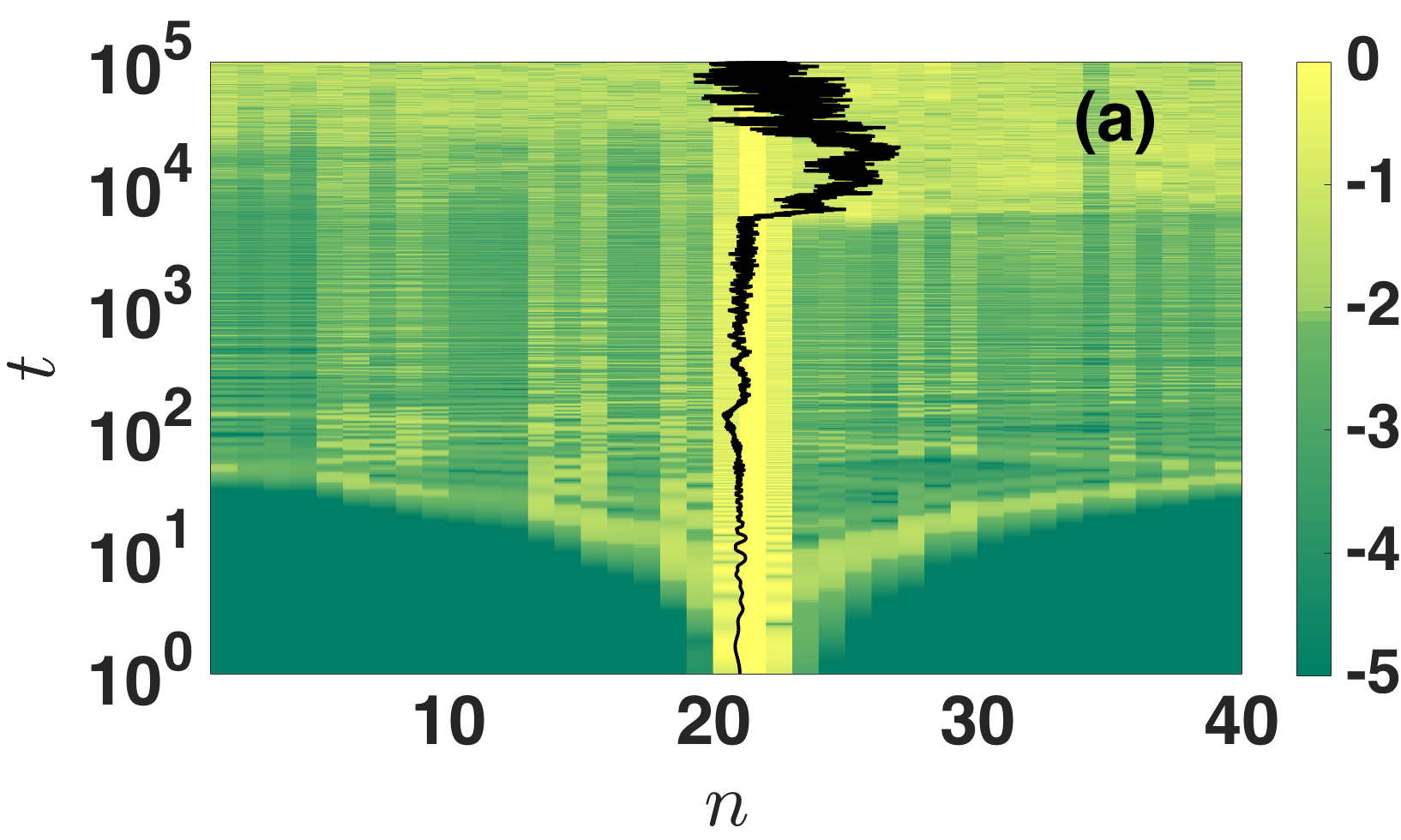}
 \includegraphics[width=4.2cm]{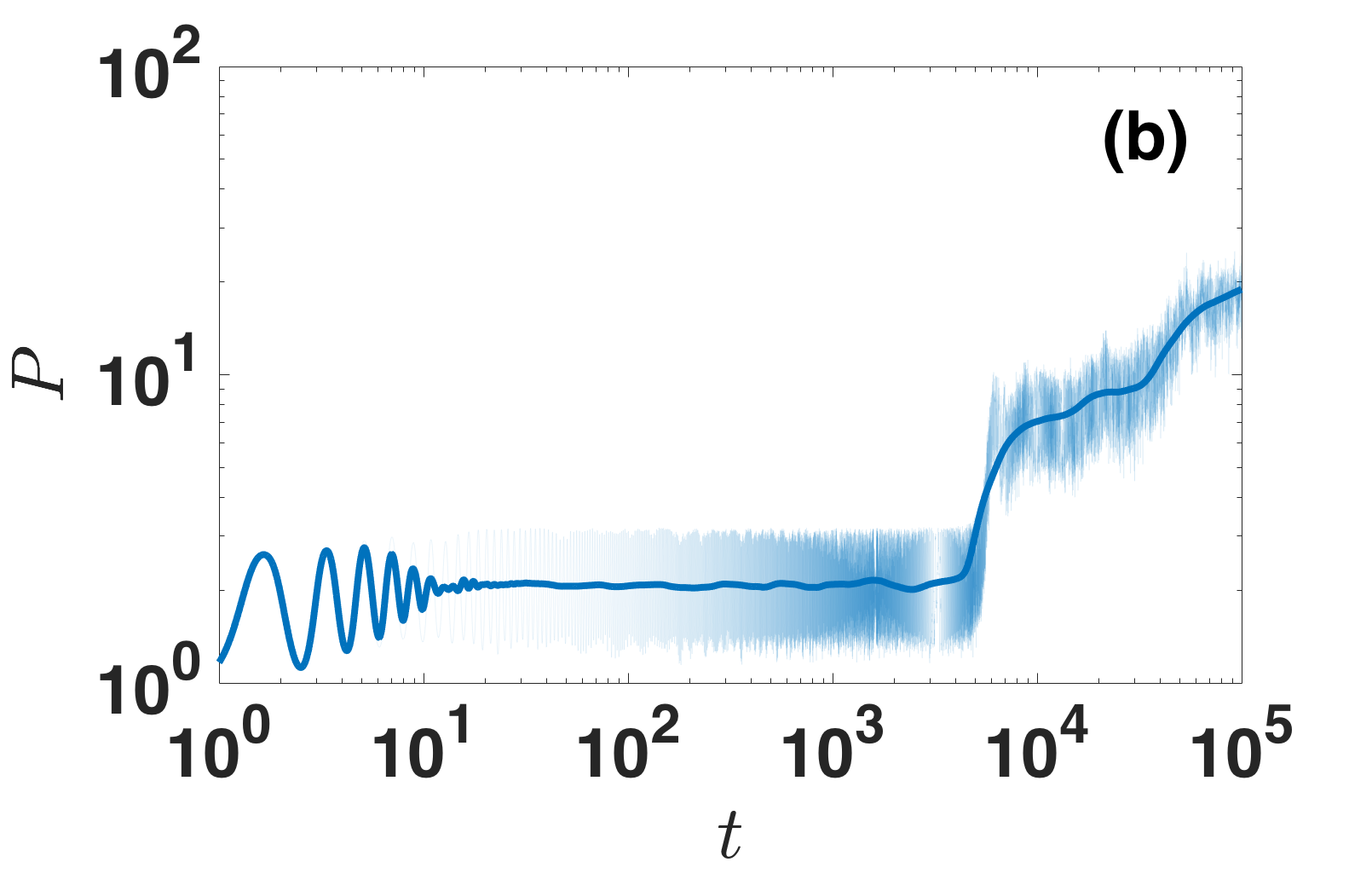}
 \includegraphics[width=4.2cm]{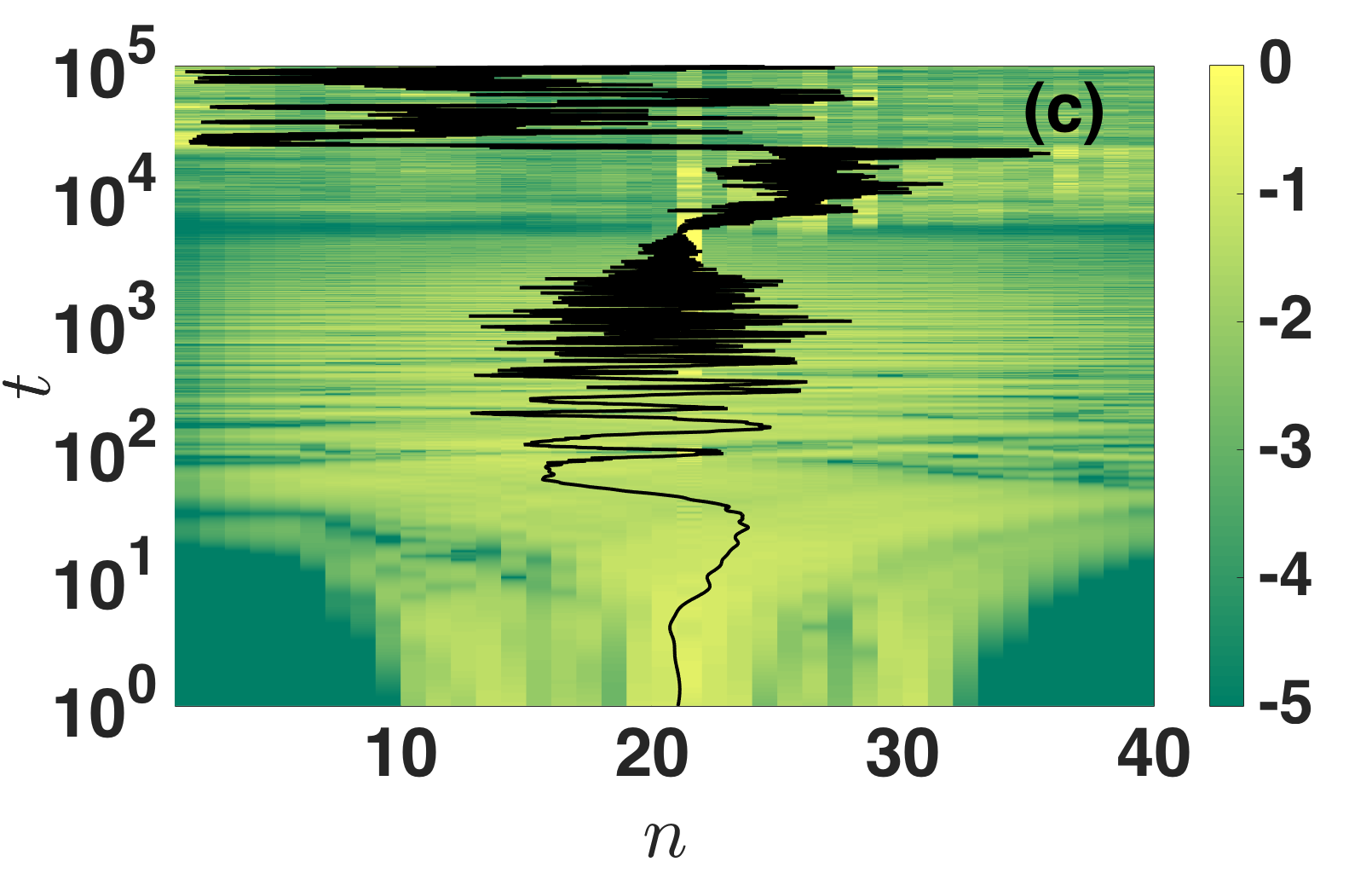}
 \includegraphics[width=4.2cm]{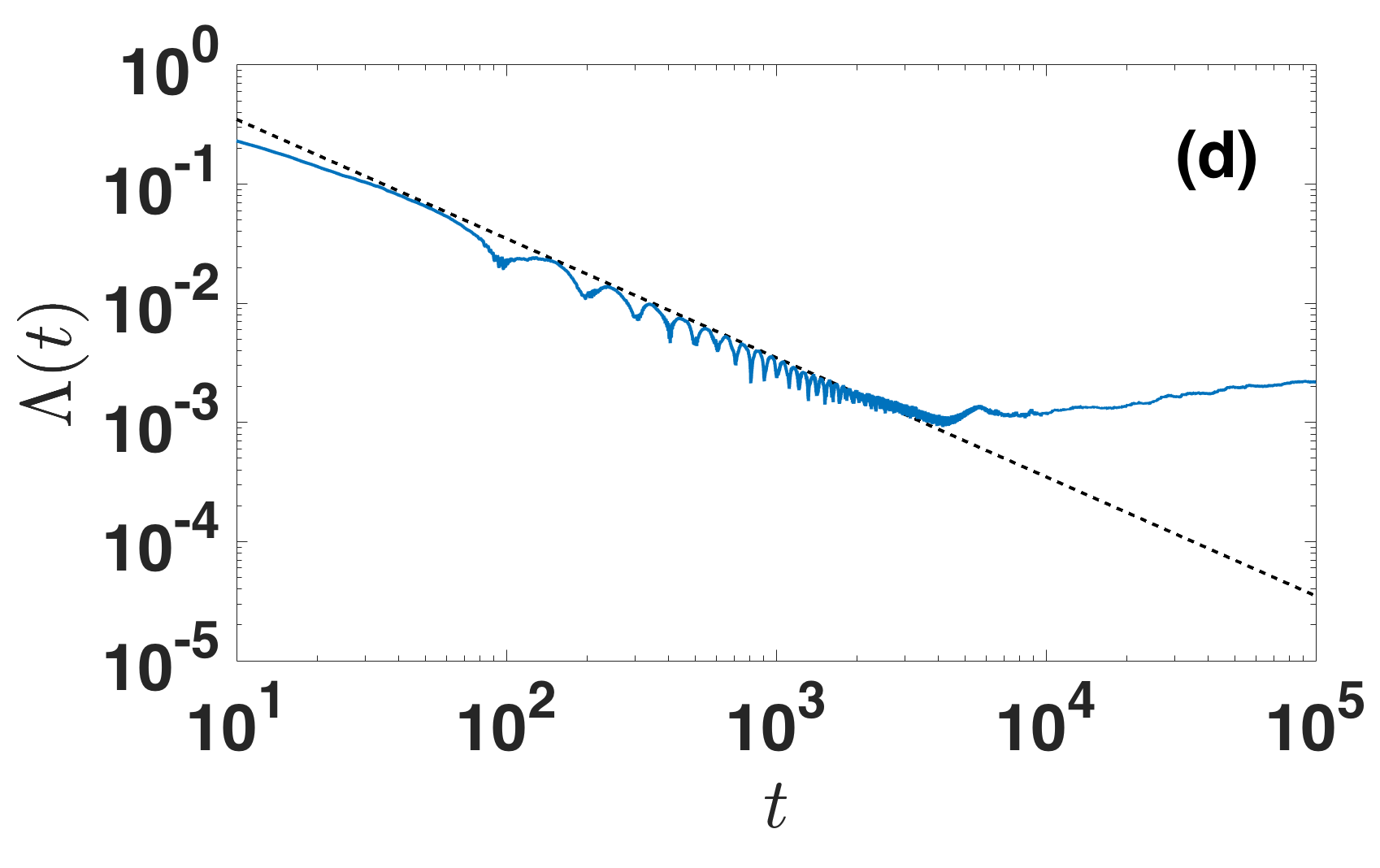} \\
 \includegraphics[width=4.2cm]{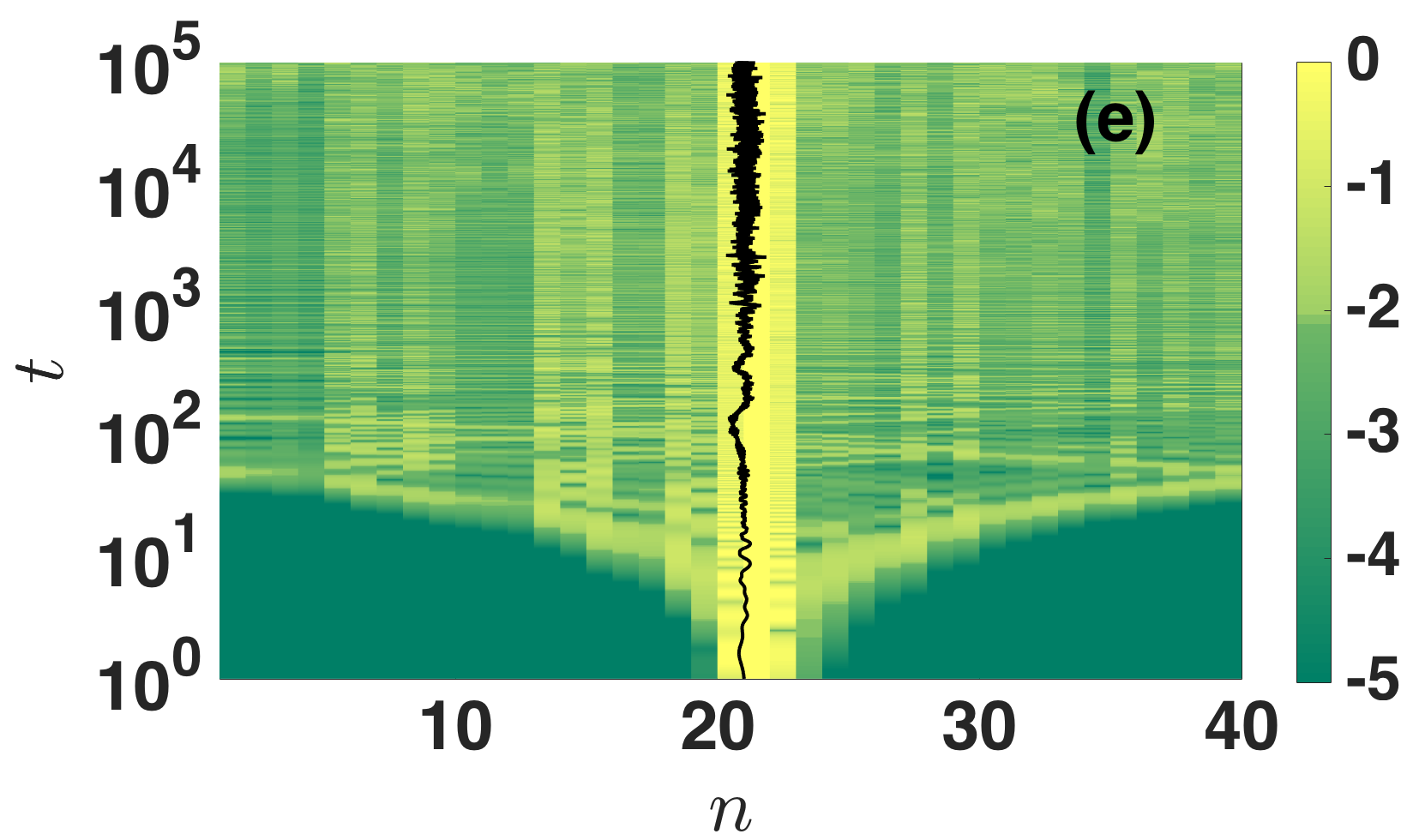}
 \includegraphics[width=4.2cm]{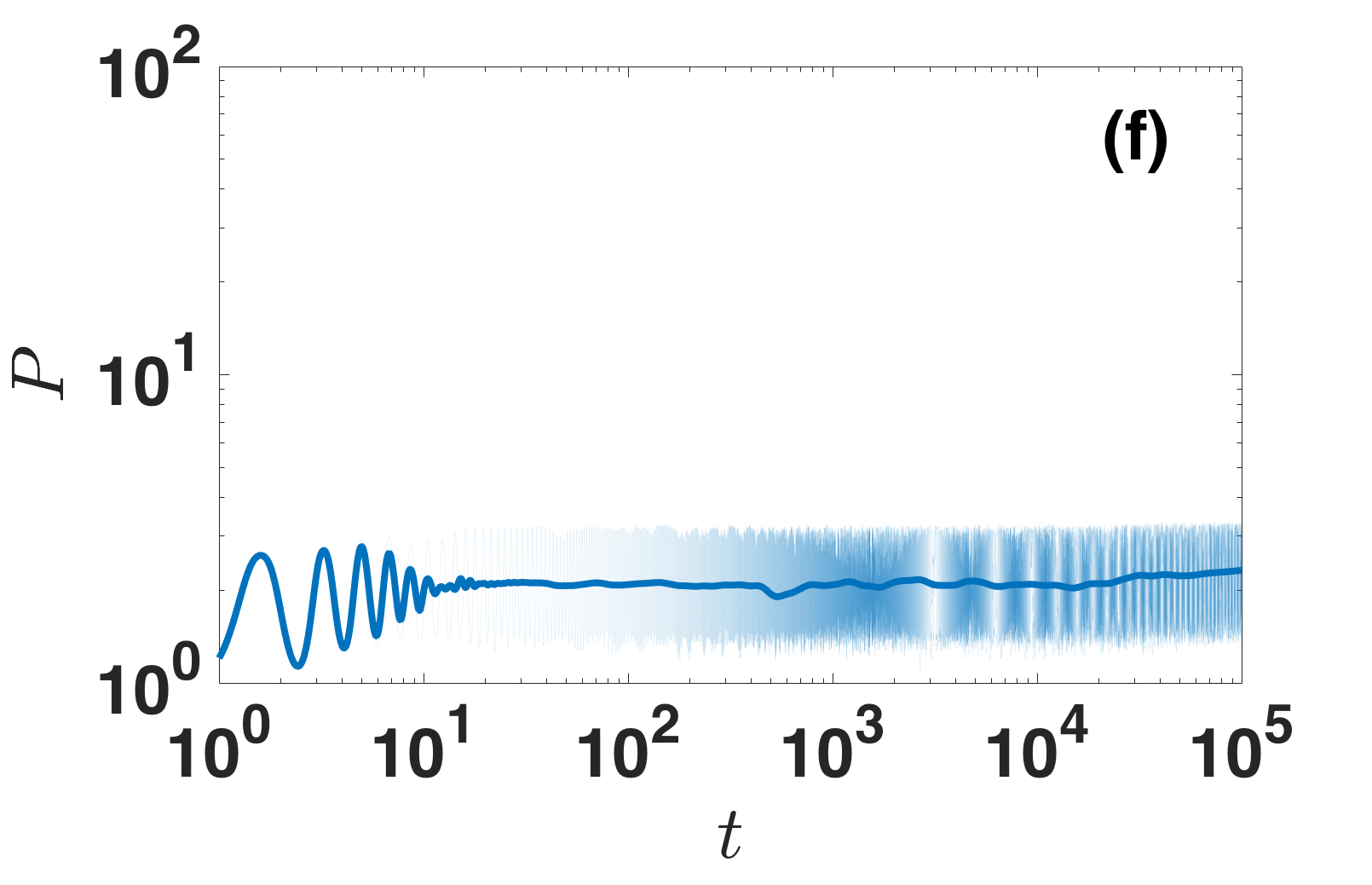}
 \includegraphics[width=4.2cm]{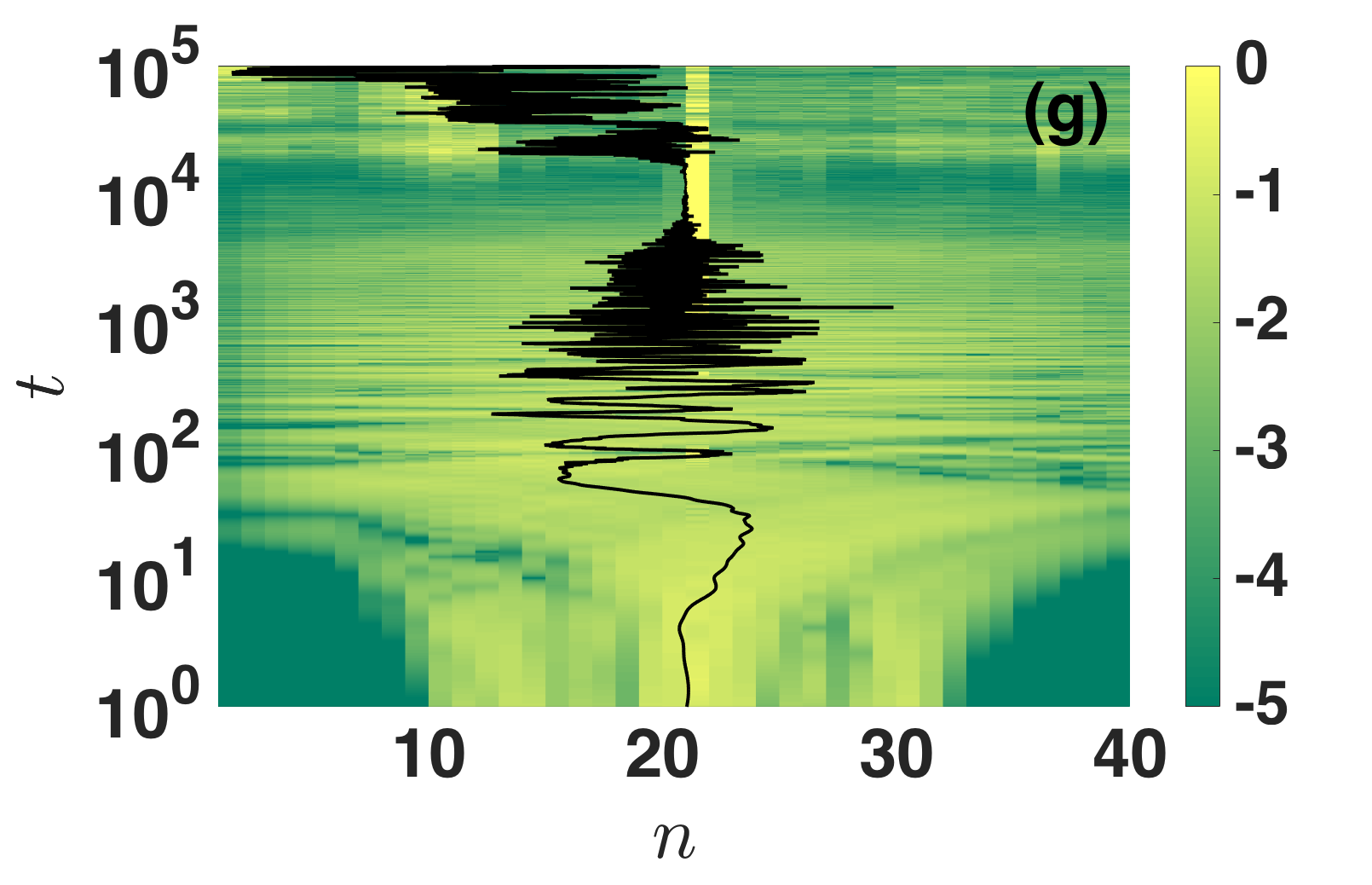}
 \includegraphics[width=4.2cm]{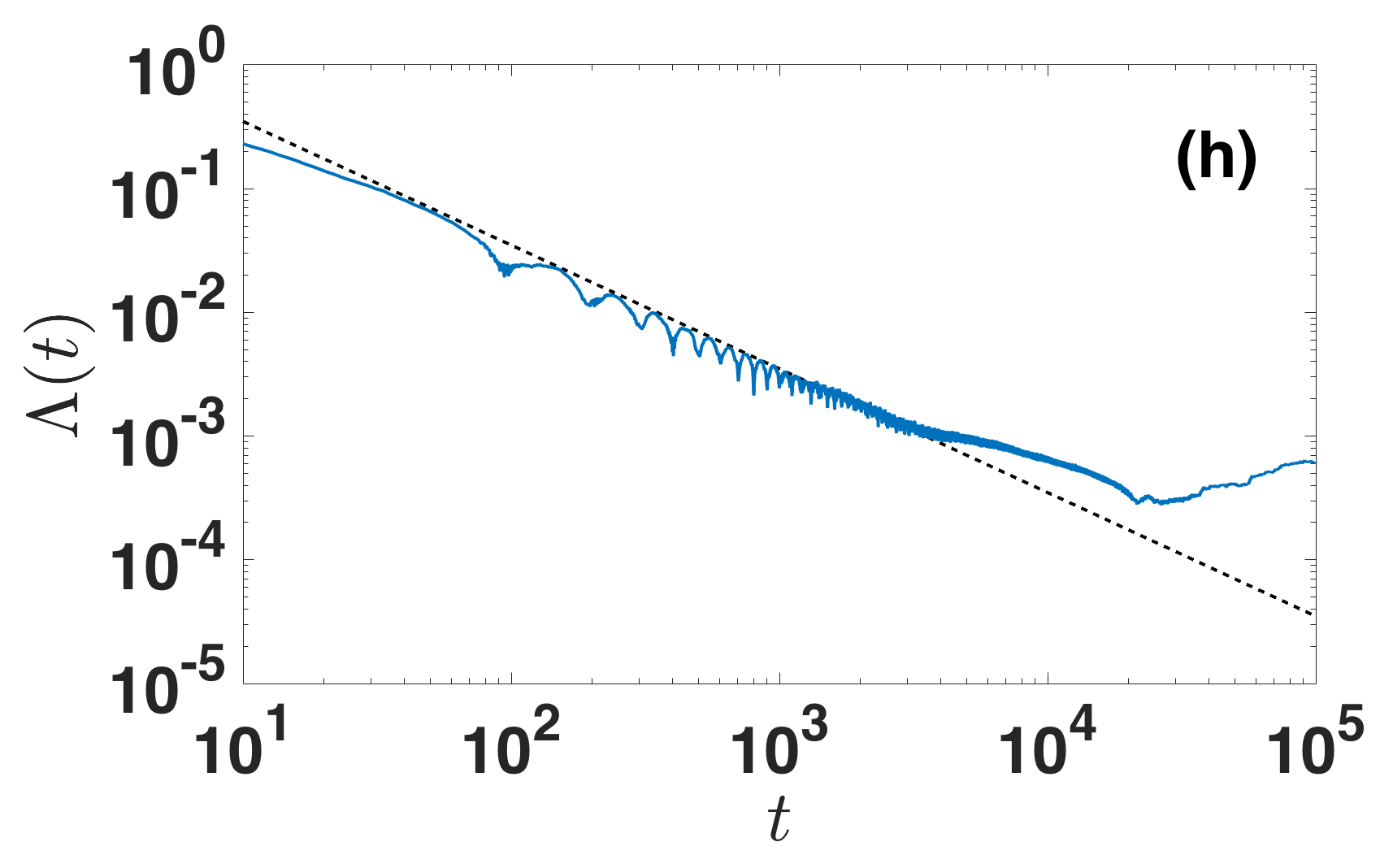} \\  
 \includegraphics[width=4.2cm]{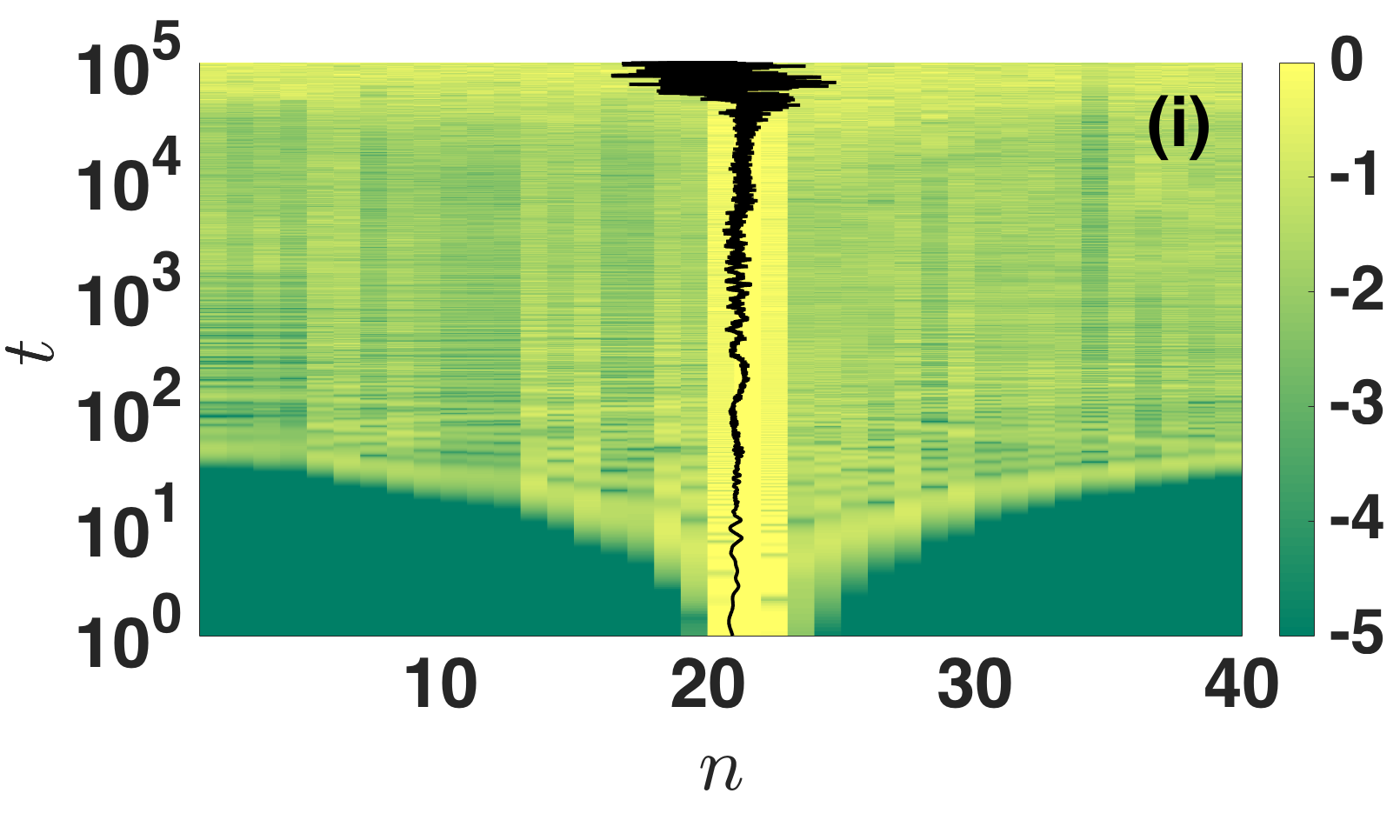}
 \includegraphics[width=4.2cm]{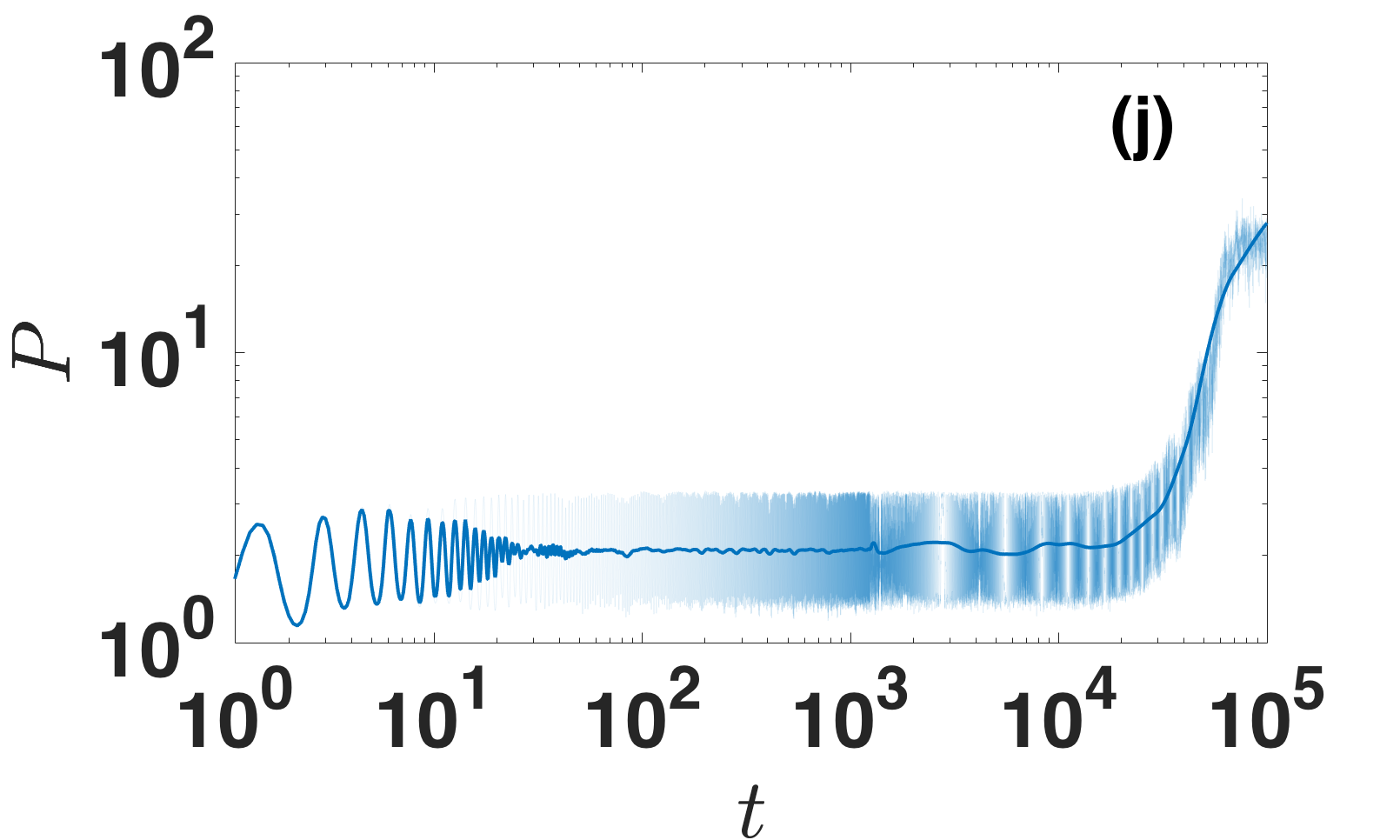}
 \includegraphics[width=4.2cm]{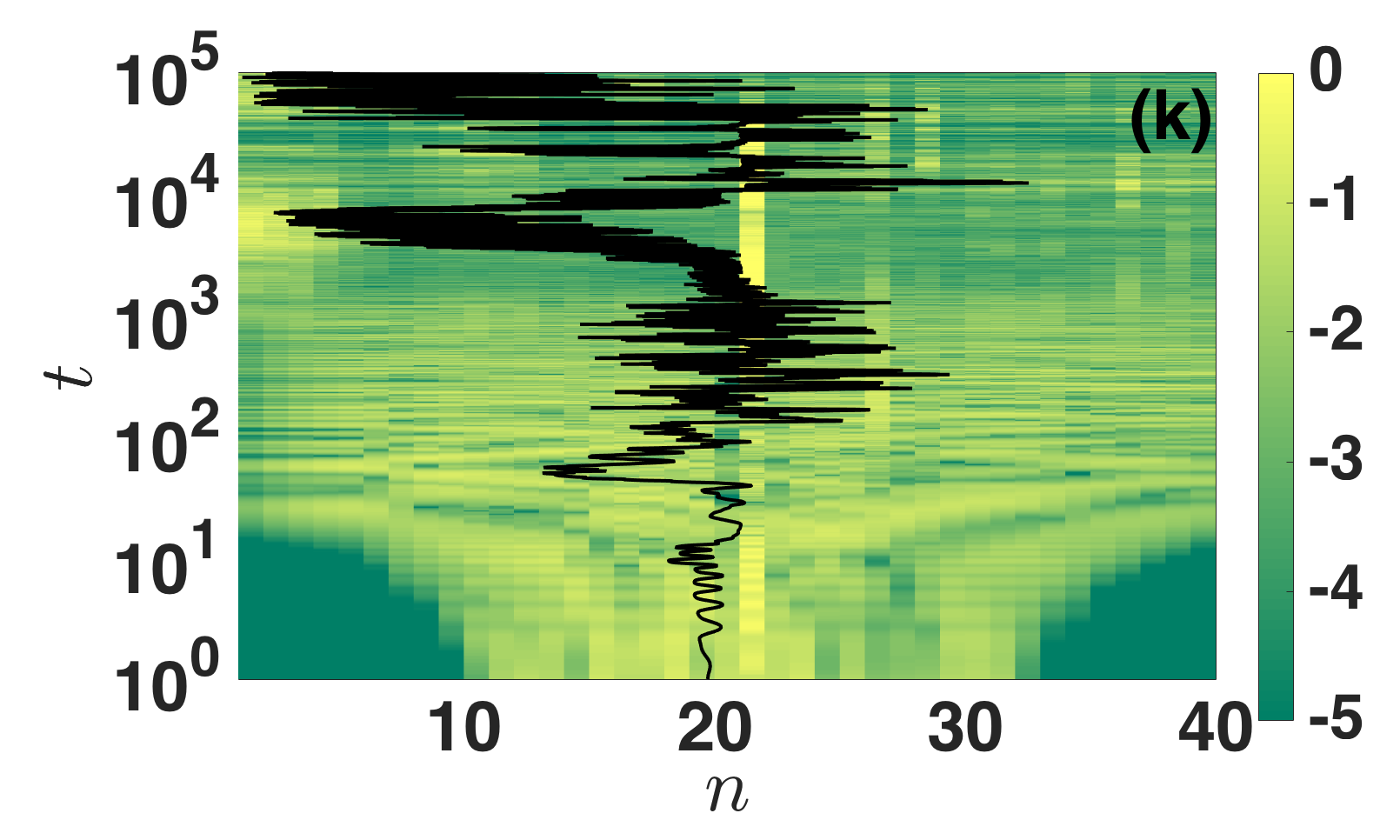}
 \includegraphics[width=4.2cm]{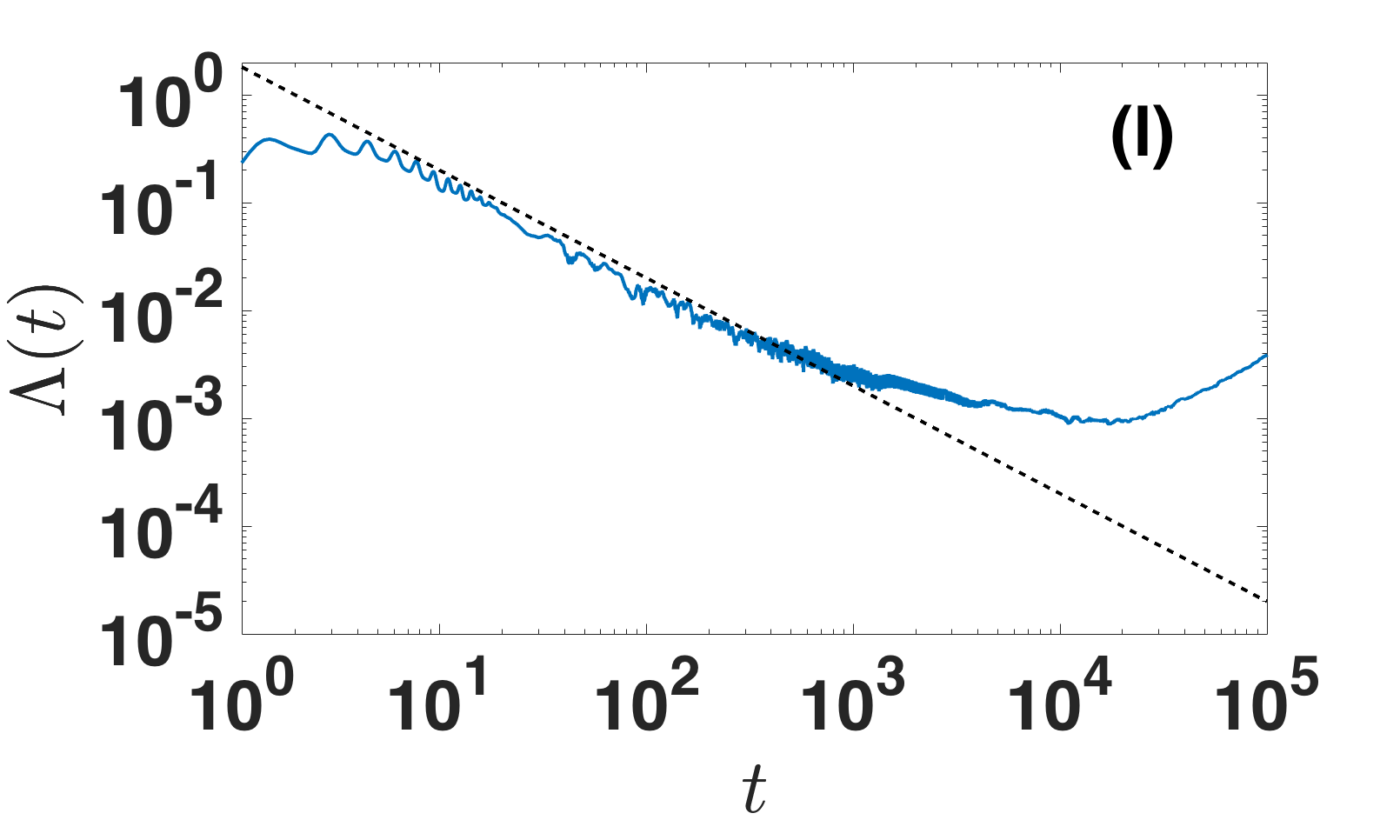} \\      
\caption{(a), (b), (c) and (d) depict the energy density, $P$, DVD, $P_{\rm D}$ and $ \Lambda (t) $ respectively 
for the FPUT with $H = 2.9$. The second, and third rows correspond to energies $H=4$ and $H=8.7381$
respectively. The color bars on the right sides of panels (a), (c), (e), (g), (i) and (k) are in logarithmic scale.} 
\label{dynms29}
\end{figure*}

\subsubsection{Spatiotemporal evolution of chaos}
%
In order to better understand the onset of chaos in the aforementioned cases, we study more closely the behavior of these DVDs.
 In Figs.~\ref{dynms05b}(a) and (c) we plot the DVDs for the Hertzian model corresponding to the energies $H=0.5$ and $H=1.8$ respectively.
Focusing on the case of $H=0.5$ we see that initially, when the system behaves regularly, the DVD exhibits an extended smoothed profile.
This is more clearly seen by the red (dotted) curve in Figs.~\ref{dynms05b}(b).
Thereafter, during a period up to $ t\approx 4\times 10^{3}$ the DVD gradually
converges around site $n=21$. A profile of the DVD in this era is shown with the magenta curve in Fig.~\ref{dynms05b}(b).
Finally for the rest of the simulation the profile of the DVD is strongly localized around site $n=21$ as also confirmed by two different profiles
during the last decade shown in Fig.~\ref{dynms05b}(b).
 Other recent studies (i.e. Refs.~\cite{SGF13, Bob2018, malcolmDNA}) also used the DVD to spatially characterize chaos. 
In these works, it was found that the profile of the DVD exhibits a peak that oscillates within a chaotic region, while in our case it remains attached to a single site indicating strongly localized chaos.


At a higher energy of  $ H = 1.8 $ the DVD initially exhibits an extended and smooth profile and an example is plotted in
Fig.~\ref{dynms05b}(d) with the red (dotted) line. Further on, the DVD  concentrates around a region close to the center of the chain and beyond this point the system is chaotic.
The evolution of the DVD during this chaotic era, is characterized by one dominant peak  along with other smaller peaks usually two orders of magnitude smaller (at most).
This is best illustrated in the profiles shown in Fig.~\ref{dynms05b}(d). We particularly choose three cases where the dominant peak is either at the center (magenta), closer to the right
edge (blue) or at the left edge (black). In contrast to the localized chaos exhibited for $H=0.5$, for this energy the chaoticity of the system is extended featuring strongly chaotic spots
throughout the whole lattice.
In these two examples, we illustrated the importance of the DVD which enables us to differentiate between localized and extended chaos.
For energies  $ H >1.8 $ we have found that the dynamics of the Hertzian model always lead to a delocalized chaotic state.

\subsubsection{Chaos and delocalization for the FPUT model}

On the other hand, the dynamics of the FPUT model appears to remain localized and regular 
up to $H=1.8$ [see Figs.~\ref{dynms05a} (c) and (d)].
In order to find cases where the FPUT model is chaotic and spreads, we perform simulations with increasing energy. Some typical 
examples are shown
in Fig.~\ref{dynms29}. The first energy at which the FPUT model's wave-packet is delocalized, exhibiting also a chaotic behavior,
is around $H = 2.9$ (first row of Fig.~\ref{dynms29}). After an initial transient time for which the wave-packet remains localized, it then 
appears to spread during the last decade of the simulation as shown by the energy density and $P$ in Figs.~\ref{dynms29}(a) and (b) 
respectively. 
The time evolution of $ \Lambda (t) $ shown in panel (d) significantly deviates from the  $ \Lambda (t) \propto t^{-1}$ line, indicating 
chaotic dynamics for $t \gtrsim 4\times 10^3$.
Furthermore the DVD shown in Fig.~\ref{dynms29}(c) exhibits peaks  at different places within the lattice when the system is chaotic, similarly to 
Fig.~\ref{dynms05b}(c) for the Hertzian model, which is associated with extended chaos.

An important difference between the two models is found for higher energy excitations. It appears that increasing the energy
for the FPUT model does not necessarily lead to delocalization, as it was the case for the Hertzian model. 
A particular example is shown in the second row of Fig.~\ref{dynms29} which corresponds to $H=4$. 
In this case the wave-packet remains well localized and the participation number hardly changes 
[compare Figs.~\ref{dynms29}(a) and (b) with Figs.~\ref{dynms29}(e) and (f)]. 
This is somewhat a surprising result and it highlights  the complexity of the phase-space of a disordered FPUT lattice.
\begin{figure*}
\includegraphics[width=8cm]{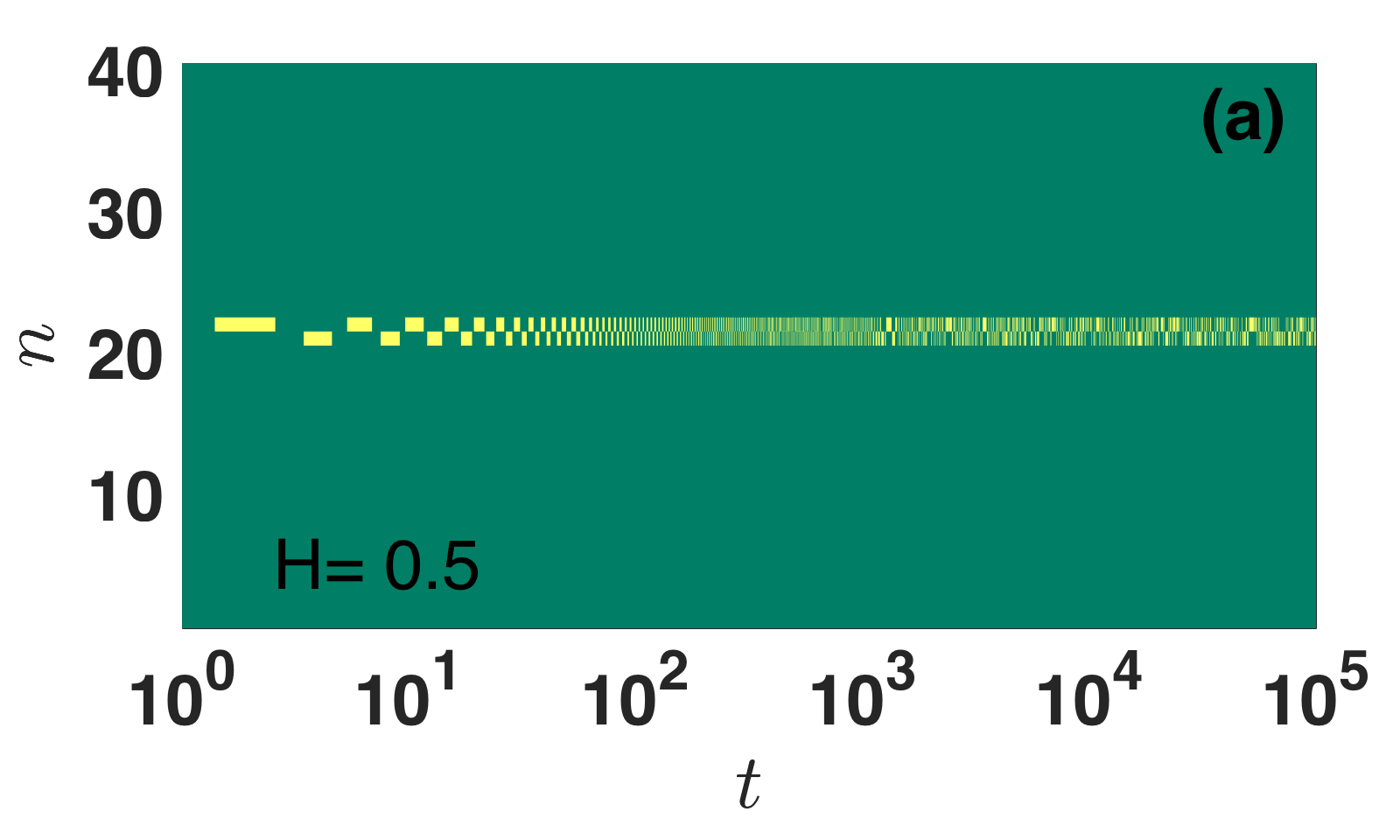}
\includegraphics[width=8cm]{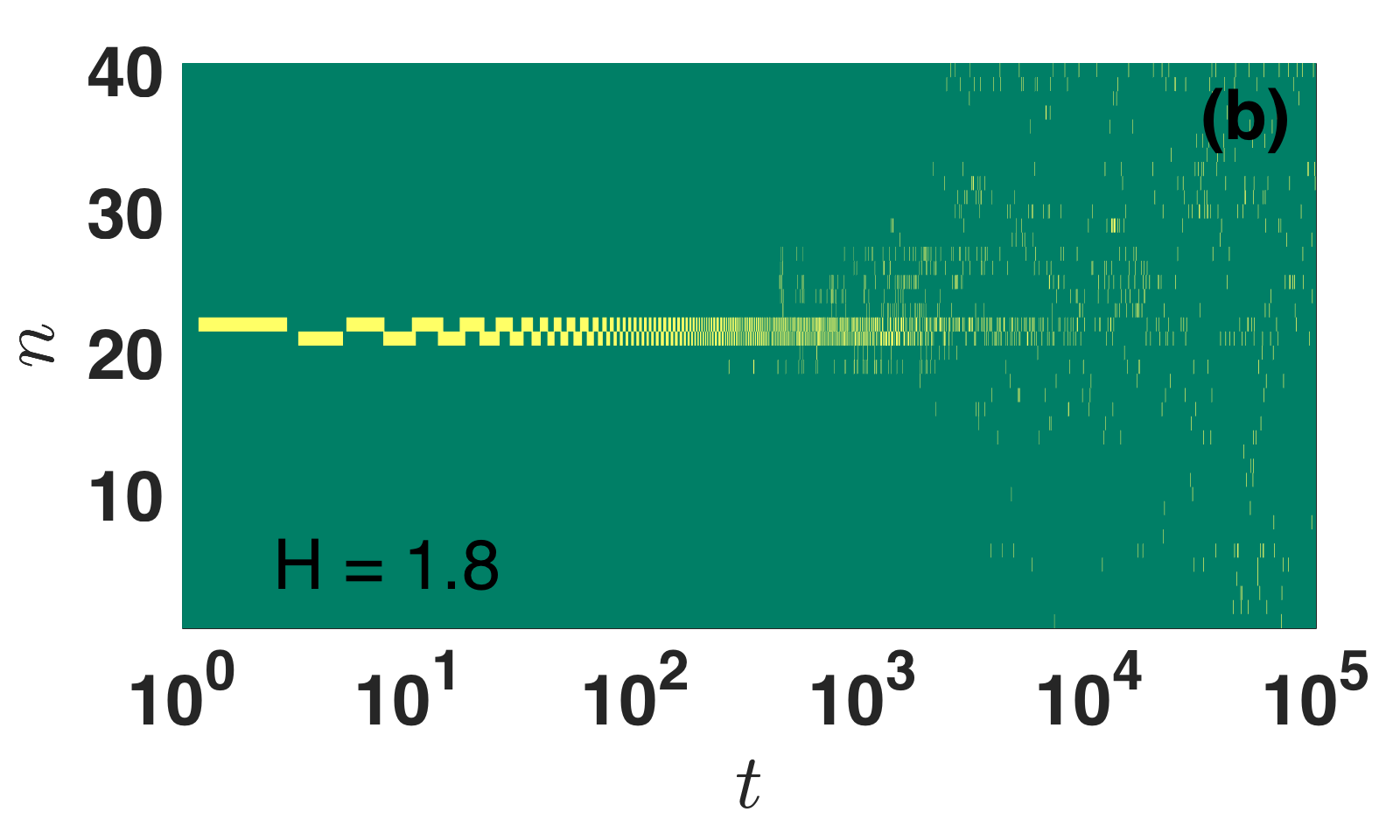} \\
\includegraphics[width=8cm]{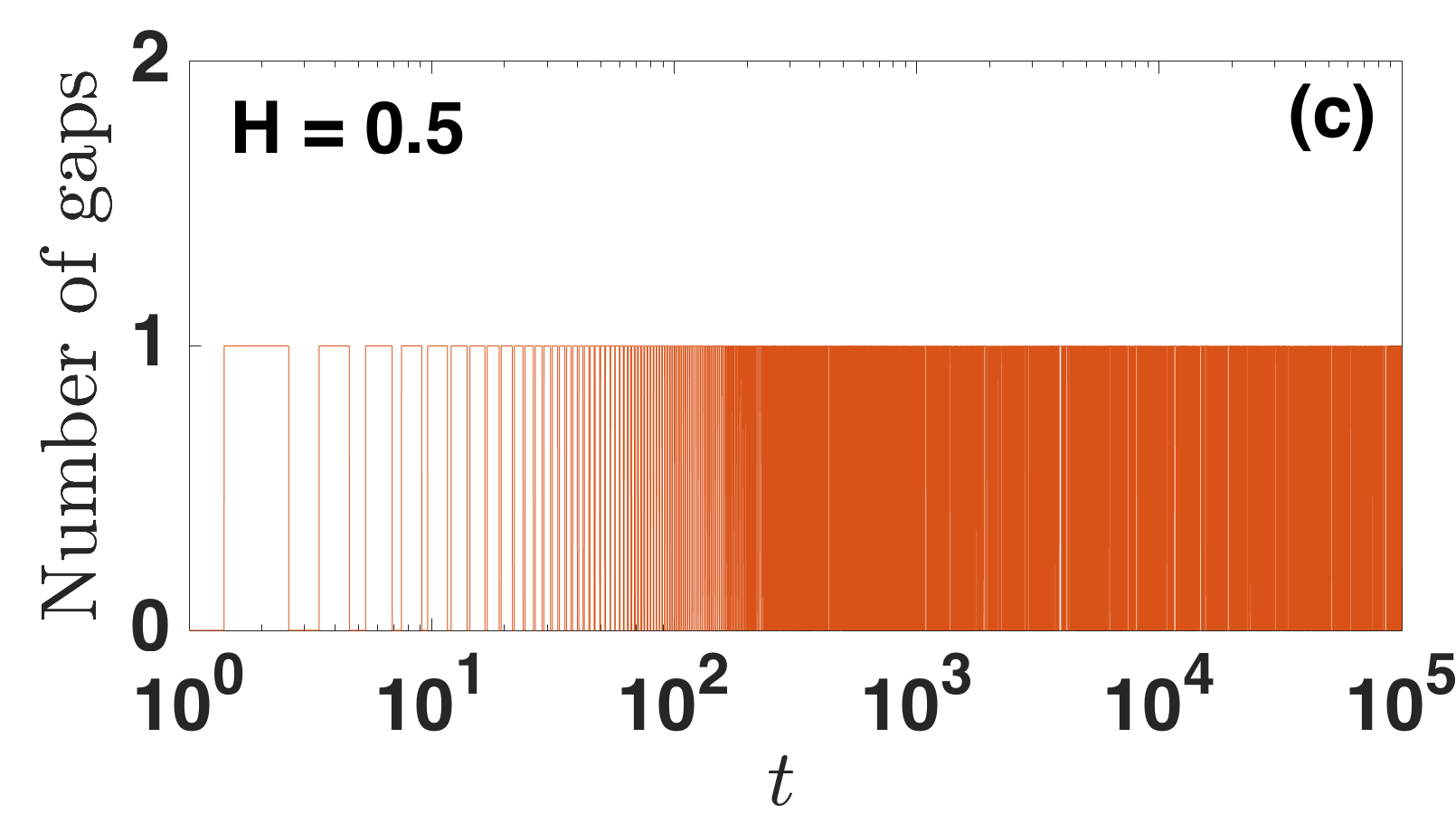}
\includegraphics[width=8cm]{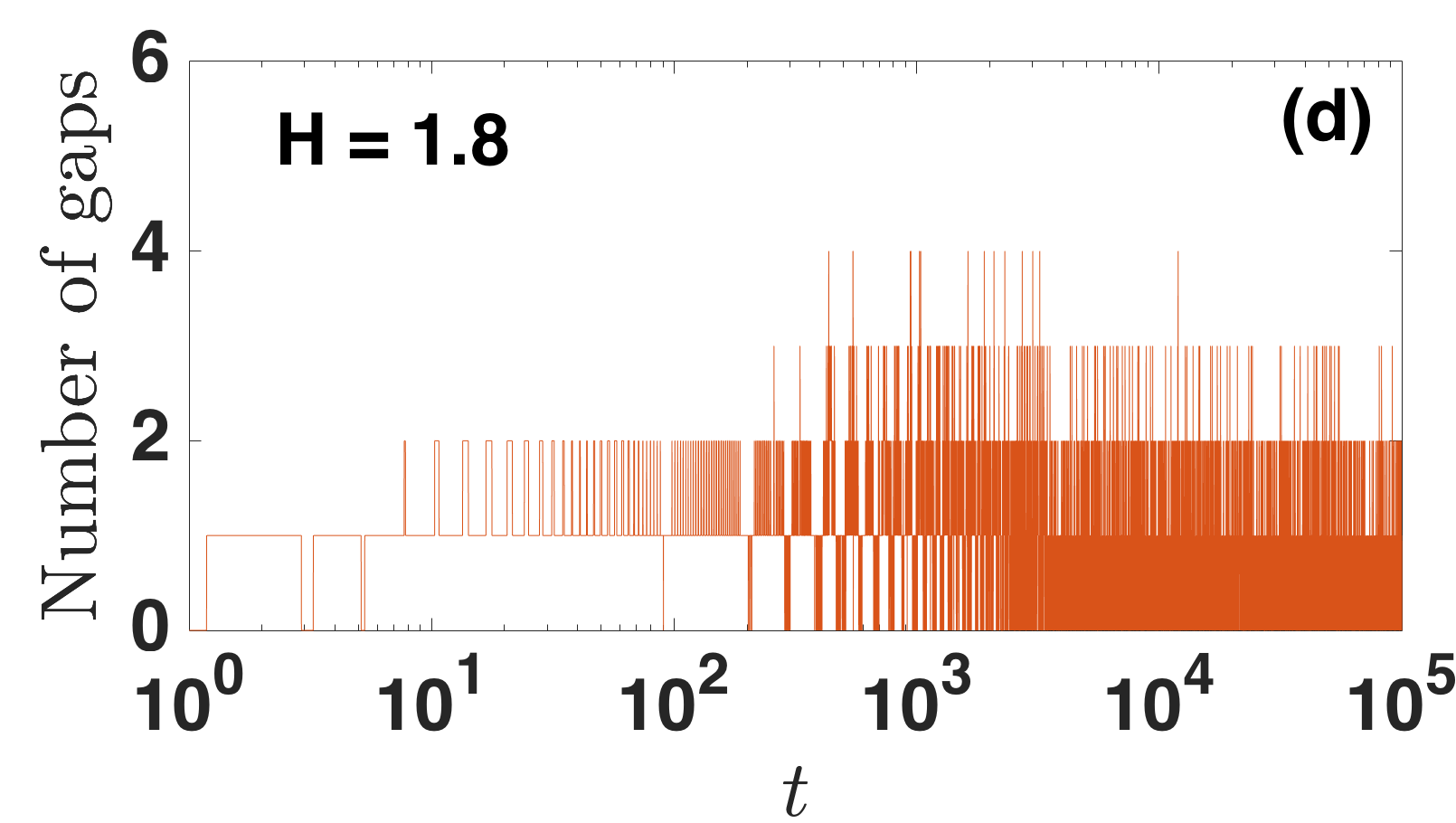} 
\caption{ The spatiotemporal evolution of the gaps in the Hertzian model for energies $H = 0.5$ (a) and  $H = 1.8$ (b). The yellow (lighter) color
corresponds to the lattice points where  $(u_n(t)-u_{n-1}(t))>\delta_n$. The instantaneous total number of gaps for the Hertzian model for energies $H= 0.5$ (c) and $H= 1.8$ (d).}
\label{gaps}
\end{figure*}
\begin{figure*}
 \includegraphics[width=4.2cm]{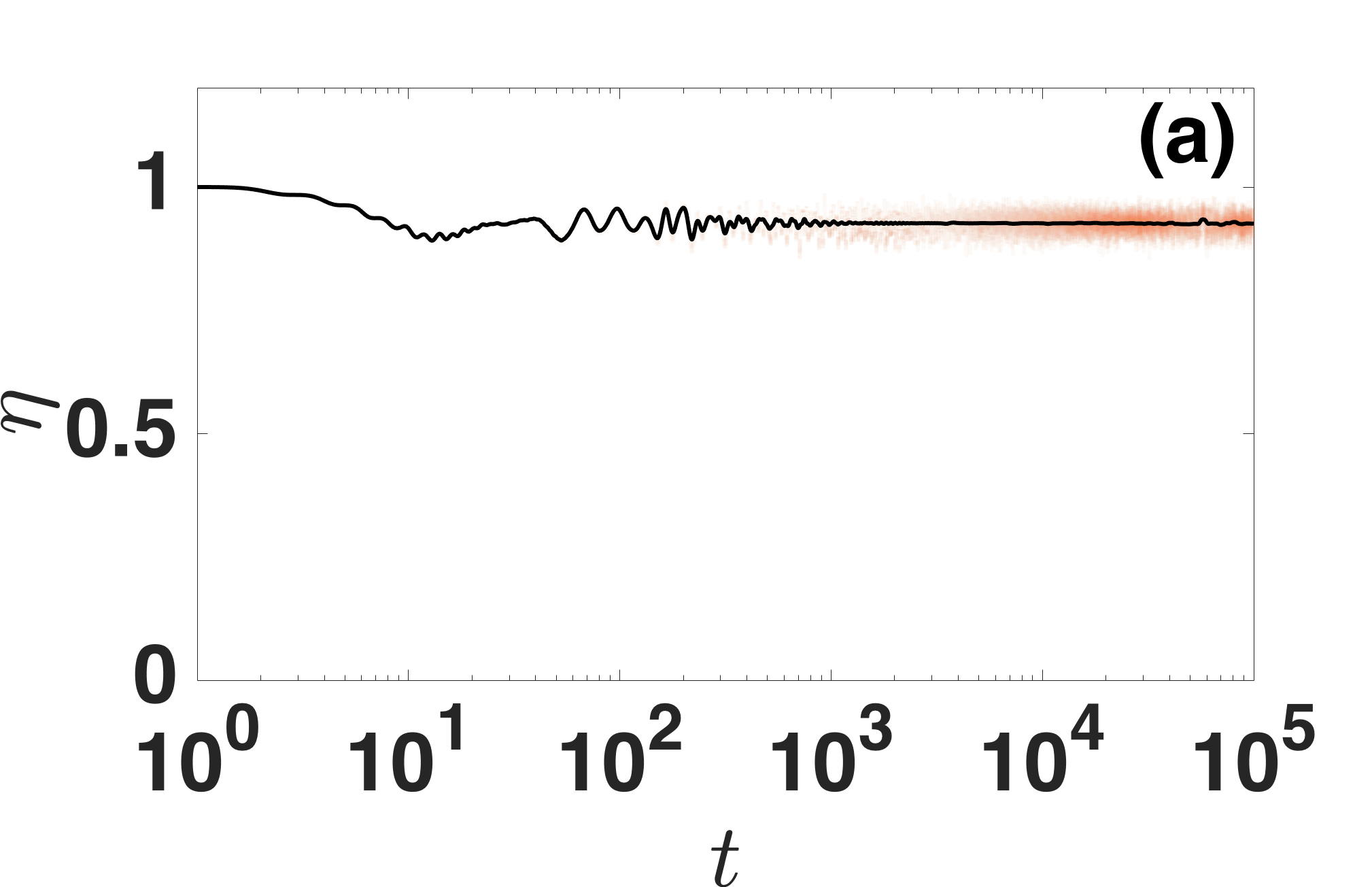}
 \includegraphics[width=4.2cm]{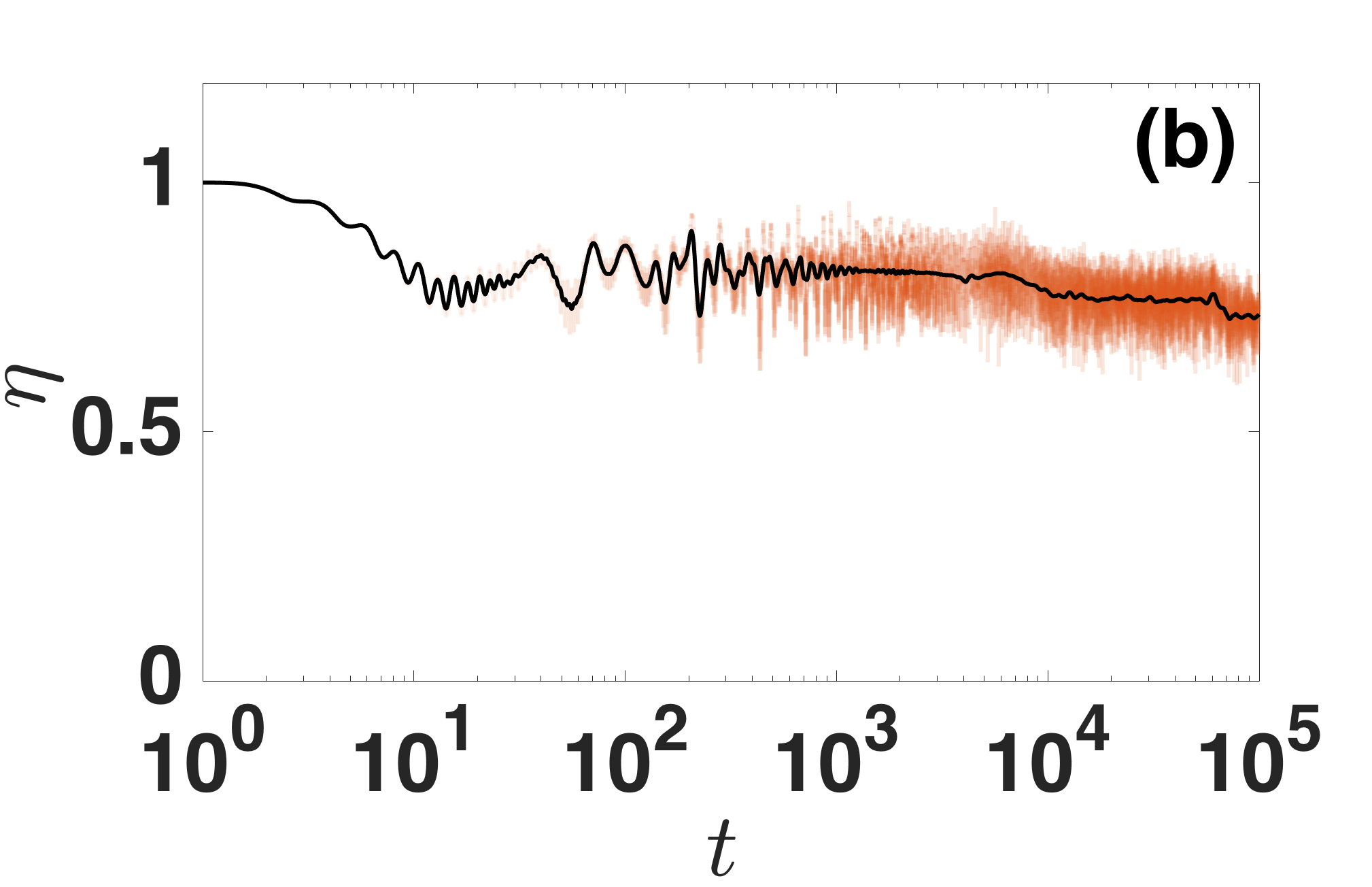}
 \includegraphics[width=4.2cm]{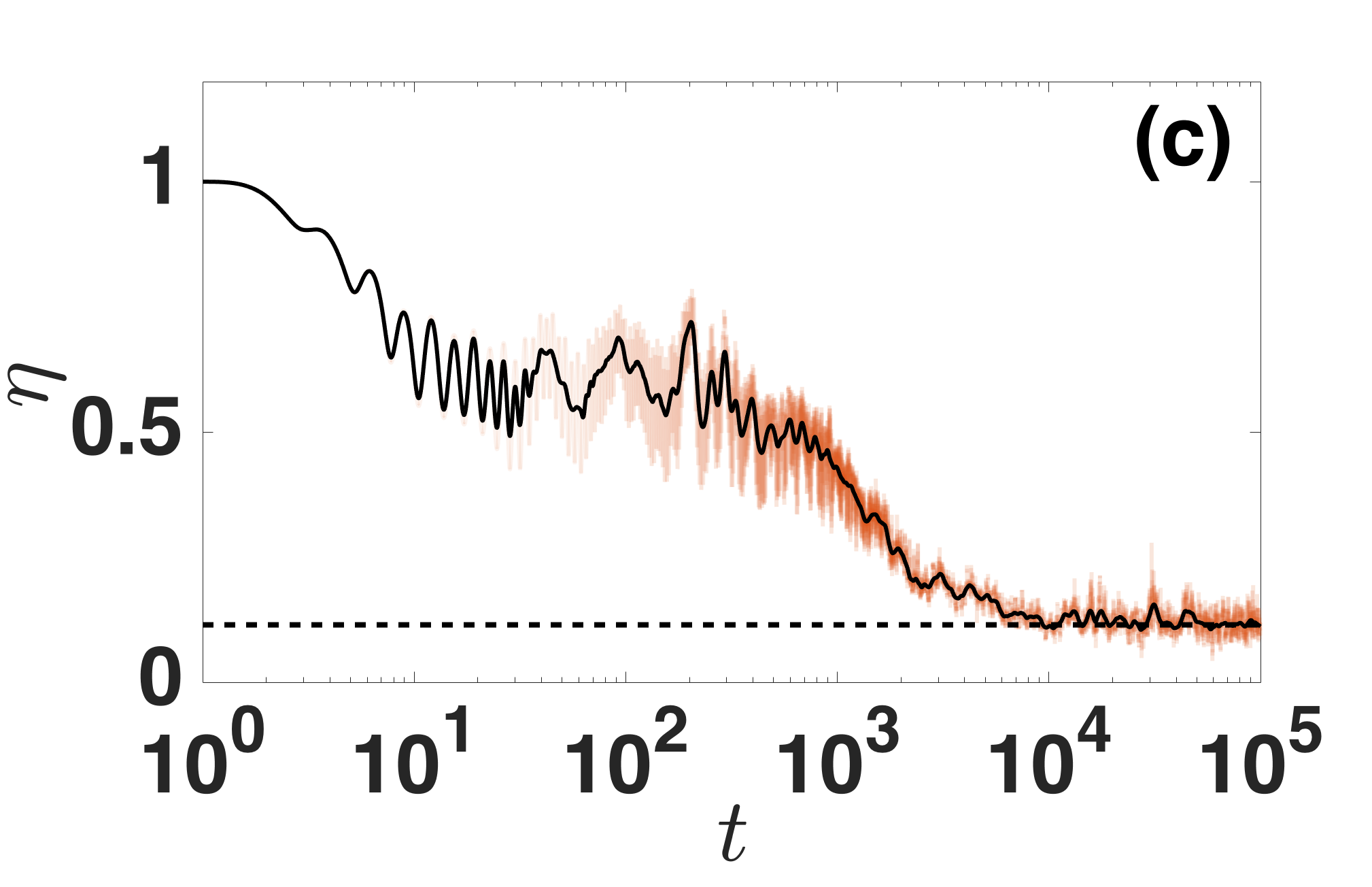}
 \includegraphics[width=4.2cm]{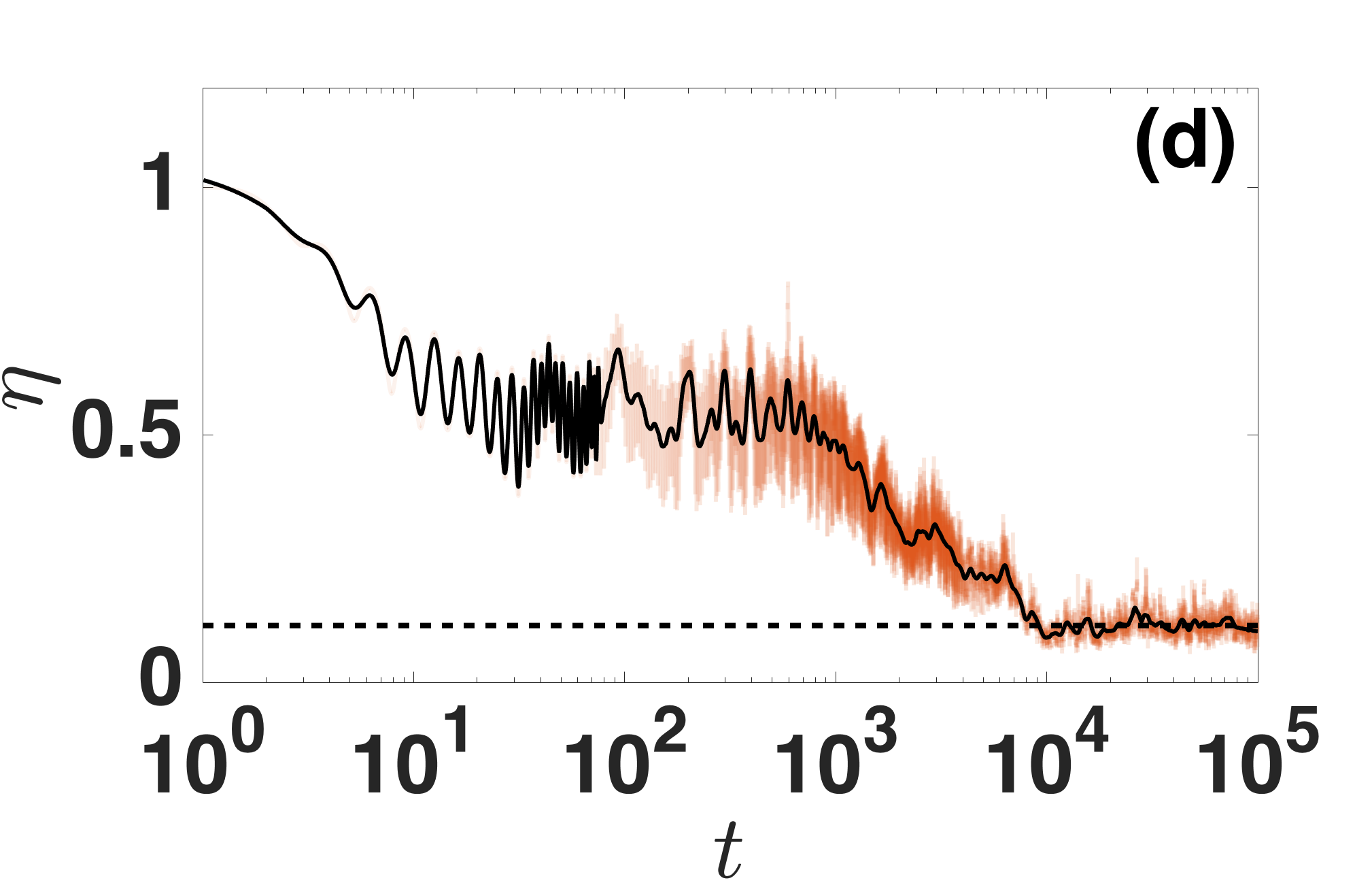}
\\
 \includegraphics[width=4.2cm]{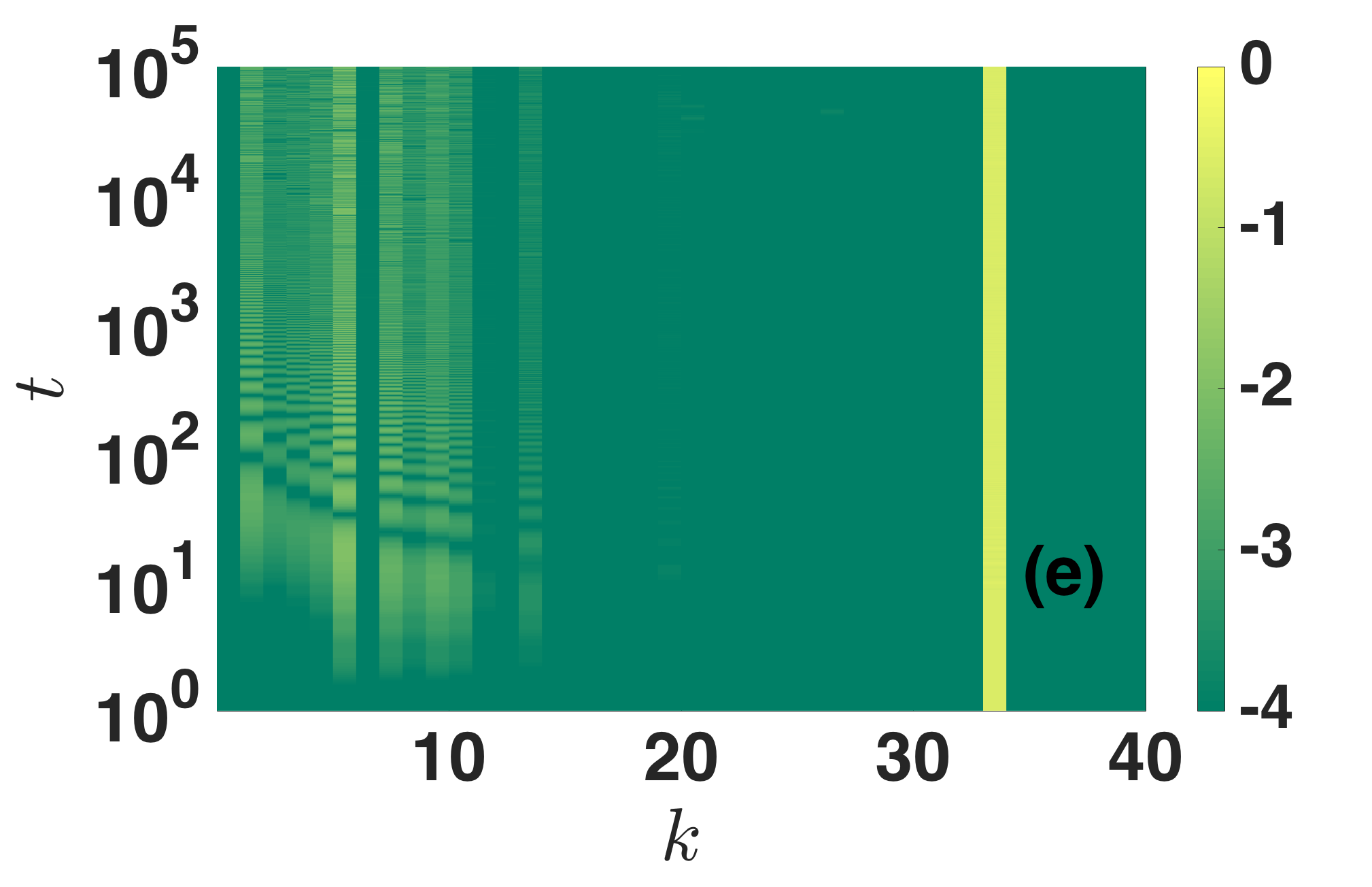}
\includegraphics[width=4.2cm]{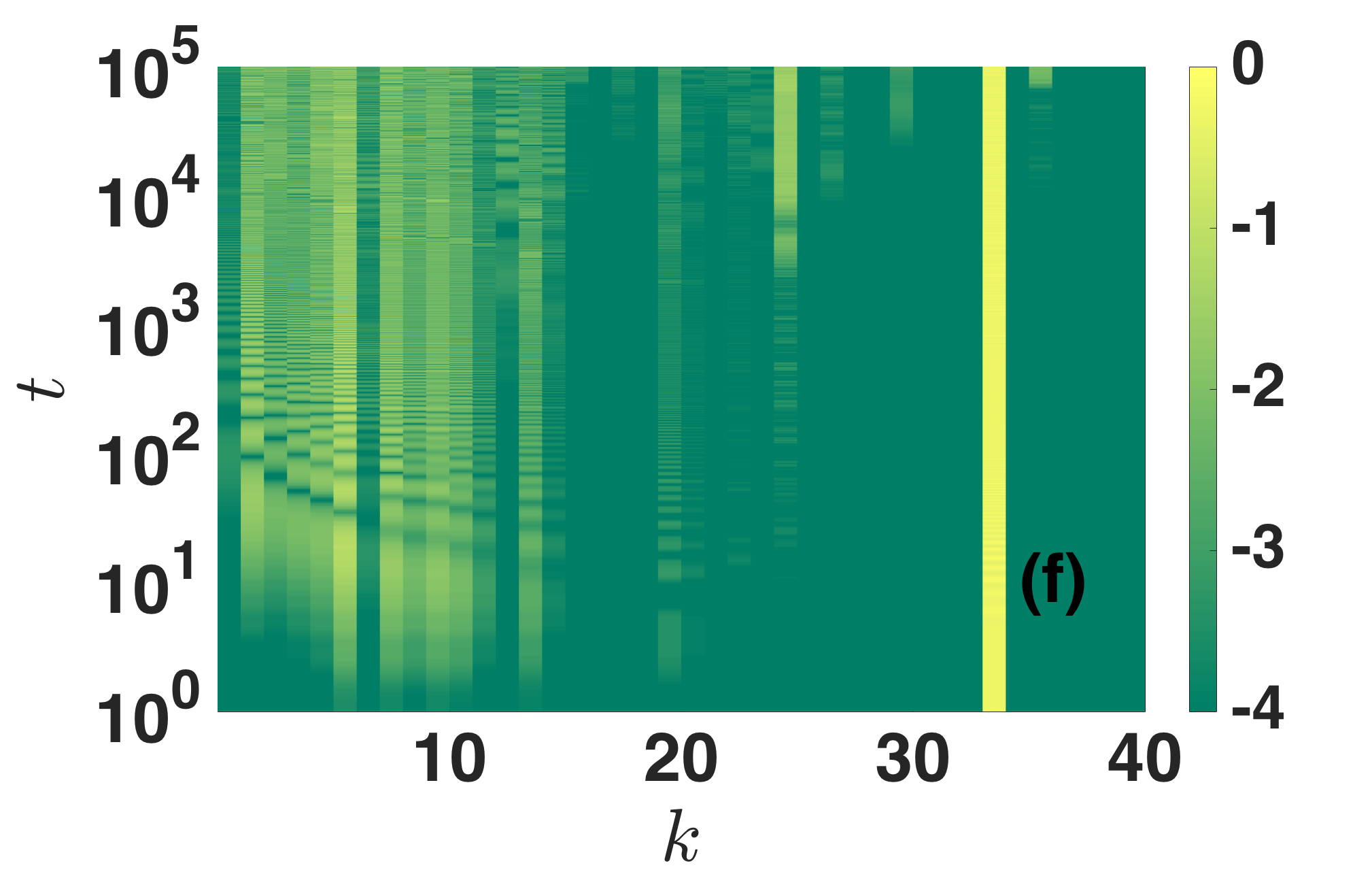} 
\includegraphics[width=4.2cm]{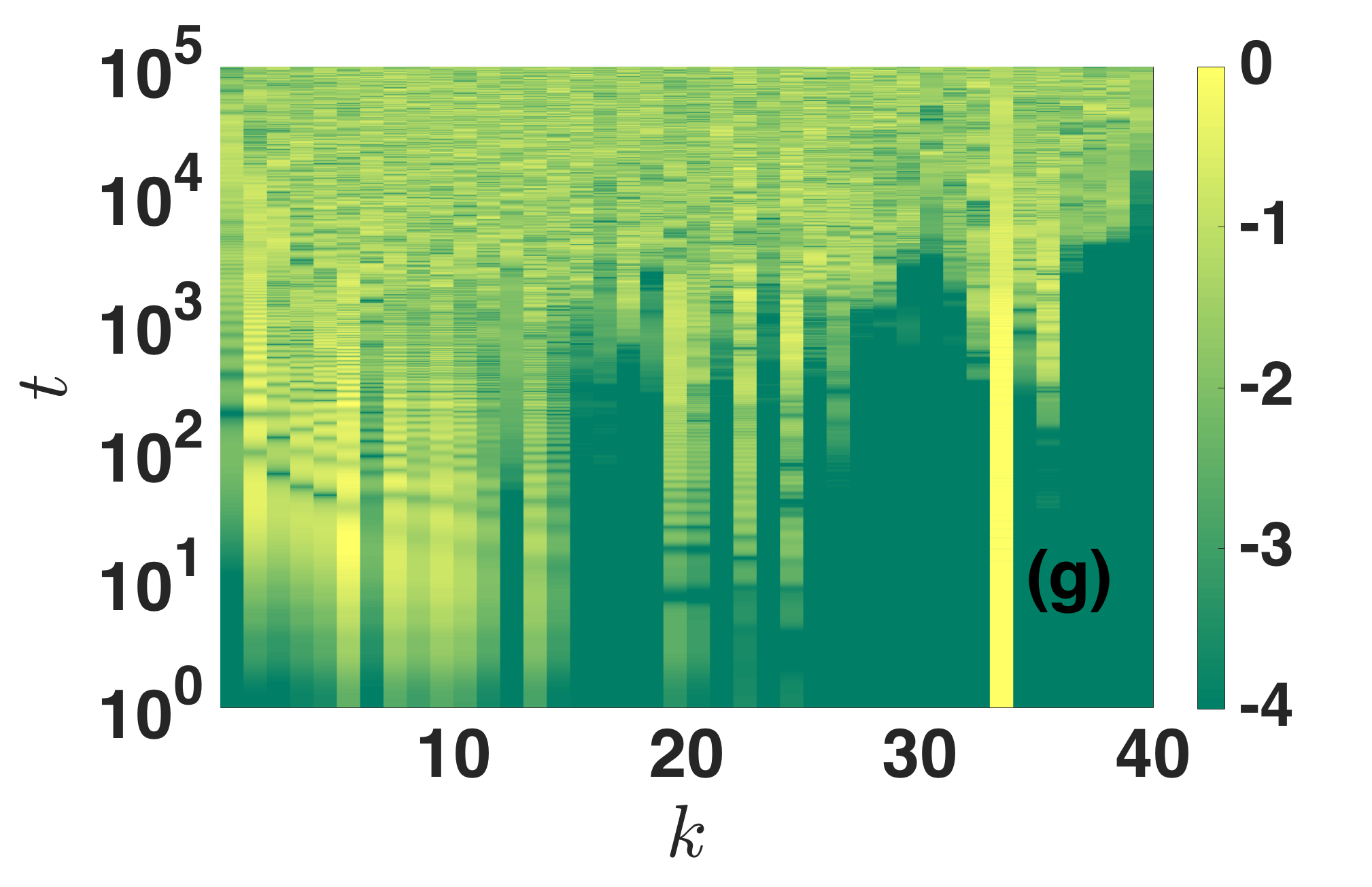}
\includegraphics[width=4.2cm]{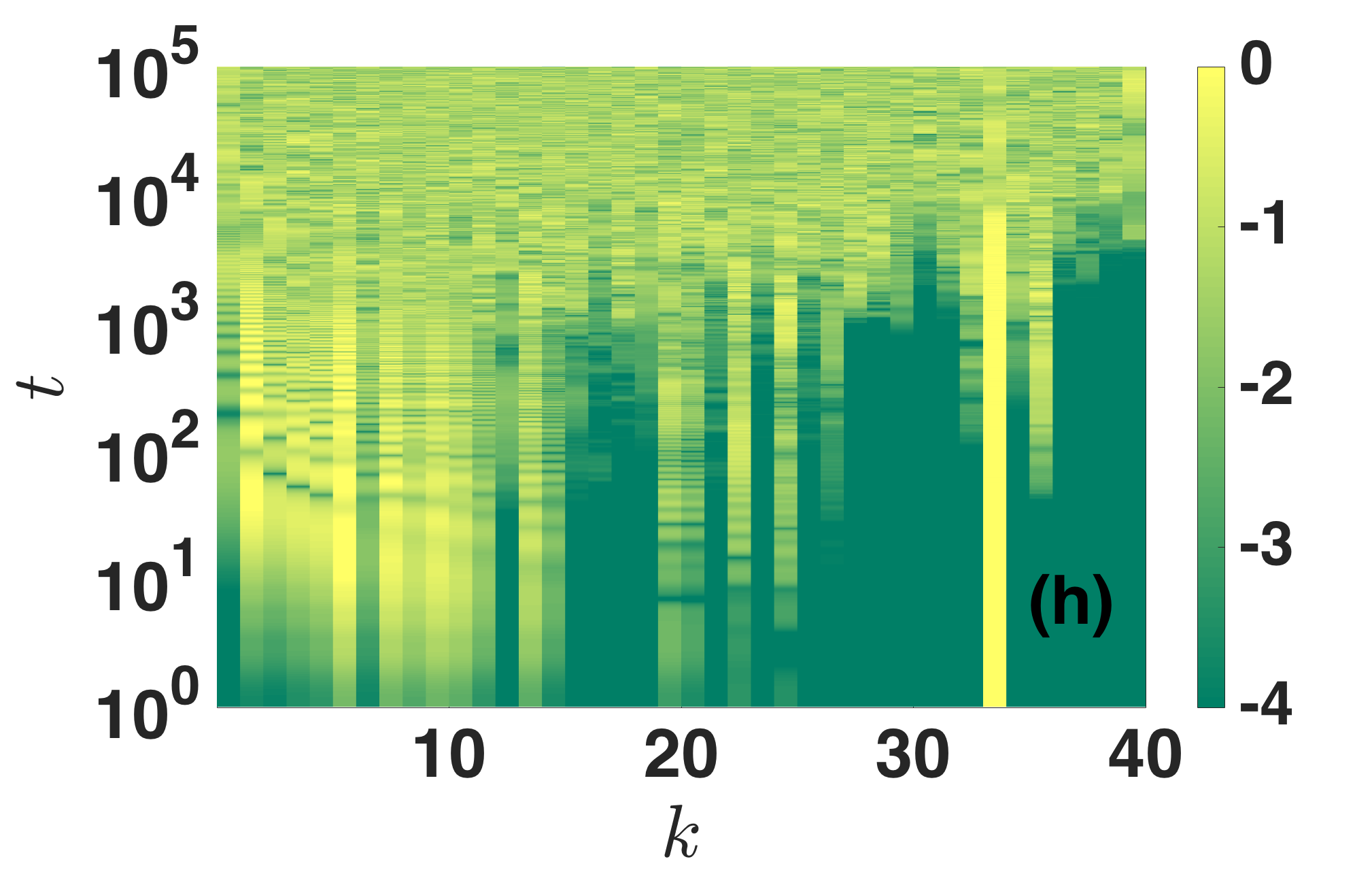}
\caption{ Top row: The time evolution of the normalized spectral entropy  $\eta (t) $ for the Hertzian model. 
The dashed horizontal line in panels (c) and (d) show the mean value $\langle \eta\rangle$ given by Eq.~(\ref{entranal}).
 Bottom row: The evolution of the weighted harmonic energy of eigenmodes as a function of time. The modes are sorted by increasing frequency
[c.f. Fig.~\ref{stats1}(b)]. The values of the energy are $ H = 0.25$ (a)-(e),  $ H = 0.5$ (b)-(f), $ H = 1.8$ (c)-(g), and $ H = 3$ (d)-(h). The 
color bars on the right sides of panels (e)-(h) are in logarithmic scale.}
\label{entropies1}
\end{figure*}
The dynamics of the DVD shown in Fig.~\ref{dynms29} (g) appear to localize around site $n=21$ up to  $t\approx 2\times 10^4$. During this time
$\Lambda(t)$ only slightly deviates from the regular orbit slope  shown in  Fig.~\ref{dynms29} (h). Further on, the mean position of the DVD departs from $n=21$ and is oscillating around the left part of the lattice. This suggests that although the wave-packet is still largely localized around site $ n =21 $, other lattice sites become quite chaotic and are able to drive the DVD away from the center.

To our surprise, by performing simulations at higher energies for the FPUT lattice, instead of an energy threshold above which all initial excitations lead 
to delocalization of the wave-packet, we observe alternations between localized and extended final states. Nevertheless, regarding the chaoticity of the FPUT system, we found
that for high energies, $H>2.8$, the system is always chaotic irrespectively of the localized or delocalized nature of the wave-packet.
To better visualize this alternate behavior of the FPUT, we show another example for $H=8.7381$ in the bottom row of Fig.~\ref{dynms29} which exhibits a delocalized and chaotic final energy profile. For this energy, the system behaves qualitatively the same as in the first row with $H=2.9$.

\subsection{Role of the non-smooth nonlinearity and energy equipartition}

\begin{figure*}
\includegraphics[width=4.2cm]{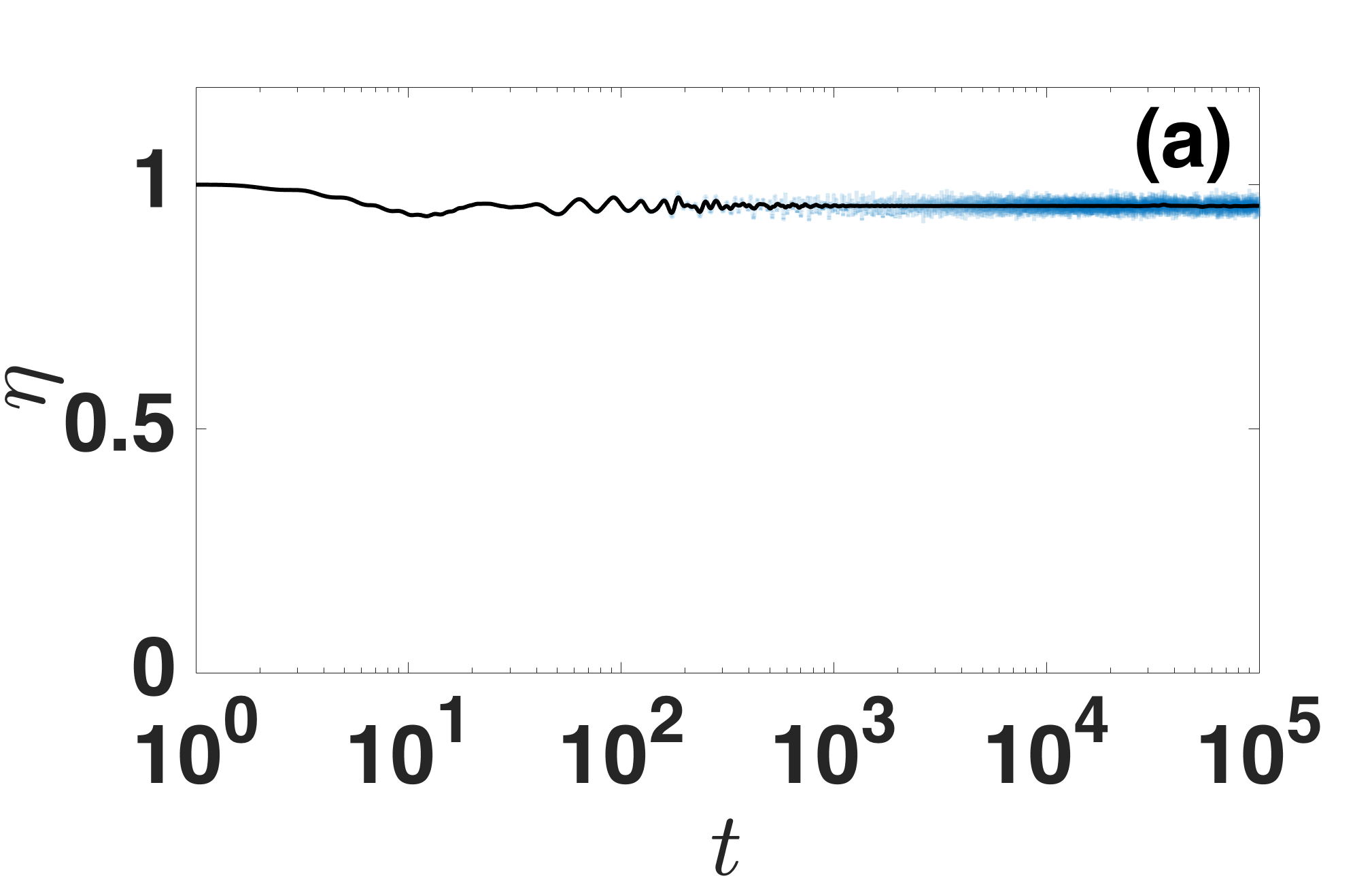}
\includegraphics[width=4.2cm]{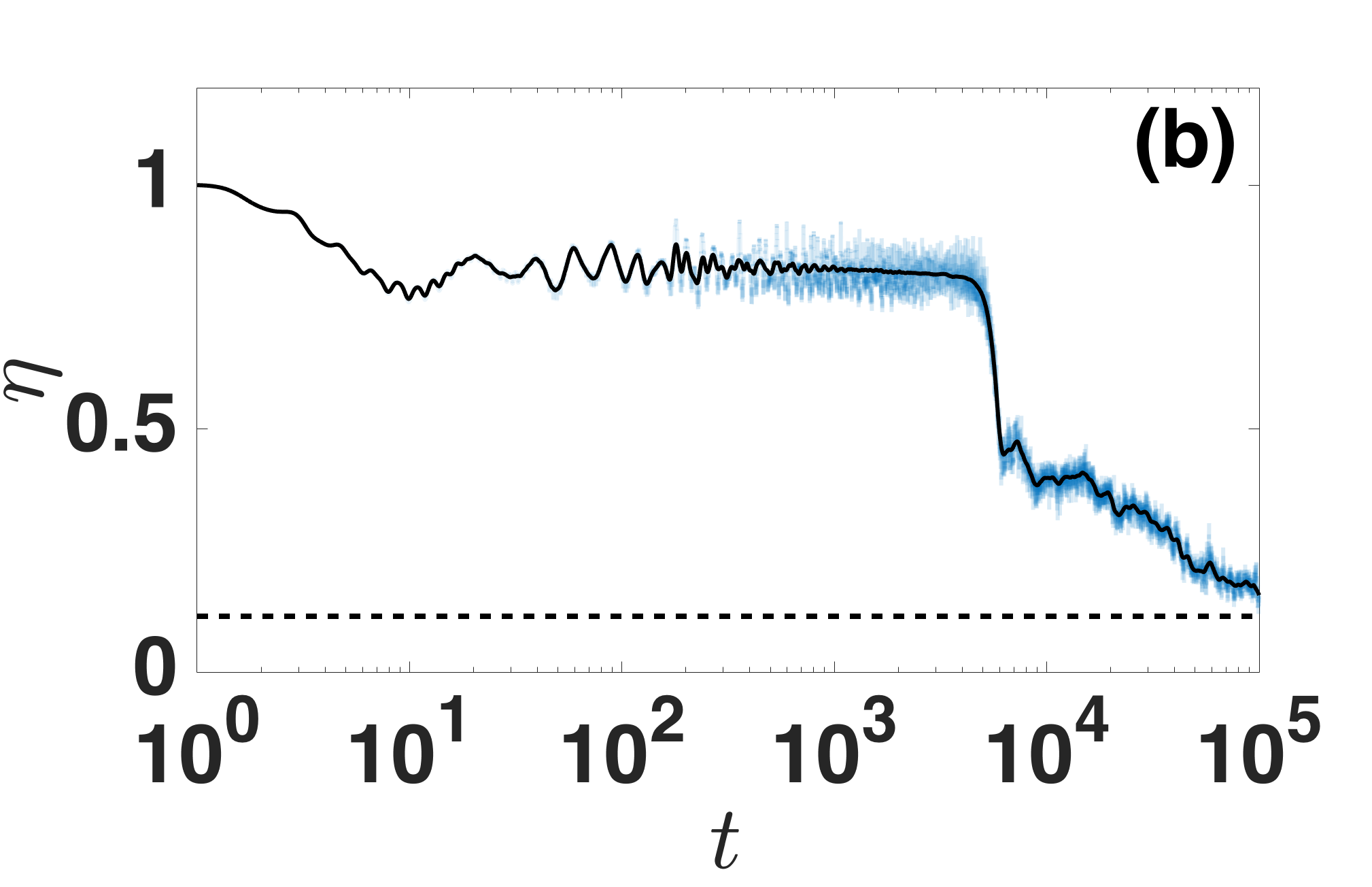}
\includegraphics[width=4.2cm]{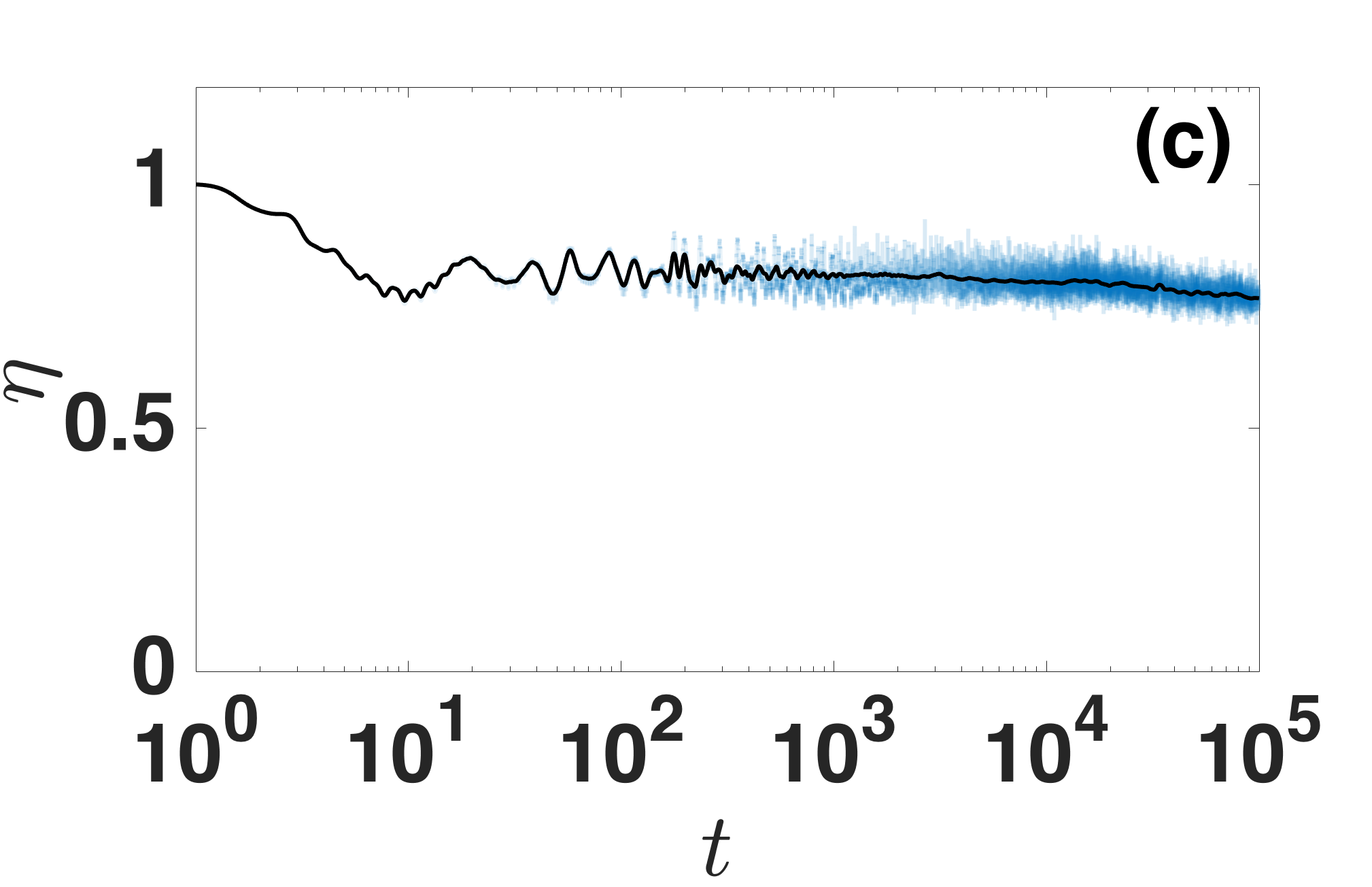}
\includegraphics[width=4.2cm]{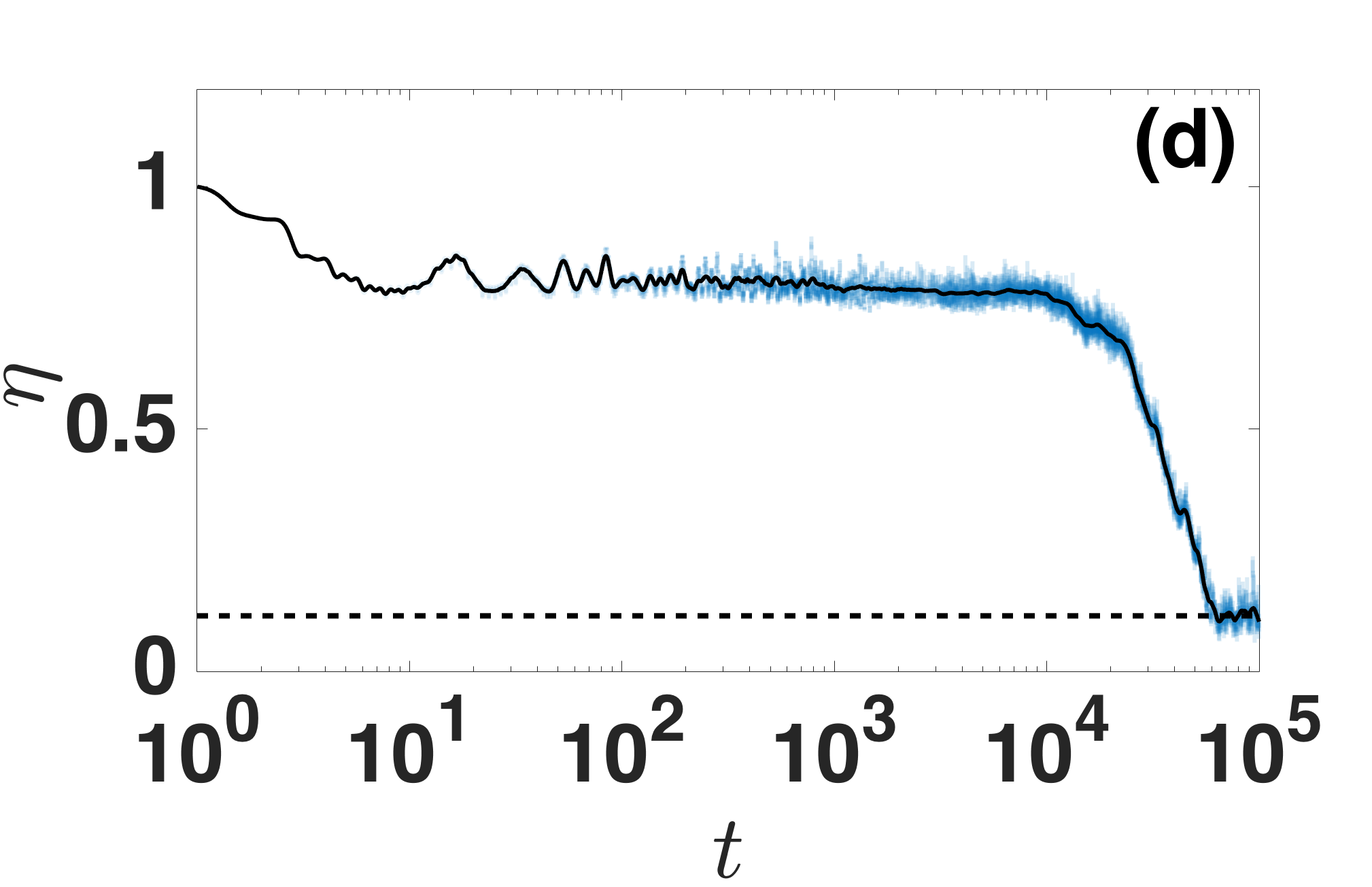}
\\
\includegraphics[width=4.2cm]{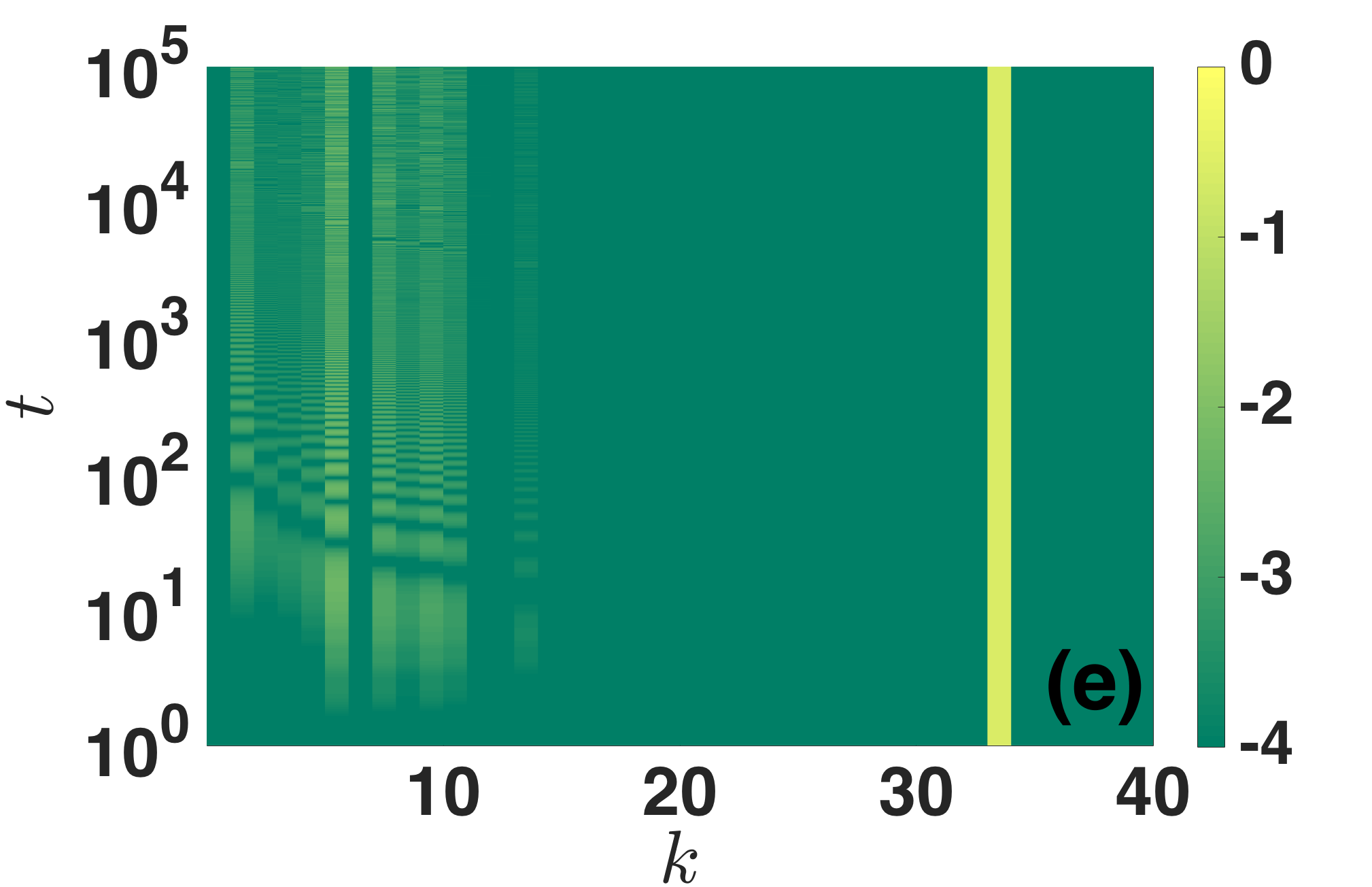}
\includegraphics[width=4.2cm]{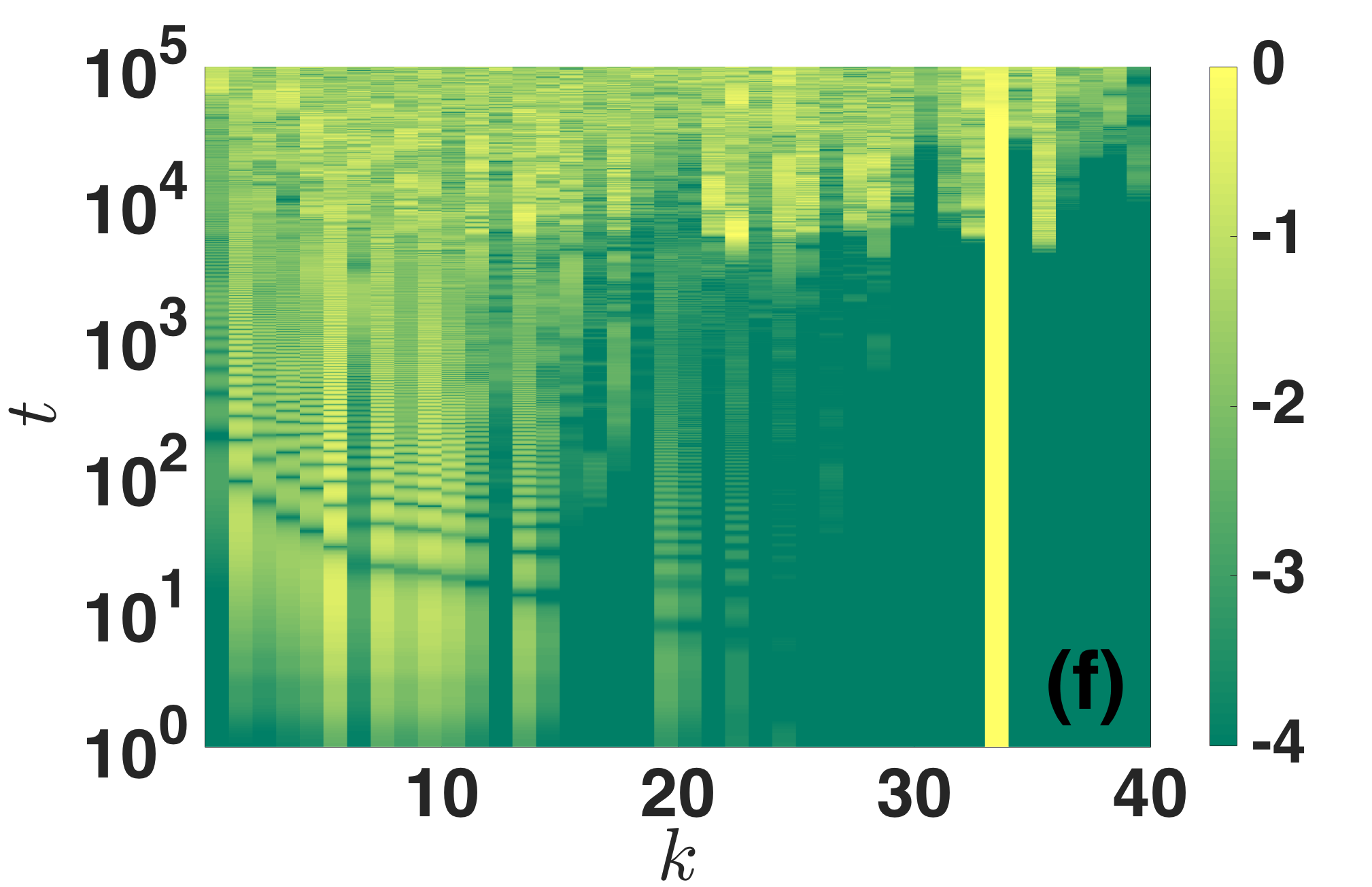}
\includegraphics[width=4.2cm]{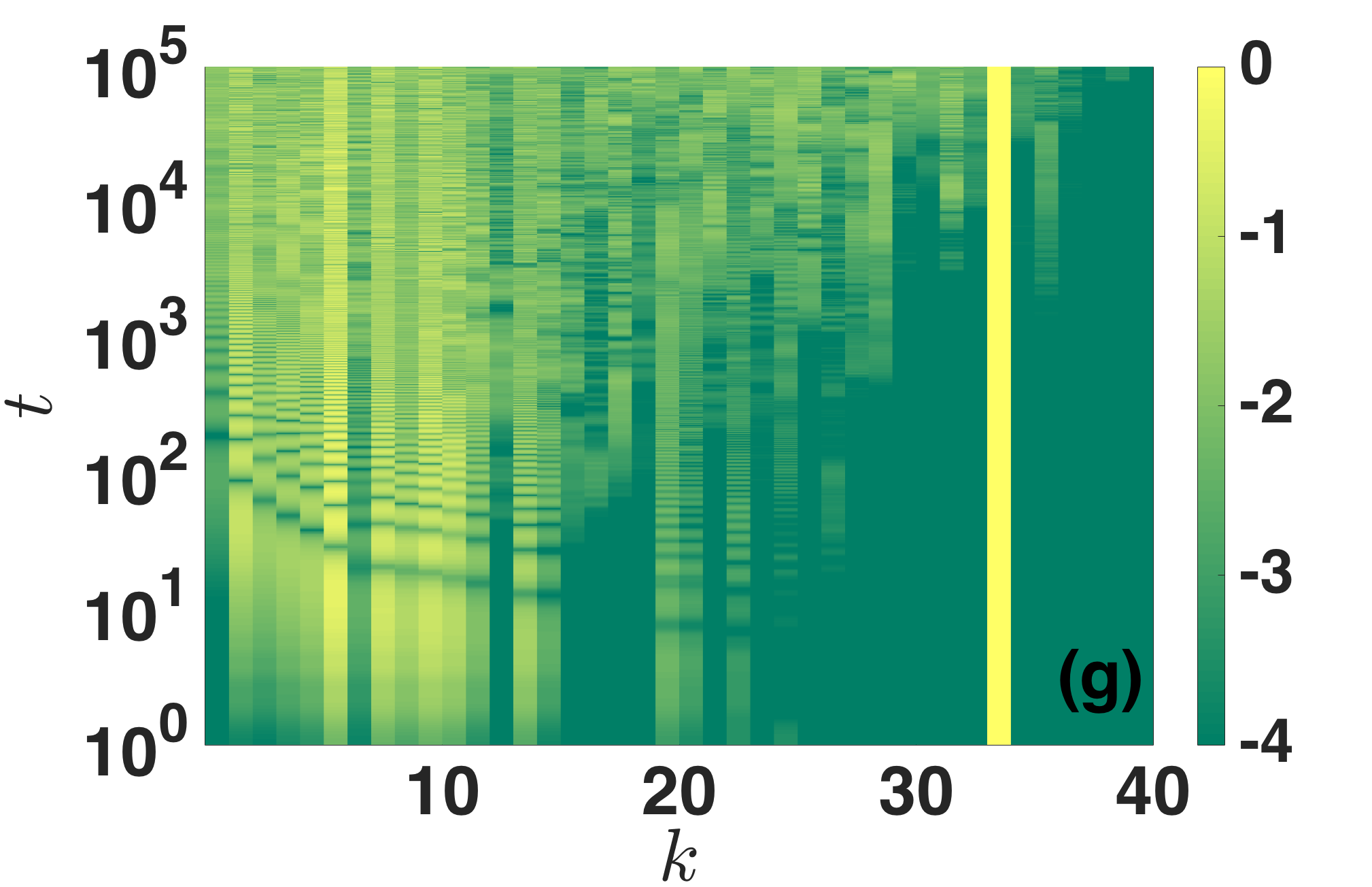}
\includegraphics[width=4.2cm]{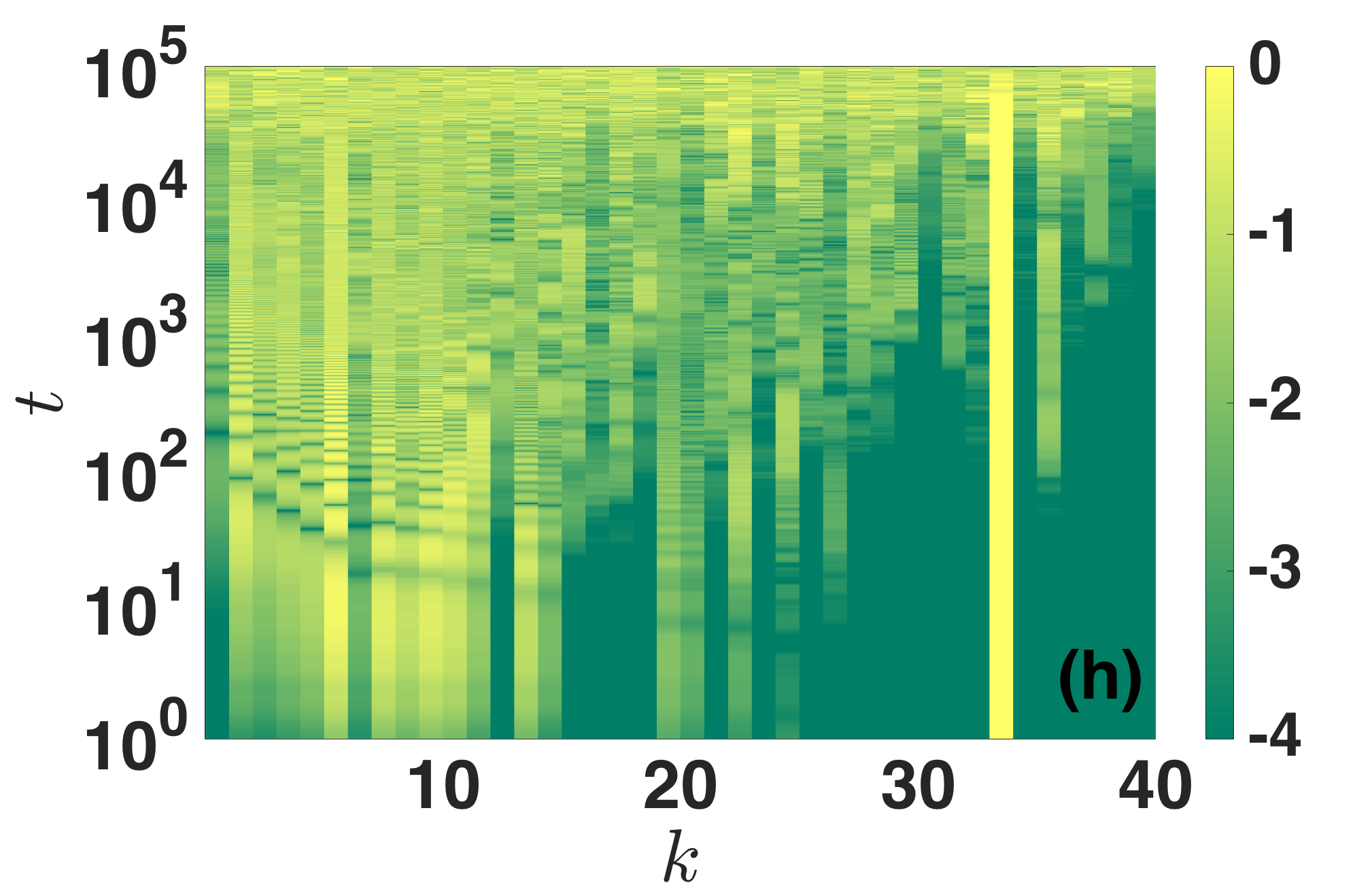}
\caption{Same as in Fig.~\ref{entropies1} but for the FPUT model. 
The dashed horizontal line in panels (b) and (d) show the mean value $\langle \eta\rangle$ given by Eq.~(\ref{entranal}).
The values of the energy in this case are $ H = 0.25$ (a)-(e),  $ H = 2.9$ (b)-(f), $ H = 4$ (c)-(g), and $ H = 8.7381$ (d)-(h).}
\label{entropies2}
\end{figure*}

In order to further track down the mechanisms responsible for the different behaviours between the two models we 
monitor the appearance of the non-smooth nonlinearity [i.e. whenever $(u_n-u_{n-1})>\delta_n$]  for the 
Hertzian model, or in other words the appearance of gaps. Fig.~\ref{gaps}(a), shows the position of gaps for the case $H=0.5$, which 
corresponds to the
panels of the first row of Fig.~\ref{dynms05a}. We clearly see that on the left and right side of site $n=21$ a gap often opens during the 
system's evolution triggering the appearance of the non-smooth nonlinearity. At this energy no more than one gap is open at any 
instant as observed in Fig.~\ref{gaps}(c) where the total number of gaps as a function of time is plotted.
Since for this energy, the dynamics of both the Hertzian and FPUT models is equivalent, and knowing that the Hertzian model appears 
to be chaotic, we identify the non-smooth nonlinearity around $n=21$ as the ingredient which induces chaos for the Hertzian model.

For  the energy $H=1.8$ shown in  Fig.~\ref{gaps}(b), we find that more gaps start to open``moving'' away from site $n=21$, covering eventually the whole lattice. In fact, for the energy region $0.5\lesssim H \lesssim1.8$ the wave-packet starts to delocalize (as quantified by $ P $) at the same time that additional gaps start to move away from site  $n=21$. For example in the case of $H=1.8$ shown in Fig.~\ref{gaps}(b), this happens around $t\approx 3\times10^2$ which is the same time that $P$ [see Fig.~\ref{dynms05a}(b)] starts to increase and the wave-packet starts to delocalize. These results, indicate a direct connection between the \textit{spreading of gaps} within the lattice and the energy threshold beyond which the Hertzian model always traverses to delocalized and extended chaos. 

To complete the comparison between the two models
we also calculate the so called ``spectral entropy'' ~\cite{PRAruffo} by monitoring the corresponding normal modes. 
We write the weighted harmonic energy of the $k$th mode as $ v_{k}  =  E_{k} /  \sum_{k=1}^{N}  E_{k} $ where $ E_{k} $ is the $k$th mode's energy. We thus obtain the spectral entropy at time $t$ as:
\begin{equation}
 S (t) = -\sum_{k=1}^{N} v_k(t) \ln v_k(t) .
\end{equation}
with $0<S\leq S_{max}=\ln N$.
It is however more convenient to use the normalized spectral entropy $ \eta (t) $ which can be written as,
\begin{equation}
\eta (t) = \frac{ S(t) - S_{max} }{ S(0) - S_{max} }.
\end{equation}
The value of $\eta$ is normalized such that $ 0 \leq \eta  \leq 1 $.
With this normalization, when $\eta$ remains close to one the dynamics does not substantially deviate from the initially
excited modes. On the other hand as more modes are excited, $\eta$ decreases towards zero. For a system at equipartition, a theoretical prediction
for the mean entropy $\langle\eta\rangle$ exists, which assumes that the modes at equipartition follow a Gibbs distribution when the nonlinearity is weak. 
The analytical form of the mean entropy $ \langle \eta\rangle$ is given by~\cite{Goedde, danieli}
\begin{equation}
\langle \eta\rangle=\frac{1-C}{\ln N - S(0)}
\label{entranal}
\end{equation}
with  $C\approx 0.5772$ being the Euler constant.

In Fig.~\ref{entropies1} we plot the time evolution of $\eta$ and of the normal modes for different values of the energy $H$.
As shown in Fig.~\ref{entropies1}(a),  for $H=0.25$ where the dynamics for both models is localized,
the normalized entropy initially has a value of $\eta=1$ and only slightly decreases from that value.
This indicates that the dynamics is dominated by the single mode initially excited along with
some weakly excited low frequency modes. 
This is also very clear in Fig.~\ref{entropies1}(e) where the time 
evolution of the weighted modes is shown. Initially only mode $34$ is visible, and after some brief transient phase, a set 
of extended (low frequency modes) are slightly excited. In fact after $t\approx 10^2$ the amplitude of each mode 
remains approximately constant and so does the time evolution of the normalized entropy $\eta(t)$.

Similar behavior for the Hertzian model is observed at $H=0.5$ [Figs.~\ref{entropies1}(b)-(f)], although in this 
case $\eta(t)$  reduces its value at different time instants. By closely inspecting panel (f) we 
see that indeed  around $t\approx 8\times10^3$ and $t\approx 8\times10^4$ new modes appear to kick in. 
For the two examples with $H\geq 1.8$ shown in panels (c),(g) and (d),(h) of Fig.~\ref{entropies1} the system is 
driven closer to equipartition. The entropy $\eta(t)$ features a plateau at a value around $\eta\approx 0.5$ and then
decreases into a minimum value. The horizontal dashed lines in Figs.~\ref{entropies1}(c) and (d) indicate the value of the mean entropy at equipartition
as given by Eq.~(\ref{entranal}). The asymptotic value of $\eta(t)$ approaches the theoretically predicted value of
 $\langle\eta\rangle$ with $H \geq 1.8$  as indicated in panels (c) and (d).
The fact that the final stages of these simulations are close to an equipartition state is also supported by the mode 
energy distribution which clearly shows that at the last decade all modes appear to participate in the dynamics.

The corresponding results for the FPUT model are shown in  Fig.~\ref{entropies2}.
As expected for $H=0.25$ [panels (a) and (e)] the behavior is the same as for the Hertzian model: $\eta(t)$ 
saturates to a finite value close to 1 and a dominant mode along with some low frequency modes are present.
For a much higher energy excitation of $H=2.9$ shown in Figs.~\ref{entropies2}(b) and (f), from the early stages of the
evolution more modes are excited and the entropy exhibits a plateau at  $\eta\approx 0.7$. Note that such a plateau 
is well known and studied in homogeneous FPUT chains and is associated with a metastable phase~\cite{danieli}. Beyond this point
the entropy abruptly falls at $t\approx 5\times 10^3$ and finally reaches a minimum value which is found to be close
to the analytical result for equipartition given by Eq.~(\ref{entranal}). As shown in Fig.~\ref{entropies2} (f) this is 
associated with the excitation of almost all linear modes.

For a larger initial energy $H=4$, i.e. the case presented in the second row of Fig.~\ref{dynms29}, the dynamics of $\eta$ is quite surprising. As shown in 
Fig.~\ref{entropies2}(c) the entropy saturates for most of the evolution around a relatively large
value $\eta\approx 0.82$. For the last two decades it starts to decrease, but with a very small slope.
This is unexpected (also in accordance to the homogeneous FPUT studies e.g. Ref.~\cite{danieli}) 
since for higher energy excitations we anticipate to have a shorter plateau (than the one for $H=0.25$) and the system to be driven faster 
towards equipartition. However, here the dynamics suggests that the contribution of modes
other than mode $34$ remains weak. This is also seen in Fig.~\ref{entropies2}(g) where not all modes have been excited
 at the end of the simulation, and in particular the highest frequency ones are still ``mute''. However, it is expected,  that for larger timescales the system will reach equipartition, and $\eta$ will eventually drop.
 
To highlight the alternate behavior found for the disorder FPUT model, in Fig.~\ref{entropies2}(d) we show the entropy for an 
even higher energy excitation of $H=8.7381$ which corresponds to the results presented in the third row of Fig.~\ref{dynms29}. Similarly to the case of $H=2.9$ the entropy saturates for a long time interval at a value $\eta\approx 0.8$.
Then at $t\approx 10^4$, $\eta$ starts to drop and at the end of the simulation reaches a minimal value well captured by the analytical
prediction of Eq.~(\ref{entranal}).
Accordingly in Fig.~\ref{entropies2} (h) we observe that as time increases more modes participate in the dynamics, 
and at the final stages of the simulation all modes are present.

\section{Summary and Conclusions}
In this work, we numerically studied energy localization/delocalization and the chaoticity of two one-dimensional disorder models:
the Hertzian model featuring a non-smooth nonlinearity, and the FPUT model. By choosing a certain strong disorder 
realization we focused on the delocalization of an initially excited localized (almost single site) mode.
We showed that for sufficiently small energies the two models behave quantitatively similar, with excitations remaining localized 
and non chaotic, at least for the time scales of our simulations. For larger energy values, a transient energy region is found for which 
the Hertzian model exhibits localized but chaotic behavior of the wave-packet. After an energy threshold, associated to the spreading of gaps in the lattice, the Hertzian model evolves into an equipartition, chaotic state independent of the particular value of the initial energy.

On the other hand, wave-packet delocalization and chaos emerge for higher energies for the FPUT model.  
We find strong numerical evidences that this difference is attributed to the non-smooth nonlinearity which is present
in the Hertzian model. Furthermore  for higher energy values the FPUT system shows an alternating behaviour between
chaotic localized and chaotic extended dynamics lacking a particular threshold beyond which equipartition is always reached.

Our results provide further insights into the chaotic dynamics of strongly disorder chains. Using additional chaos indicating tools 
such as the deviation vector densities, we are able to clearly separate localized (in space) from extended chaotic behavior. In addition we show that non-smooth nonlinearities can not only induce the destruction of  Anderson localization but also
they provide a mechanism to drive the system into equipartition for lower energies than for the FPUT model. 
Our results pave the way for probing the interplay between disorder and nonlinearity and the resultant effect on the system's equipartition time scales from a statistical 
point of view.

\section*{acknowledgments}
Ch.~S.~was supported by the National Research Foundation of South Africa (Incentive Funding for Rated Researchers, IFFR and Competitive Programme for Rated Researchers, CPRR) and thanks LAUM for its hospitality during his visits when part of this work was carried out. We also thank the Center for High Performance Computing (\url{https://www.chpc.ac.za}) for providing computational resources for  performing  part of this paper's  computations.

\end{document}